\documentclass[pdflatex,sn-mathphys-num]{sn-jnl}% Math and Physical Sciences Numbered Reference Style
\usepackage{adjustbox}
\usepackage{graphicx}%
\usepackage{multirow}%
\usepackage{amsmath,amssymb,amsfonts}%
\usepackage{amsthm}%
\usepackage{mathrsfs}%
\usepackage[title]{appendix}%
\usepackage{xcolor}%
\usepackage{textcomp}%
\usepackage{manyfoot}%
\usepackage{booktabs}%
\usepackage{algorithm}%
\usepackage{algorithmicx}%
\usepackage{algpseudocode}%
\usepackage{listings}%
\usepackage{array}
\newcolumntype{L}[1]{>{\raggedright\arraybackslash}m{#1}}
\newcolumntype{C}[1]{>{\centering\arraybackslash}m{#1}}
\usepackage{subcaption}
\usepackage{tikz}
\usetikzlibrary{patterns, arrows.meta, decorations.pathreplacing, calc}
\newif\ifrevisionmarkup
\revisionmarkupfalse
\definecolor{revcol}{RGB}{0,0,205}
\definecolor{runcol}{RGB}{214,90,0}
\definecolor{needcol}{RGB}{200,0,40}
\definecolor{rescol}{RGB}{125,0,170}
\ifrevisionmarkup
  \newcommand{\rev}[1]{\textcolor{revcol}{#1}}
  \newcommand{\revcolor}{\color{revcol}}
  \newcommand{\res}[1]{\textcolor{revcol}{#1}} % was rescol with provisional numbers
  \newcommand{\rescolor}{\color{revcol}} % was rescol with provisional numbers
  \newenvironment{revblock}{\par\color{revcol}}{\par}
  
  \newcommand{\runs}[1]{\textcolor{runcol}{\textbf{[Final runs: #1]}}}
  \newcommand{\need}[1]{\textcolor{needcol}{\textbf{[Need further input: #1]}}}
\else
  \newcommand{\rev}[1]{#1}
  \newcommand{\revcolor}{}
  \newcommand{\res}[1]{#1}
  \newcommand{\rescolor}{}
  \newenvironment{revblock}{\par}{\par}
  
  \newcommand{\runs}[1]{}
  \newcommand{\need}[1]{}
\fi
\theoremstyle{thmstyleone}%
\theoremstyle{thmstyletwo}%

\theoremstyle{thmstylethree}%

\begin{document}

\title{Shift or curtail? How much data-center flexibility is worth depends on the host power grid}


\author[1]{\fnm{Saroj} \sur{Khanal}}\email{skhanal8@jhu.edu}

\author[2]{\fnm{Geon} \sur{Roh}}\email{rohgeon1224@kentech.ac.kr}

\author[3]{\fnm{Boyu} \sur{Yao}}\email{byao3@jhu.edu}

\author[4]{\fnm{Abraham} \sur{Silverman}}\email{asilve39@jh.edu}

\author[5]{\fnm{Dennice} \sur{Gayme}}\email{dennice@jhu.edu}
\author[6]{\fnm{Charalambos} \sur{Konstantinou}}\email{charalambos.konstantinou@kaust.edu.sa}

\author[2]{\fnm{Jip} \sur{Kim}}\email{jipkim@kentech.ac.kr}

\author*[1,3]{\fnm{Yury} \sur{Dvorkin}}\email{ydvorki1@jhu.edu}

\affil*[1]{\orgdiv{Department of Electrical and Computer Engineering}, \orgname{Johns Hopkins University}, \orgaddress{\street{3400 North Charles Street}, \city{Baltimore}, \postcode{21218}, \state{MD}, \country{USA}}}

\affil[2]{\orgdiv{Department of Energy Engineering}, \orgname{Korea Institute of Energy Technology}, \orgaddress{\street{21 Kentech-gil}, \city{Naju-si}, \postcode{58330}, \country{Republic of Korea}}}

\affil*[3]{\orgdiv{Department of Civil and Systems Engineering}, \orgname{Johns Hopkins University}, \orgaddress{\street{3400 North Charles Street}, \city{Baltimore}, \postcode{21218}, \state{MD}, \country{USA}}}
\affil[4]{\orgdiv{Ralph O’Connor Sustainable Energy Institute}, \orgname{Johns Hopkins University}, \orgaddress{\street{3400 North Charles Street}, \city{Baltimore}, \postcode{21218}, \state{MD}, \country{USA}}}
\affil[5]{\orgdiv{Department of Mechanical Engineering}, \orgname{Johns Hopkins University}, \orgaddress{\street{3400 North Charles Street}, \city{Baltimore}, \postcode{21218}, \state{MD}, \country{USA}}}
\affil[6]{\orgdiv{Division of Computer, Electrical and Mathematical Science and Engineering}, \orgname{King Abdullah University of Science and Technology}, \orgaddress{\street{4700 King Abdullah University of Science and Technology}, \city{Thuwal}, \postcode{23955-6900}, \country{Saudi Arabia}}}

%% Nature Energy: abstract <= 150 words, single unreferenced paragraph %%
%%==================================%%

\abstract{Data-center growth risks overbuilding grid infrastructure and stranding capital. \rev{Flexible data centers, which shift or interrupt load, could avoid this. We classify data-center load into firm, shiftable and interruptible tiers and embed them in capacity-expansion models of  PJM and Korea. Under  the reference mix (30\% firm, 50\% shiftable, 20\% interruptible), this flexibility lowers system cost by 0.8\% in 2028 and 6.0\% in 2038 in PJM, and by 0.2\% and 2.5\% in Korea. By 2038, two thirds of the saving is capacity credit earned in stress hours, avoiding 44~GW of natural gas capacity  in PJM, and 6~GW of 4-hour batteries in Korea. Spatial shifting helps satisfy renewable obligations in PJM  and temporal shifting defers load into midday solar hours under Korea's carbon cap. Without a carbon limit, PJM's 2038 emissions increase 33\% as coal replaces avoided gas. Shadow prices and price variation computed without flexibility indicate which mechanism a grid will value.}}

 % Note: Some Springer Nature journals treat the keyword order as meaningful, with the first term carrying most weight for indexing



\maketitle

\section{Introduction}\label{sec1}

Electricity demand from data centers has moved from a marginal planning consideration to a leading driver of power system infrastructure investment \cite{shehabi20242024, norris2025rethinking}. Individual hyperscale campuses  request gigawatt-scale interconnection capacity, often concentrated in a few locations, and recent forecasts attribute most near-term peak-load growth to data centers \cite{shehabi20242024}. %These loads are also concentrated in a few locations. 
However, whether \rev{any particular} data-center project materializes is \rev{hard to predict}, with local opposition \cite{rauh2026local} and equipment supply chains among the reasons \cite{yao2026grid, boyu_npj_supply_chain}. Power system planners use capacity expansion models (CEM) to decide \textit{how much} new generation, storage and transmission infrastructure to build and \textit{where} to build it for a given target year. Rapid growth of data centers exacerbates a current-practice limitation of treating \rev{data-center load} as fixed, inelastic demand. Under this assumption, CEM outcomes are planned for a hypothetical peak demand, thus risking stranded investments whose costs are \rev{eventually} recovered from \rev{not only data-center but all} consumers through \rev{higher rates. Besides energy costs, the rates consumers pay include charges that recover investment, such as capacity payments, interconnection charges and, for a regulated utility, the cost of the capacity it builds. Stranded investment raises these charges.} \rev{Data-center load growth has also contributed to resource-adequacy risks \cite{pjm2027brarep, pjm2028brarep, brancucci2025flexible}, as it increases the peak demand that accredited  capacity  must supply. For instance, PJM's 2027/2028 and 2028/2029 capacity auctions cleared at the price cap and about 6.5 and 6.8~GW short of the reliability requirement \cite{pjm2027brarep, pjm2028brarep}}.

\rev{Operating data centers flexibly is a techno-economically viable mechanism to alleviate both affordability concerns and resource-adequacy risks. A flexible data center adjusts its power consumption in response to grid conditions during peak or other operationally constrained hours (\emph{stress hours}), which reduces the capacity that must be built to serve it. However, data-center flexibility is not costless, because expensive data-center hardware has an opportunity cost to its owner and to society, and is not a single product,  because it can be delivered through different physical mechanisms. In turn, these mechanisms affect flexibility value to the power grid. By \textit{mechanism} we mean the operational form of flexibility, not a market or commercial arrangement, which} can be exercised by shifting workload to later hours (\textit{\rev{temporal shifting}}), \rev{relocating} it to another data center (\textit{\rev{spatial shifting}}), or drawing less power from the grid\rev{, by pausing computation or by running on-site generation resources co-located with the data center (\textit{interruption}, also often called \textit{curtailment}).} %These operational levers reshape the load within the fixed planning year, and are distinct from the model's separate planning decisions of how much to build and where.
Hence, a central question is  \textit{how} data-center flexibility creates value for the grid and \rev{whether a mechanism that is most valuable on one grid is also the most valuable on another.} 

Data-center flexibility can be described from two perspectives. A data-center operator classifies computing tasks by whether they must run now, can wait, can run at another site, or can be dropped. \rev{The grid operator, however, only sees the power withdrawn at the point of interconnection and cannot differentiate, for example, whether a data center drew less power because it paused a training job or because it kept the job running and supplied it from on-site generation resources}. \rev{We therefore classify each data center's load into three \emph{tiers}, which are shares of its load defined by how their power withdrawal from the grid may change}. Data-center load is \rev{\textit{firm}} if it must be served \rev{constantly}, \rev{\textit{shiftable}} if it can be \rev{moved to later hours}  or to another eligible data center, and \rev{\textit{interruptible}} if the power withdrawal can be reduced \rev{nearly immediately,} either by stopping computations or by \rev{switching to} on-site \rev{generation resources (see Table~\ref{tab:modes})}. \rev{The shiftable and interruptible tiers are the \emph{flexible tiers}, and each} has its limits. \rev{For instance, the shiftable tier is limited by a make-up time for deferred data-center workloads and a shift cap for relocated ones}, and the interruptible tier \rev{is limited by the} depth, duration, ramp and recovery \rev{of each interruption}.

\begin{table}[!htbp]
\centering
\scriptsize
\caption{\textbf{Mapping data-center flexibility options onto grid-facing tiers
and mechanisms.} Operational options reported by the Electric Power Research
Institute (EPRI) Data Center Flexible Load Initiative (DCFlex) and related work
\cite{epri2025gridflexibility, epri2026powering}, each classified by the tier and
mechanism it produces under the grid-operator view. Many distinct options collapse
onto three tiers, because workload management and on-site supply shape a facility's
power withdrawal from the grid in the same ways. Entries marked (approx.) shape
power withdrawal in the same direction but are not represented explicitly in the
model.}
\label{tab:modes}
\setlength{\tabcolsep}{3pt}
\renewcommand{\arraystretch}{1.15}
\begin{tabular}{
>{\raggedright\arraybackslash}p{0.30\linewidth}
>{\raggedright\arraybackslash}p{0.13\linewidth}
>{\raggedright\arraybackslash}p{0.30\linewidth}
>{\raggedright\arraybackslash}p{0.20\linewidth}
}
\toprule
Operational option & Tier & Mechanism & Representative studies\\
(data-center view) & \multicolumn{2}{l}{(grid-operator view)} & \\
\midrule
\multicolumn{4}{@{}l}{\emph{Workload management}}\\
\cmidrule(r){1-4}
Defer batch or training jobs & \rev{Shiftable} & \rev{Temporal shifting} & \cite{papavasiliou2013large, liu2013data}\\
Pre-cool or use thermal storage & \rev{Shiftable} & \rev{Temporal shifting} (approx.) & \cite{epri2025gridflexibility}\\
Migrate jobs to another site & \rev{Shiftable} & \rev{Spatial shifting} & \cite{liu2011greening, qureshi2009cutting, radovanovic2022carbon, williams2026power}\\
Throttle or pause compute & Interruptible & \rev{Interruption} & \cite{colangelo2026ai, williams2026power, acun2026investigating}\\
Curtail under an interruptible tariff & Interruptible & \rev{Interruption} & \cite{chao1987priority, nerc2026assessment, acun2026investigating}\\
\addlinespace
\multicolumn{4}{@{}l}{\emph{On-site (co-located) supply}}\\
\cmidrule(r){1-4}
Discharge on-site battery or UPS & \rev{Shiftable} & \rev{Temporal shifting} (approx.) & \cite{epri2026powering}\\
Run on-site or backup generation & Interruptible & \rev{Interruption} & \cite{epri2026powering}\\
Draw on co-located generation & Interruptible & \rev{Interruption} & \cite{ferc2026largeload}\\
\addlinespace
\multicolumn{4}{@{}l}{\emph{No flexibility}}\\
\cmidrule(r){1-4}
Serve all workload as it arrives & Firm & None & \cite{ferc2026largeload, epri2026powering}\\
\bottomrule
\end{tabular}
\end{table}

Two trends make the data-center operational options  in Table~\ref{tab:modes} appealing for  integration with CEM. First, a growing share of data-center workloads is schedulable \cite{nerc2026assessment, brancucci2025flexible, epri2026powering, epri2025gridflexibility, norris2025rethinking}. Hardware measurements put the inherent power flexibility of representative training and inference workloads at 18--55\% of average consumption \cite{acun2026investigating}. The compute industry has also begun to price urgent and delay-tolerant workloads differently \cite{loher2026gemini}, in part to temporarily reduce consumption from the  power grid \rev{and to protect} latency-critical jobs \cite{williams2026power, acun2026investigating}. Thus, the EPRI DCFlex initiative classifies workloads by how far they can move in time or space, from real-time (and streaming) tasks that must be served immediately through interactive and batch tasks that can wait hours or run at another site, to training (and pure-batch) tasks that can be switched off within annual limits \cite{epri2025gridflexibility}. Second, the same change in power withdrawal from the power grid can be achieved  without altering the workload. On-site resources can reshape net power withdrawal \cite{epri2025gridflexibility, vaidhynathan2025vulcan} by reducing or shifting it in time. Defining {the data-center load tiers} by power withdrawal from the power grid therefore allows for assessing the value of data-center flexibility without committing to a specific delivery mechanism. 

The three \rev{mechanisms} in Table~\ref{tab:modes} \rev{map} to \rev{existing} demand-management practices. Shifting compute to later hours is a data-center analogue of deferrable-demand scheduling \cite{papavasiliou2013large} and \rev{avoiding} coincident peaks \cite{liu2013data}. It \rev{is beneficial} when the marginal cost of energy varies  within a day, and the same intraday reshaping can be produced by on-site resources. \rev{Shifting workload to data centers with cheaper power is geographical load balancing \cite{liu2011greening, qureshi2009cutting} and} has been used to route compute toward renewable-surplus, lower-emission regions \cite{radovanovic2022carbon}. 

\rev{This paper integrates the data-center flexibility mechanisms with power system  planning.} A recent field demonstration rerouted part of a live inference workload from a Virginia cluster to one in Illinois during a Dominion winter-peak event, holding user-facing latency essentially unchanged \cite{williams2026power}.  \rev{Similarly, training workloads can also be moved. For instance, Meta's global scheduler ran about 30\% of its training workloads in more than one region over 30 days \cite{choudhury2024mast}. It moves them from day to day to balance GPU availability. Building on these promising demonstrations, this paper considers the effect of large-scale, hourly spatial shifting for synchronous training workloads.} \rev{Interruption} is the cheapest mechanism, but is limited by how much power withdrawal a data center can actually give up. %Priority tariffs offer a lower price in exchange for a lower grade of service \cite{chao1987priority}. 
A hyperscale data center can potentially curtail  roughly a quarter of its power for several hours through workload and chip-frequency scaling  \cite{colangelo2026ai}.  Field experiments  have shown 10--40\% reductions for hours and met emergency dispatch within tens of seconds while protecting \rev{high} priority jobs \cite{williams2026power}. \rev{Interruption} is therefore shape-limited, and Fig.~\ref{fig:eventshape} defines the limits a credible interruptible tier must specify, i.e., how deeply power \rev{withdrawals} may be reduced, how long \rev{interruptions} may last, how fast \rev{they} may change, and how soon another event may follow.

\begin{figure}[!htbp]
\centering
\begin{tikzpicture}[xscale=0.62, yscale=0.045, >=Latex]
  % axes
  \draw[->,thick] (0,0) -- (14.2,0);
  \draw[->,thick] (0,0) -- (0,124);
  \node[rotate=90, anchor=south] at (-1.0,60) {\footnotesize Power withdrawal (\% of nominal)};
  \node[anchor=north] at (7,-6) {\footnotesize Time (hours)};
  % nominal reference
  \draw[dashed,gray] (0,100) -- (14,100);
  \node[anchor=east] at (-0.15,100) {\footnotesize 100};
  \node[anchor=east] at (-0.15,0) {\footnotesize 0};
  % shaded interruption window
  \fill[blue!8] (2.5,0) rectangle (6,100);
  % load curve: nominal, ramp down, hold (duration), ramp up, recovery, smaller 2nd event
  \draw[very thick, blue!70!black]
     (0,100)--(2.5,100)--(3.3,45)--(6,45)--(6.8,100)--(10.6,100)
     --(11.4,72)--(12.3,72)--(13.0,100)--(14,100);
  % depth
  \draw[<->] (4.5,100)--(4.5,45);
  \node[anchor=west] at (4.4,72) {\footnotesize depth};
  % duration
  \draw[<->] (3.3,37)--(6,37);
  \node[anchor=north] at (4.8,35) {\footnotesize duration};
  % ramp
  \node[anchor=south east] at (3.3,50) {\footnotesize ramp};
  \draw[->] (3.3,58)--(3.45,49);
  % recovery
  \draw[<->] (6.8,111)--(10.7,111);
  \node[anchor=south] at (8.8,111) {\footnotesize recovery};
\end{tikzpicture}
\caption{\textbf{Event-shape limits on an interruptible data-center load.}
Each \rev{interruption} is bounded by its depth, duration, ramp and recovery, and the
annual interrupted energy and interruption hours are capped as well.}
\label{fig:eventshape}
\end{figure}

Utilities, grid operators and regulators are \rev{developing tariffs and interconnection products} to capture one or \rev{more flexible tiers}, allowing data centers to trade operational flexibility for faster or more favorable interconnection terms \cite{epri2025gridflexibility, brancucci2025flexible, ferc2026largeload}. Large-load customers have historically been offered a single firm-service level, with demand-response programs as a notable and often rare exception. Texas Senate Bill~6 (2025) requires large loads to install remote-disconnect capability and to curtail during grid emergencies, with implementation beginning in 2026 \cite{TX-SB6-2025}. In June 2026 the Federal Energy Regulatory Commission (FERC) issued show cause orders to all six jurisdictional Regional Transmission Organizations (RTOs) and Independent System Operators (ISOs) to justify or reform their large-load tariffs and to develop new transmission services for flexible large loads \cite{ferc2026largeload}. \rev{Within PJM, FERC accepted in October 2026 a definition of large loads as those of 50~MW or more, but rejected PJM's proposal to curtail new large loads that do not bring their own capacity before other customers \cite{pjm2026iras, ferc2026pjmlargeload}}. These proposals and reforms remain contested, and \rev{data-center} loads may instead rely on on-site generation to preserve firm uptime. This underlines the importance of  characterizing data-center \rev{load} tiers by their grid value rather than by how the flexibility is provided.

Despite empirical evidence that the data-center flexibility \rev{mechanisms} vary in their grid value, most CEMs reduce data-center flexibility to a single mechanism or study these mechanisms in isolation \rev{although they compete for the same capacity and compliance margins} \cite{zuluaga2027nodal, liu2026watts}. \rev{For example, deferring and interrupting load both reduce demand in the few stress hours that set capacity needs, and relocating and interrupting load both reduce the energy subject to renewable-generation obligations, so the value of one mechanism depends on how much the others already provide.} Standards and industry bodies, including the North American Electric Reliability Corporation (NERC) and the EPRI DCFlex consortium, increasingly frame flexibility as a spectrum rather than a binary firm or non-firm choice \cite{nerc2026assessment, epri2026powering, epri2025gridflexibility}. System-level studies estimate how much new flexible load existing grids can absorb, i.e., the headroom available without new generation or transmission. Norris et al.\ \cite{norris2025rethinking} find that the 22 largest U.S.\ balancing authorities could host 76--126~GW of additional load at annual curtailment rates of 0.25--1\%, with roughly 18~GW of headroom in PJM alone at 0.5\% curtailment. This assessment considers the new load as a fixed demand block, curtails it only when total demand exceeds each region's historical seasonal peak, and leaves out transmission limits, power plant start-up and ramping constraints, and any changes in generation or transmission. It therefore assesses \emph{how much} flexible load fits under present peaks, but not \emph{where} on the network flexibility is needed, \emph{when} the additional capacity it defers would  otherwise be built, or \emph{which} data-center flexibility \rev{mechanisms} create value once generation, storage and transmission are co-planned.

A smaller set of studies embeds data-center flexibility directly into capacity expansion or interconnection planning, and each resolves part of the problem. Kim et al.\ \cite{kim2026flexibility} propose a data-center siting framework that represents each candidate as a small set of standardized flexibility profiles, pre-screens locations through system-wide market impacts, and identifies pre-certified sites that can connect without necessitating  major grid upgrades. On a synthetic 2{,}000-bus Texas system with a 2~GW data center,  it expands viable siting options by up to 21\% and shortens interconnection from years to months. That framework is a short-run siting tool for fast-track interconnection, and long-run generation, storage and transmission investments are not considered. Senga et al.\ \cite{senga2026flexible} quantify the value of \rev{temporal shifting} across Western Electricity Coordinating Council (WECC), Texas and mid-Atlantic testbeds, but use a single aggregate flexibility representation, with no interruptible tier and no event-shape limits. Zuluaga et al.\ \cite{zuluaga2027nodal} co-optimize data-center placement with transmission and generation investment, but their only mechanism is siting the data center itself, which is a one-time investment decision without considering operational detail. % rather than an operating one.  
Gu et al.\ \cite{gu2025role} examine interruptible and temporal-shifting flexibility for distribution-grid interconnection under limits on utility-controlled interventions. Chen et al.\ \cite{chen2026defer} model both \rev{temporal shifting} and \rev{spatial shifting} within a generation expansion model, but do not consider storage investment and resource adequacy and simplify transmission expansion. Hence, the model cannot capture how flexibility defers investment jointly in generation, storage and transmission, and its single spatial mechanism attributes all locational value to congestion relief. Shao et al.\ \cite{shao2025stochastic} co-optimize generation, storage and transmission for data-center siting on a 21-zone PJM system, but the flexibility is modeled using a limited spatial-shifting mechanism in which workload is relocated across zones but must still be served in the hour it arrives. Several of these models \cite{kim2026flexibility, zuluaga2027nodal, gu2025role, chen2026defer} are, moreover, demonstrated on reduced-order, synthetic  test systems with limited policy realism. No prior study comprehensively accounts for  the three data-center \rev{load} tiers  within a CEM  that co-optimizes generation, storage and transmission subject to resource adequacy and policy constraints, and then applies that model to \rev{two systems that differ in generation mix, planning institutions, policy instruments and network structure.}

This paper therefore seeks to characterize and identify the  factors driving the value of data-center flexibility and to integrate the three data-center \rev{load} tiers into CEM-based planning. Data-center load at each site is divided into firm, \rev{shiftable} and interruptible tiers. The \rev{shiftable tier} can shift load to later hours, creating a backlog that must be made up within a fixed horizon \cite{papavasiliou2013large}, and can move it to other eligible locations \cite{liu2011greening}. The interruptible tier can be curtailed within an annual energy budget, optionally subject to limits on total interruption hours and to the shape of each event (as illustrated in Fig.~\ref{fig:eventshape}). \rev{Both flexible tiers earn capacity credit only for the load they actually reduce in stress hours, so their contribution to resource adequacy follows from the model's operating decisions and is not assumed in advance. We apply the same CEM formulation to PJM and Korea and use \textit{leave-one-out tests}, which disable one flexibility mechanism at a time, to isolate each mechanism's contribution within a grid. We also compute indicators from a model without data-center flexibility, namely the adequacy shadow price, the variation of prices between hours, the compliance-adjusted price spread between load locations, and the level of the delivered price.}

\rev{Comparing the PJM and Korean systems is timely because rules for data-center flexibility are being written now, often from the experience of a single system, and a mechanism that is valuable on one grid may not be as valuable elsewhere. Both systems face rapid, geographically concentrated data-center demand growth, but they differ in four dimensions that can materially determine how flexibility creates value.} First, the generation mixes differ, with gas and coal plus proposed gas and renewable additions in PJM, against solar PV expansion and coal phase-out in  Korea.  Second, the planning institutions differ, with decentralized wholesale markets and multi-party stakeholder processes in PJM, against centralized national planning in  Korea where a single plan sets load forecasts and candidate resources. \rev{Third}, the policy constraints differ, with \rev{state renewable portfolio standards (RPS)} in PJM, against a single national carbon cap in  Korea. \rev{Fourth, the networks differ in resolution, with PJM represented as 21 zones linked by interzonal transmission interfaces and Korea as 193 buses linked by 746 AC transmission lines and two 500~kV HVDC lines. The systems also differ in data-center penetration, which we measure as data-center load as a share of system peak demand, at 31\% in PJM and 7\% in Korea in 2038, and each is modeled with its own resource-adequacy rules. To compare the two systems on a common basis, we express value per MWh of data-center demand, so that it does not simply scale with penetration, and we vary penetration, policy regime, accreditation and candidate resources within each grid. We also test, over 11 target years from 2028 to 2038, whether indicators computed without data-center flexibility anticipate each mechanism's contribution in both systems.}

PJM is a focal point of US data-center growth. Most projected peak-demand growth over the next decade is attributed to data centers, with Northern Virginia's Dominion zone hosting the world's largest hyperscale data-center concentration \cite{shehabi20242024}. Recent capacity auctions have cleared at sharply higher prices, partly due to large-load additions, while debate over flexible interconnection intensifies \cite{pjm2027brarep, pjm2028brarep, brancucci2025flexible}. PJM, and Dominion in particular, is  a policy-relevant test of whether flexibility can defer or relocate the supply-side capacity additions, transmission network upgrades and compliance investments that firm hyperscale load requires. By contrast,  Korea plans centrally through the Basic Plan for Electricity Supply and Demand (BPESD). The 11th BPESD projects data-center peak demand of 6.2~GW by 2038, or about 8.9~GW of contract capacity \cite{MOTIE202511th}. The demand is geographically concentrated, with 60.4\% of data centers in the metropolitan area in 2024, where new loads face congestion and grid-stability constraints \cite{KEEI2026DomesticDC}.  Korea's approach has focused on siting incentives that steer new facilities toward non-metropolitan regions \cite{DistributedEnergyAct2024, MOTIE2023DCDispersion}, and it has yet to develop a non-firm interconnection product \cite{ferc2026largeload}.

\rev{This study finds that data-center flexibility lowers total system cost in both systems, by different amounts and through pathways that are partly shared. In 2028, when accredited capacity is still relatively cheap in PJM and not yet needed in Korea, a reference data-center load tier mix (30\% firm, 50\% shiftable and 20\% interruptible load) lowers cost by \res{0.8\%} in PJM and \res{0.2\%} in Korea, relative to a baseline in which all data-center load is firm. By 2038 the cost saving is \res{6.0\%} in PJM and \res{2.5\%} in Korea, and in each system about two thirds of it comes from capacity credit that the shiftable and interruptible tiers earn by reducing load during stress hours. That credit replaces the need for building \res{44~GW} of natural gas generation capacity in PJM and \res{6~GW} of batteries in Korea. On each grid, it displaces the cheapest source of accredited capacity. In PJM this is gas-fired capacity, because a 4-hour battery is rated at only 22\% of installed capacity, and in Korea it is batteries, because storage is credited at its full installed capacity. Beyond capacity, the grids differ in where flexibility creates operating value. In PJM, spatial shifting adds value almost entirely by moving RPS obligations between states, not by relieving congestion or lowering energy cost. In Korea, spatial shifting adds almost no value, and temporal shifting instead creates value by deferring load into midday solar hours under the carbon cap. Cost saving per MWh of data-center demand in 2038 is \res{\$10.0/MWh} in Korea and \res{\$5.4/MWh} in PJM. Solving each year from 2028 to 2038 shows that, in both systems, the operation cost savings of each mechanism rise with hourly electricity prices computed from a model without data-center flexibility, and the capacity value rises with the adequacy shadow price. Fully interruptible load yields the lowest cost in both systems, but only because it removes the data-center load altogether. It shows the upper-bound value that interruption could provide to the grid, and that value decreases when realistic interruption limits and event-shape constraints apply.}

\section{Results}\label{sec2}

\subsection{Modeling framework and cases}

We represent data-center demand at each location using the firm, \rev{shiftable and interruptible tiers}.  \rev{Work deferred by the shiftable tier must be made up within a make-up horizon of 24~h, and this tier may also move load to other eligible data-center locations, within limits at both the sending and the receiving site.} The interruptible tier may curtail up to 5\% of its own annual energy in the base case \rev{(the annual interruptible-energy cap, $\phi^{\mathrm{inter}}_{\max}$)}.  \rev{Additional sensitivity to the limits on the shape of each interruption event (Fig.~\ref{fig:eventshape}) is analyzed in Sec.~\ref{sec:inter-event-shape}.} Firm \rev{(non-data-center)} load may also be shed involuntarily, at \rev{a penalty equal to} the value of lost load.

The complete CEM is given in Methods (Sec.~\ref{sec:methods}). Resource adequacy is enforced through \rev{accredited capacity. The default stress hours are the ten highest net-load hours of the modeled days, and the shiftable and interruptible tiers are credited only with the load they actually reduce in those hours, so their credit follows from the operational decisions the model chooses and is not assumed in advance (Sec.~\ref{sec:ra}).} Data-center demand follows each system's own forecast. In PJM it rises to 16.3~GW by 2028 and 74.8~GW by 2038 \rev{(9.4\% and 31\% of system peak, 14\% and 38\% of annual electricity demand)}, and in  Korea to 4.9 and 8.9~GW \rev{(4.5\% and 6.9\% of peak, 6.6\% and 10\% of annual demand)}. Candidate generation, storage and transmission are drawn from the PJM interconnection queue and the  BPESD in Korea (Table~\ref{tab:case_study_design}).

\rev{For each system we consider two target years.} The 2028 scenario captures near-term constraints on interconnection and equipment supply \cite{yao2026grid}, and the 2038 scenario  allows  \rev{for} longer-term co-planning of generation, storage and transmission. \rev{For the transferable indicator analysis (Sec.~\ref{sec:indicators}) we also consider each year from 2028 to 2038.} In each \rev{target year,  we} solve four data-center cases\rev{. Three are single-tier cases, with all data-center load firm, shiftable or interruptible. The fourth is the reference tier mix of 30\% firm, 50\% shiftable and 20\% interruptible load. It follows industry estimates that a minority of workloads are must-run, about half can wait or run elsewhere, and a smaller share can be curtailed \cite{epri2025gridflexibility}, with an additional sensitivity analysis in Sec.~\ref{sec:sensitivity} and Sec.~\ref{sec:robust}.} In the all-interruptible case the \rev{annual interruptible-energy cap} is relaxed to 100\% of the interruptible tier's own annual energy ($\phi^{\mathrm{inter}}_{\max}=1$), so the tier may be curtailed in full.

\subsection{\rev{2028 results: flexibility is worth little while capacity is not scarce}}

\begin{table}[!htbp]
\centering
\caption{\textbf{The 2028 outcomes for PJM and  Korea.} Columns are the
all-firm baseline, the two remaining single-tier cases (\rev{All-shift}, \rev{All-inter})
and the \rev{reference tier mix (Ref., 30\% firm, 50\% shiftable and 20\% interruptible load)}. Percentages are changes relative to the all-firm baseline of the same system. Costs are annualized \rev{in 2023 \$B}; energy \rev{in}
TWh. \rev{Operation includes the cost of using data-center flexibility and any firm load shed, so the three cost rows add to the total. Policy \& adequacy aggregates RPS compliance payments and resource-adequacy shortfall payments; Korea's carbon cap and adequacy requirement are hard constraints and carry no payment. All-inter has no cap on interruptible energy ($\phi^{\mathrm{inter}}_{\max}=1$).}}
\label{tab:s1_2028}
\begingroup
\scriptsize
\setlength{\tabcolsep}{2.0pt}
\renewcommand{\arraystretch}{1.05}
\begin{tabular}{@{}lrrrrrrrr@{}}
\toprule
& \multicolumn{4}{c}{\textbf{PJM (2028)}}
& \multicolumn{4}{c}{\textbf{Korea (2028)}}\\
\cmidrule(lr){2-5}\cmidrule(lr){6-9}
Metric & All-firm & \rev{All-shift} & All-inter & \rev{Ref.} & All-firm & \rev{All-shift} & All-inter & \rev{Ref.}\\
\midrule
Total cost (\$B) & \res{22.4} & \res{$-1.0\%$} & \res{$-23.5\%$} & \res{$-0.8\%$} & \res{18.9} & \res{$-0.1\%$} & \res{$-12.6\%$} & \res{$-0.2\%$}\\
Investment (\$B) & \res{2.27} & \res{$-12.0\%$} & \res{$-19.7\%$} & \res{$-11.0\%$} & \res{1.18} & \res{$+2.7\%$} & \res{$-32.0\%$} & \res{$+0.9\%$}\\
Operation (\$B) & \res{18.9} & \res{$+1.3\%$} & \res{$-22.1\%$} & \res{$+0.9\%$} & \res{17.8} & \res{$-0.2\%$} & \res{$-11.3\%$} & \res{$-0.2\%$}\\
Policy \& adequacy (\$B) & \res{1.31} & \res{$-14.8\%$} & \res{$-51.3\%$} & \res{$-8.2\%$} & \res{${\sim}0$} & \res{${\sim}0$} & \res{${\sim}0$} & \res{${\sim}0$}\\
Temporally shifted (TWh) & \res{${\sim}0$} & \res{4.10} & \res{${\sim}0$} & \res{2.83} & \res{${\sim}0$} & \res{0.24} & \res{${\sim}0$} & \res{0.13}\\
Spatially shifted (TWh) & \res{${\sim}0$} & \res{19.3} & \res{${\sim}0$} & \res{9.64} & \res{${\sim}0$} & \res{0.20} & \res{${\sim}0$} & \res{0.09}\\
\rev{Interrupted} (TWh) & \res{${\sim}0$} & \res{${\sim}0$} & \res{142} & \res{1.42} & \res{${\sim}0$} & \res{${\sim}0$} & \res{42.6} & \res{0.43}\\
\bottomrule
\end{tabular}
\endgroup
\end{table}

\begin{figure*}[!htbp]
\centering
\begin{subfigure}[t]{0.96\textwidth}
\centering
    \includegraphics[width=\linewidth]{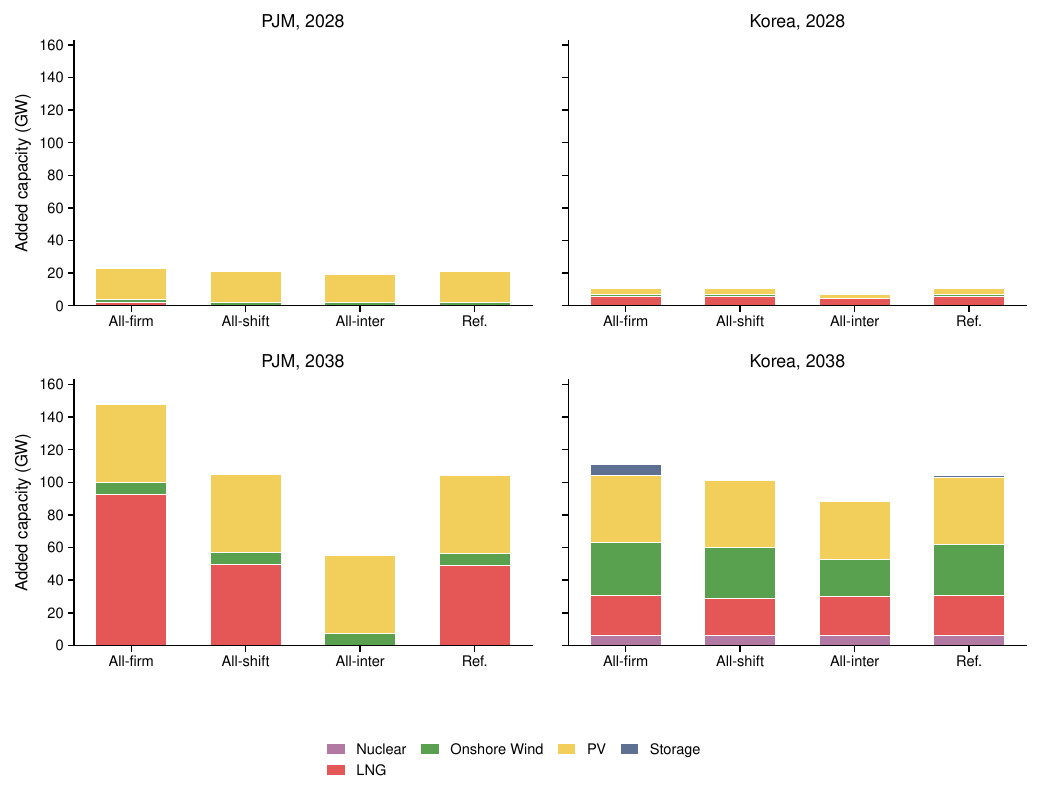}
\end{subfigure}
\caption{\textbf{Generation and storage expansion.} \rev{New capacity by technology in the four data-center cases; LNG (Liquefied Natural Gas) denotes gas-fired capacity in both systems. Data-center flexibility reduces gas generator expansion in PJM in both years, and most in 2038, when it avoids \res{44~GW}. In Korea it slightly increases solar PV in 2028 and, in 2038, removes most of the \res{7.0~GW} of batteries that the all-firm case builds.}}
\label{fig:buildout_updates}
\end{figure*}

\rev{In PJM, the reference tier mix lowers total annual cost by \res{0.8\%} (\res{\$0.19B}) relative to the all-firm baseline, with investment cost down \res{11\%} and operating cost up \res{0.9\%} (Table~\ref{tab:s1_2028}). New firm capacity is barely needed in 2028. The all-firm case builds \res{2.0~GW} of natural gas combined-cycle generation capacity, and the all-shiftable case and the reference tier mix build none (Fig.~\ref{fig:buildout_updates} and supplementary Fig.~\ref{fig:PJMexpS1}). Solar PV  (\res{18.9~GW}) and onshore wind generation capacity (\res{2.1~GW}) are built up to the limit of the interconnection queue in these three cases, so enabling data-center flexibility does not change the renewable build. The adequacy constraint is therefore relatively cheap, with a shadow price of about \res{\$13,000/MW-yr} in the all-firm case, about one eighth of the capacity-shortfall price. The \res{2.0~GW} of natural gas generation capacity that the all-firm case builds gets accredited for \res{1.6~GW}. The reference tier mix builds no natural gas and meets the same requirement by reducing its load by \res{1.6~GW} in every stress hour. The interruptible tier alone can provide this, since it interrupts \res{2.9--3.3~GW} in each of those hours.}

\rev{In 2028, spatial shifting accounts for the largest volume of exercised flexibility in PJM, where \res{9.6~TWh} is relocated between zones, or \res{13.5\%} of shiftable-tier energy, against \res{2.8~TWh} (\res{4.0\%}) deferred in time (supplementary Fig.~\ref{fig:flex_mechanisms_S1}). Dominion is the source of \res{92\%} of the relocated energy, and AEP receives \res{65\%} of it, with ComEd (\res{1.2~TWh} net) and PPL (\res{0.8~TWh}) next (supplementary Fig.~\ref{fig:spatial_s1}). This direction follows compliance cost, rather than energy prices. RPS obligations follow retail sales, so a relocated MWh also moves its obligation with it, and Dominion's load carries a 35\% Tier~1 requirement in 2028 while Ohio has no requirement after 2026. When obligations are instead held at each zone's nominal load, spatial shifting falls from \res{9.6} to \res{1.6~TWh} and the saving reduces from \res{\$0.19B} to \res{\$0.10B}.}

The 2028 results for Korea reveal a different pattern. The \rev{reference tier mix} saves only \rev{\res{0.2\%} (\res{\$0.03B})} in total cost, raising annualized investment by \rev{\res{0.9\%}} while lowering operating cost by \rev{\res{0.2\%}}. This is not a contradiction. Data-center flexibility allows the model to build and use more solar PV, which is cheap to operate but contributes little firm capacity, so it is attractive only if demand can be reshaped to use it. \rev{With shiftable load able to absorb midday output, the fuel saved by an additional MW of solar PV exceeds its annualized cost. Solar PV expansion rises from \res{3.6~GW} in the all-firm case to \res{3.8~GW} in the reference tier mix case and \res{4.0~GW} in the all-shiftable case} (supplementary Fig.~\ref{fig:KPGexpS1}). \rev{The system is not yet capacity-tight. In 2028 therefore the adequacy requirement is not binding, so the data-center tiers do not earn capacity credit and the shiftable tier is used lightly. Temporal shifting defers \res{0.13~TWh} and spatial shifting relocates \res{0.09~TWh}, each about \res{0.5\%} of shiftable-tier energy. Most of the cost saving comes from interruption, which mainly occurs in the  stress hours. For instance, with both shifting mechanisms disabled the saving is still \res{\$0.025B} of \res{\$0.031B}.} The all-interruptible case is the cheapest at \rev{\res{13\%}} below the all-firm baseline, but it builds the least solar PV at \rev{\res{2.4~GW}}, because curtailing the load eliminates the demand that would justify new solar PV. %The \rev{reference tier mix} produces the results in between these extremes, with solar PV at 8.20~GW. Its main near-term effect is to make additional solar PV economically usable rather than to avoid capacity.

\rev{Overall, data-center flexibility lowers cost only slightly in 2028 in both systems, but in different ways. In PJM, investment is cut by \res{11\%} and RPS compliance cost by \res{8\%}, while operating cost rises slightly (\res{$+0.9\%$}). In Korea, investment rises by \res{0.9\%} to build more solar PV, and this extra investment is recovered through lower operating cost (\res{$-0.2\%$}). Per MWh of data-center demand the saving is \res{\$1.3/MWh} in PJM and \res{\$0.7/MWh} in Korea. In neither system, however, is firm capacity scarce  in 2028.}

\subsection{2038 results: data-center flexibility \rev{replaces gas-fired capacity in PJM and batteries in Korea}}

\begin{table}[!htbp]
\centering
\caption{\textbf{The 2038 outcomes for PJM and  Korea.} Columns are the
all-firm baseline, the two remaining single-tier cases (\rev{All-shift}, \rev{All-inter})
and the \rev{reference tier mix (Ref., 30\% firm, 50\% shiftable and 20\% interruptible load)}. Percentages are changes relative to the all-firm baseline of the same system. Costs are annualized in \rev{2023} \$B; energy is in TWh. \rev{Rows are defined as in Table~\ref{tab:s1_2028}.}}
\label{tab:s2_2038}
\begingroup
\scriptsize
\setlength{\tabcolsep}{2.0pt}
\renewcommand{\arraystretch}{1.05}
\begin{tabular}{@{}lrrrrrrrr@{}}
\toprule
& \multicolumn{4}{c}{\textbf{PJM (2038)}} & \multicolumn{4}{c}{\textbf{Korea (2038)}}\\
\cmidrule(lr){2-5}\cmidrule(lr){6-9}
Metric  & All-firm & \rev{All-shift} & All-inter & \rev{Ref.} & All-firm & \rev{All-shift} & All-inter & \rev{Ref.}\\
\midrule
Total cost (\$B) & \res{59.1} & \res{$-6.6\%$} & \res{$-57.4\%$} & \res{$-6.0\%$} & \res{31.7} & \res{$-2.9\%$} & \res{$-25.4\%$} & \res{$-2.5\%$}\\
Investment (\$B) & \res{14.7} & \res{$-34.0\%$} & \res{$-72.3\%$} & \res{$-34.6\%$} & \res{14.5} & \res{$-8.5\%$} & \res{$-19.5\%$} & \res{$-5.9\%$}\\
Operation (\$B) & \res{35.6} & \res{$+8.8\%$} & \res{$-44.8\%$} & \res{$+7.9\%$} & \res{17.3} & \res{$+1.8\%$} & \res{$-30.4\%$} & \res{$+0.5\%$}\\
Policy \& adequacy (\$B) & \res{8.76} & \res{$-23.5\%$} & \res{$-83.4\%$} & \res{$-14.8\%$} & \res{${\sim}0$} & \res{${\sim}0$} & \res{${\sim}0$} & \res{${\sim}0$}\\
Temporally shifted (TWh) & \res{${\sim}0$} & \res{14.7} & \res{${\sim}0$} & \res{12.2} & \res{${\sim}0$} & \res{15.5} & \res{${\sim}0$} & \res{9.04}\\
Spatially shifted (TWh) & \res{${\sim}0$} & \res{94.9} & \res{${\sim}0$} & \res{47.6} & \res{${\sim}0$} & \res{0.79} & \res{${\sim}0$} & \res{0.51}\\
\rev{Interrupted} (TWh) & \res{${\sim}0$} & \res{${\sim}0$} & \res{655} & \res{6.55} & \res{${\sim}0$} & \res{${\sim}0$} & \res{77.9} & \res{0.78}\\
\bottomrule
\end{tabular}
\endgroup
\end{table}

\begin{figure*}[!htbp]
\centering
\begin{subfigure}[t]{0.96\textwidth}
\centering
\includegraphics[width=\linewidth]{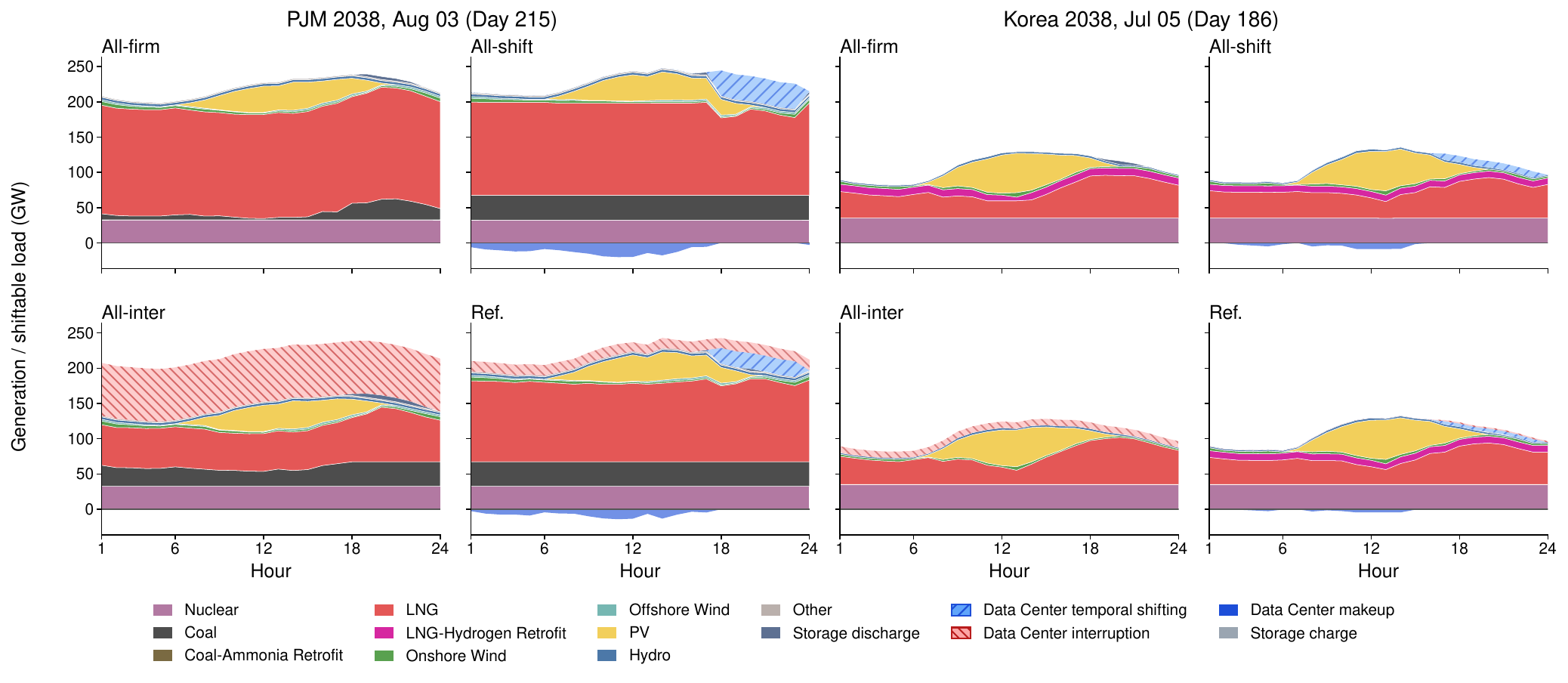}
\caption{Generation dispatch \rev{on the peak net-demand day} (2038)
}
\end{subfigure}
\vspace{2mm}
\begin{subfigure}[t]{0.96\textwidth}
\centering
\includegraphics[width=\linewidth]{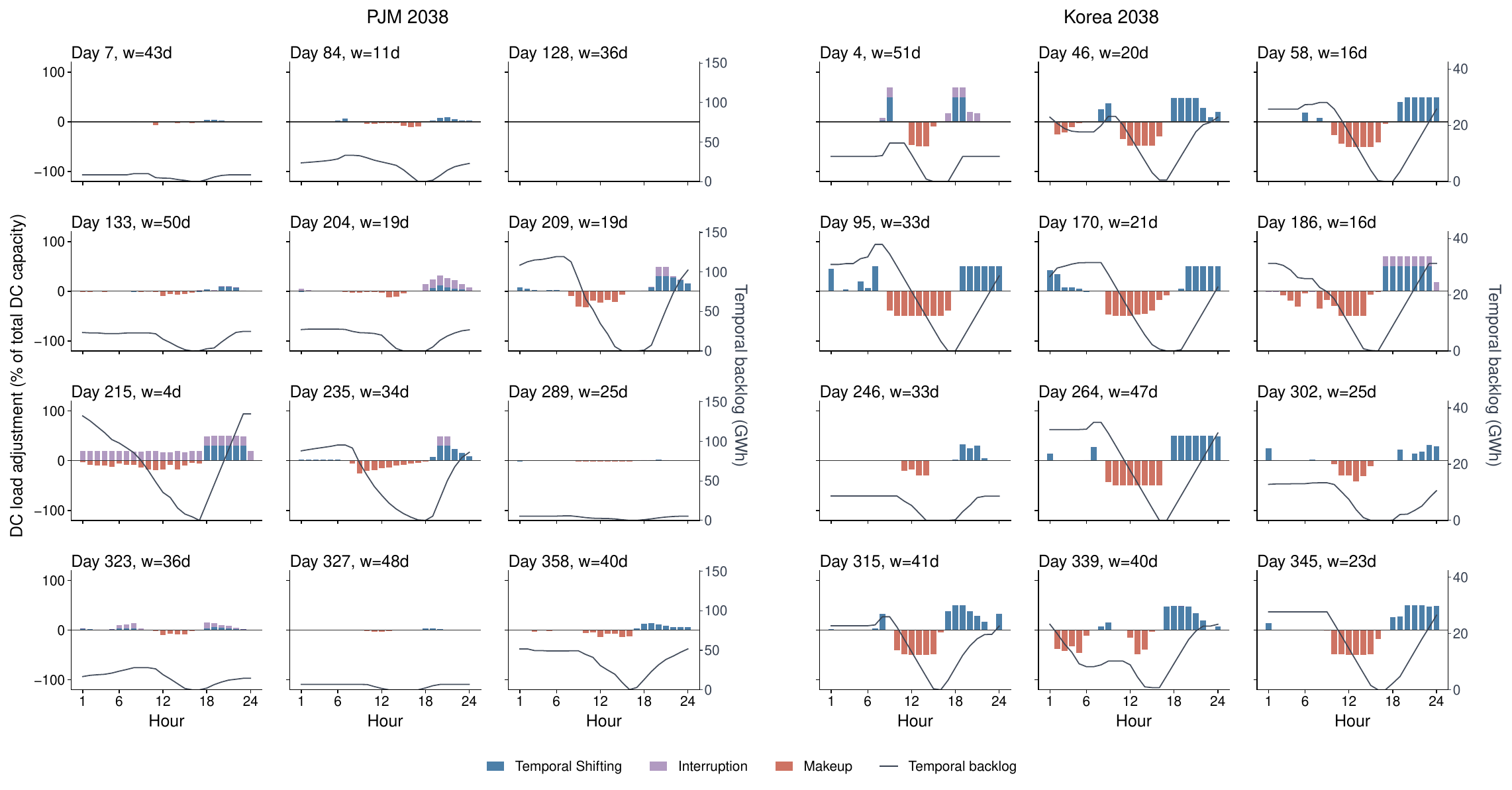}
\caption{\rev{Temporal shifting (deferral)}, make-up and interruption \rev{on the 12 representative days} (2038).}
\end{subfigure}
\caption{\textbf{\rev{Flexible data-center operation} in 2038.} \rev{(a) Generation stacks on each grid's peak net-demand day, with deferred and interrupted data-center load hatched above the stack and make-up below the axis. (b) Temporal shifting, make-up and interruption in the reference tier mix on each representative day, as a percentage of total data-center (DC) load capacity, with the backlog in black (right axis, GWh); $w$ is the day's weight in days. Korea's shiftable tier acts as virtual storage on almost every day, deferring \res{9.0~TWh} from evening and night hours and making it up in midday solar hours. PJM defers \res{12.2~TWh}, most heavily in the evening stress hours of its peak days (days 215, 235 and 209), where deferral earns capacity credit, and it shifts \res{47.6~TWh} between zones (Fig.~\ref{fig:spatial_s2}).}}
\label{fig:flex_mechanisms_S2}
\end{figure*}

\rev{By 2038 data-center flexibility becomes more valuable in both systems. In PJM, the reference tier mix lowers total annual cost by \res{6.0\%} (\res{\$3.6B}, Table~\ref{tab:s2_2038}). Investment cost declines by \res{35\%} (\res{\$5.1B}). Notably, new natural gas generation capacity is  around half of that in the all-firm case (\res{48.9~GW} out of the 92.8~GW), while solar PV  (\res{48.0~GW}) and onshore wind (\res{7.2~GW}) stay at their interconnection queue limits and no nuclear, batteries or transmission are added in either case (Fig.~\ref{fig:buildout_updates} and supplementary Fig.~\ref{fig:PJMexpS2}). RPS and adequacy costs fall by \res{15\%} (\res{\$1.3B}). Operating cost, including the cost of using data-center flexibility, increases by \res{7.9\%} (\res{\$2.8B}).}

\rev{Most of the cost saving comes from capacity credit. The reference tier mix earns credit for \res{36.5~GW}, of which \res{14.0~GW}  is from 15.0~GW of the interruptible tier and \res{22.5~GW} from 37.4~GW of the  shiftable tier, which defers \res{22.5~GW} in each of the ten stress hours. These hours are between 17:00 and 23:00, six of them on 3~August and the rest on two other summer evenings, and the deferred work is made up in the morning and midday hours of the same modeled day (Fig.~\ref{fig:flex_mechanisms_S2}). Without this credit the cost saving decreases from \res{\$3.6B} to \res{\$1.2B}, the reference tier mix builds all the gas-fired capacity in the interconnection queue, and it is still \res{3.3~GW} short of the resource-adequacy requirement. The credit is therefore worth two thirds of the saving.}

\begin{figure}[!htbp]
\centering
\includegraphics[width=\linewidth]{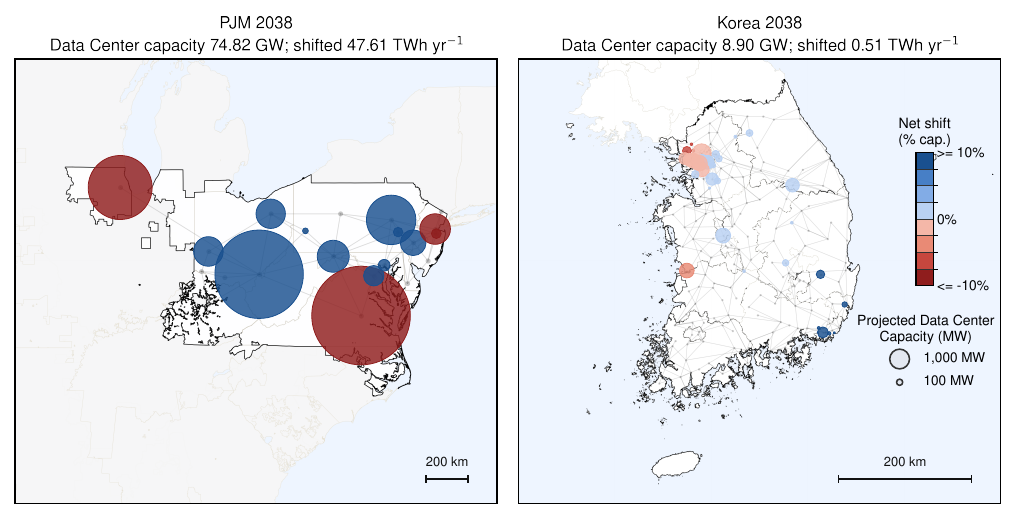}
\caption{\textbf{\rev{Spatial shifting} of data-center load with the 2038 \rev{reference tier} mix.} Circle size is the projected data-center load at each node; color is the net
spatial shift relative to that node's capacity\rev{, blue for net inflow and red for net outflow. In PJM, load leaves Dominion and ComEd, whose states have demanding RPS, mainly for AEP and PPL. In Korea \res{0.5~TWh} is shifted, mostly out of the Seoul metropolitan area, about \res{1\%} of  PJM's volume.}}
\label{fig:spatial_s2}
\end{figure}

\rev{Spatial shifting relocates \res{47.6~TWh}, or \res{14.5\%} of shiftable-tier energy, and temporal shifting defers \res{12.2~TWh} (\res{3.7\%}), so the volume shifted in space is four times the volume shifted in time. Dominion is the source of \res{69\%} of the relocated energy (\res{32.7~TWh}), and ComEd is a net source of \res{11.1~TWh}. AEP receives \res{26.3~TWh} and PPL \res{8.0~TWh}, with smaller inflows to APS, Dayton, ATSI and PECO (Fig.~\ref{fig:spatial_s2}). As in 2028, the flows run from states with demanding RPS (Virginia and Illinois) toward Ohio, whose RPS ends after 2026, and Pennsylvania, which requires 18\% of retail sales across its two tiers. The relocated volume is set by the shift cap, which is an assumption. The caps on what Dominion may send and on what AEP may receive are reached in every hour, and the volume changes in proportion when the cap is varied (Sec.~\ref{sec:robust}).}

\rev{The operating cost increase stems from building less gas-fired capacity. With \res{43.8~GW} fewer combined-cycle generators, existing coal-fired generators are used more. In this case, emissions from coal-fired generation rise from \res{23} to \res{216~Mt}~CO\textsubscript{2}, and PJM's total emissions rise from \res{411} to \res{547~Mt} (\res{$+33\%$}). The 2028 increase is smaller, from \res{258} to \res{263~Mt} (\res{$+2.0\%$}). The model has no carbon limit in PJM and applies no retirements, so this result is conditional on existing coal-fired generators remaining available \cite{DOE2025GridReliability}. It is nonetheless a direct cost of valuing flexibility against capacity alone. In Korea the binding carbon cap rules this outcome out.}

\rev{By 2038 both the carbon cap and the resource-adequacy constraints are active in the Korean system. The carbon cap tightens by 56\% compared to 2028 (189.9 to 83.1 MtCO\textsubscript{2}/year), and its shadow price increases from \res{\$10/tCO\textsubscript{2}} in 2028 to \res{\$160/tCO\textsubscript{2}}. The resource-adequacy requirement, which in 2028 is met with extra accredited capacity and has no shadow price, has a shadow price of \res{\$92,000/MW-yr} in 2038. The reference tier mix lowers total cost by \res{2.5\%} (\res{\$0.78B}). Investment decreases by \res{5.9\%} (\res{\$0.86B}), and almost all of the capacity expansion reduction is in batteries. The all-firm case builds \res{7.0~GW} of 4-hour batteries and the reference tier mix \res{1.2~GW}. New LNG capacity decreases slightly, from \res{24.5} to \res{24.2~GW}, onshore wind from \res{32.2} to \res{31.6~GW} and hydrogen co-firing retrofits from \res{10.4} to \res{9.5~GW}, while transmission line additions rise from \res{76} to \res{85~GW}. Operating cost, including the cost of using data-center flexibility, increases by \res{0.5\%}. Solar PV expansion is \res{41.1~GW} in the all-firm and all-shiftable cases and in the reference tier mix, which is the limit set by the BPESD, so data-center flexibility no longer changes the solar PV expansion.}

\rev{The shiftable tier does two things in Korea. It acts as virtual storage, deferring \res{9.0~TWh} (\res{23\%} of shiftable-tier energy) from evening and night hours to midday solar hours, while relocating only \res{0.5~TWh}. It also earns capacity credit by deferring its full \res{4.45~GW} in the evening stress hours. Together with the interruptible tier, the credit is \res{6.2~GW}, the full capacity of the two tiers, and it replaces most of the batteries that the all-firm case builds to meet the adequacy requirement. Without the credit the saving decreases from \res{\$0.78B} to \res{\$0.27B} and the reference tier mix builds \res{6.9~GW} of batteries, so, as in PJM, about two thirds of the saving comes through the credit. Batteries are the marginal adequacy resource in Korea because the Korean resource adequacy rules credit storage at its full capacity (Sec.~\ref{sec:ra}), whereas PJM's 2038/39 ELCC class rating for a 4-hour battery is only 22\%. As a result, PJM does not build batteries at scale. Flexible load therefore displaces whichever resource a grid's own accreditation rules make marginal, i.e. gas-fired capacity in PJM and batteries in Korea.}

\rev{Temporal shifting reshapes demand around the renewable fleet without reducing it. Interruption is the only mechanism that reduces demand, and the all-interruptible case decreases solar PV expansion from \res{41.1} to \res{35.5~GW} and onshore wind from \res{32.2} to \res{23.2~GW}, because removing the load removes the demand that justified them. Per MWh of data-center demand, the 2038 saving of the reference tier mix is \res{\$10.0/MWh} in Korea and \res{\$5.4/MWh} in PJM. Korea's effect on total cost is nonetheless smaller, \res{2.5\%} against \res{6.0\%}, because data-center load is 10\% of Korean demand and 38\% of PJM's.}

The dispatch and backlog profiles make the \rev{two uses of temporal shifting} directly visible (Fig.~\ref{fig:flex_mechanisms_S2} and supplementary Fig.~\ref{fig:general_dispatch}). In  Korea the shiftable tier is held back in the \rev{evening and at night} and made up during midday solar hours, so the backlog fills and empties on a daily cycle \rev{on almost every representative day}. \rev{In PJM deferral is small on most days and is concentrated in the evening stress hours of the peak net-demand days, so its purpose is to lower the load counted at the system's stressed hours, not to follow an energy price.}

\subsection{\rev{Capacity and compliance drive PJM outcomes, whereas the carbon cap and resource adequacy drive Korean outcomes}}

The \rev{capacity} expansion differences \rev{between the PJM and Korean systems} explain why the same flexibility portfolio produces different grid value. PJM is shaped by a geographically concentrated data-center load \rev{and by large differences between states in clean-energy compliance. Relocating workload to states with lower requirements lowers compliance cost even when it does not lower energy cost, and reducing load in the system's stress hours reduces the generation capacity that must be built.} This spatial value is not about relieving congested transmission lines, for two reasons. First, transmission is a small part of PJM's total system cost.  \rev{Annualized transmission investment is \res{\$0.03B} in 2028, and none is added in 2038 in the all-firm or reference tier mix cases.} Furthermore, \rev{changing every line's usable rating from 70\% to 60\% or 80\% changes PJM's cost by at most \res{0.2\%} in 2028 and \res{0.02\%} in 2038} (see supplementary Fig.~\ref{fig:tx_sweep}: PJM's total cost is nearly flat as the contingency factor $K$ falls from 0.7 to 0.6, whereas  Korea's rises). \rev{Second, the marginal value of relocating a MWh between PJM's zones is set mainly by the price of the compliance obligation that goes with it. The load-weighted price spread between zones is \res{\$2.7--2.9/MWh}, and \res{\$7.5--11.6/MWh} once compliance prices are included (Sec.~\ref{sec:indicators}).} The \rev{main} driver is therefore \rev{where the load counts toward clean-energy requirements and how much of it can be reduced during stress hours,} not transmission \rev{congestion by itself}. This is also why \rev{the reference tier mix in PJM in} 2038 increases its operating cost by \rev{\res{7.9\%}} while \rev{reducing} total cost by \rev{\res{6.0\%}}, with the value concentrated in avoided capacity and compliance cost. 

\rev{On the other hand, o}utcomes in  Korea are shaped by a different constraint set. It has a nuclear-heavy base fleet, large solar PV expansion potential and the  CO\textsubscript{2}e cap that tightens from 189.9~Mt in 2028 to 83.1~Mt in 2038, a 56\% reduction following the 11th BPESD. \rev{In 2028 the cap is comparatively loose and capacity is not scarce, so data-center flexibility only raises the value of additional solar PV, and investment cost increases while total cost decreases slightly.} The tighter 2038 cap forces the system to decarbonize regardless of data-center flexibility. Solar PV is driven to its expansion limit in the all-firm, \rev{all-shiftable} and \rev{reference tier mix} cases, and the \rev{remaining fossil fleet must operate within the carbon cap}. Data-center flexibility therefore no longer changes the solar PV expansion and \rev{instead}  changes how the carbon-constrained system balances that solar PV. \rev{However, this is a planning limit, not a resource or physical constraint. It is the 11th BPESD's national solar PV total of 77.2~GW, which the model takes as the upper bound on installed solar PV, leaving \res{41.1~GW} of new build on top of the existing fleet. The limit is not reached in the all-interruptible case, because curtailing the data-center load removes about 10\% of annual demand, so the carbon cap can be met with less solar PV. With solar PV expansion fixed at its limit, the system needs firm capacity for the evening hours after solar PV output disappears, and resources that let midday solar PV energy serve evening demand. Flexible data-center load supplies both, which is why it replaces batteries. Korea is also more sensitive to its network than PJM. Tightening the usable line rating to 60\% raises the reference tier mix cost by \res{1.9\%} in 2028 and \res{2.6\%} in 2038 (supplementary Fig.~\ref{fig:tx_sweep}).}

\subsection{\rev{Leave-one-out tests itemize impacts of the three mechanisms}}\label{sec:leave-one-out-result}

\begin{table}[!htbp]
\centering
\caption{\rev{\textbf{Contribution of each flexibility mechanism to the saving of the reference tier mix.} Each contribution is the cost increase when that mechanism alone is disabled in the reference tier mix, per MWh of original data-center demand (2023 \$/MWh). In the upper block the CEM is reoptimized, so the contribution includes avoided investment. In the lower block investment is held at the all-firm solution, so only operating and compliance value remains. Contributions need not add to the total saving because the mechanisms overlap. 
}}
\label{tab:leave-one-out}
\begingroup\rescolor
\footnotesize
\setlength{\tabcolsep}{5pt}
\begin{tabular}{@{}lrrrr@{}}
\toprule
& \multicolumn{2}{c}{PJM} & \multicolumn{2}{c}{Korea}\\
\cmidrule(lr){2-3}\cmidrule(lr){4-5}
& 2028 & 2038 & 2028 & 2038\\
\midrule
Total saving against all-firm (\%)  & 0.8 & 6.0 & 0.2 & 2.5\\
Total saving (\$/MWh)               & 1.31 & 5.44 & 0.72 & 9.98\\
\addlinespace
\multicolumn{5}{@{}l}{\emph{Investment reoptimized} (\$/MWh)}\\
Spatial shifting                    & \textbf{0.70} & \textbf{1.18} & 0.01 & 0.10\\
Temporal shifting                   & 0.02 & 1.57 & \textbf{0.08} & \textbf{6.35}\\
Interruption                        & 0.48 & 0.62 & \textbf{0.60} & \textbf{2.69}\\
Capacity credit of both tiers       & \textbf{0.12} & 3.62 & 0.00 & \textbf{6.50}\\
\addlinespace
\multicolumn{5}{@{}l}{\emph{Investment fixed at the all-firm solution} (\$/MWh)}\\
\rev{Spatial shifting}                    & \textbf{0.69} & \textbf{1.26} & 0.01 & 0.05\\
\rev{Temporal shifting}                   & 0.02 & 0.02 & \textbf{0.08} & \textbf{1.63}\\
Interruption                        & 0.47 & 0.47 & \textbf{0.60} & \textbf{1.17}\\
\bottomrule
\end{tabular}
\endgroup
\end{table}

\rev{To attribute value to each mechanism, we disable one mechanism at a time in the reference tier mix, reoptimize investment and operation, and report the cost increase per MWh of original data-center demand (Table~\ref{tab:leave-one-out} and Sec.~\ref{sec:leave-one-out-method}). In PJM, spatial shifting contributes most in 2028 (\res{\$0.70/MWh}), while in 2038 temporal shifting contributes most (\res{\$1.57/MWh}), ahead of spatial shifting (\res{\$1.18/MWh}) and interruption (\res{\$0.62/MWh}). In Korea, spatial shifting contributes almost nothing in either year (\res{\$0.01} and \res{\$0.10/MWh} in 2028 and 2038). Interruption contributes most in 2028 (\res{\$0.60/MWh}), and temporal shifting contributes most in 2038 (\res{\$6.35/MWh}). The contributions do not add up to the total, because each mechanism can stand in for part of another. In PJM in 2038 they sum to \res{\$3.37/MWh} of a \res{\$5.44/MWh} saving, and interruption alone, with both shifting mechanisms disabled, still saves \res{\$2.62/MWh}.}

\rev{The data-center flexibility mechanisms lower different parts of system cost. In PJM in 2038, disabling temporal shifting removes the shiftable tier's \res{22.5~GW} of credit and raises new gas from \res{48.9} to \res{77.7~GW}, so its value comes from avoided capacity, net of higher operating cost. Disabling spatial shifting leaves new gas almost unchanged at \res{48.7~GW} and raises RPS compliance cost by \res{\$0.83B}, more than the \res{\$0.77B} rise in total cost. The same holds in 2028 (\res{\$0.10B} on both counts). The value of spatial shifting in PJM is therefore almost entirely avoided RPS compliance cost. When every relocated MWh keeps its RPS obligation in its home state, while relocation itself remains allowed, the value of spatial shifting decreases by \res{\$1.14} of its \res{\$1.18/MWh} in 2038 and by \res{\$0.56} of its \res{\$0.70/MWh} in 2028. RPS obligations do follow retail sales, so this value is real for the load-serving entities that pay it. However, alternative compliance payments go to state funds. Avoiding them lowers what load-serving entities pay but saves little for society as a whole, because no generation capacity, fuel or capital cost is avoided. Relocating load also changes which state's requirement applies to it, but not how much renewable capacity is built, and renewable builds are the same in these cases. With all obligations held at nominal load, the 2038 saving decreases from \res{\$3.57B} to \res{\$2.75B}, and the relocated volume decreases from \res{47.6} to \res{8.4~TWh}. We report both, because the appropriate basis depends on whether a state regulator would let an obligation leave with the load.}

\subsection{\rev{Indicators computed without data-center flexibility anticipate which flexibility mechanism is valuable}}\label{sec:indicators}

\begin{table}[!htbp]
\centering
\caption{\rev{\textbf{Transferable indicators.} Price indicators are computed from economic dispatch with investments fixed at the all-firm solution, so no data-center flexibility is modeled. The 2026 column is a benchmark on the 2026 fleet with no new investment. Compliance-adjusted prices add to each location's price the compliance cost of a MWh of load there; Korea has no RPS, so its two price measures coincide. Shadow prices of the planning constraints come from the all-firm CEM, because the adequacy dual is not unique once investments are fixed.}}
\label{tab:indicators}
\begingroup\rescolor
\footnotesize
\setlength{\tabcolsep}{3pt}
\begin{tabular}{@{}llrrrrrr@{}}
\toprule
Mechanism & Indicator & \multicolumn{3}{c}{PJM} & \multicolumn{3}{c}{Korea}\\
\cmidrule(lr){3-5}\cmidrule(lr){6-8}
& & 2026 & 2028 & 2038 & 2026 & 2028 & 2038\\
\midrule
Spatial & Price spread between load locations (\$/MWh) & 2.7 & 2.7 & 2.9 & 1.2 & 1.5 & 3.2\\
 & Compliance-adjusted spread (\$/MWh) & 7.6 & 7.5 & 11.6 & 1.2 & 1.5 & 3.2\\
 & Congestion markup on line flows (\$/MWh) & 0.9 & 1.0 & 1.0 & 1.3 & 1.2 & 1.9\\
\addlinespace
Temporal & Hourly s.d.\ of system price (\$/MWh) & 4.0 & 4.2 & 6.3 & 3.4 & 3.9 & 22.9\\
 & Wind + solar PV capacity / peak (\%) & 23 & 34 & 39 & 38 & 41 & 88\\
\addlinespace
Interruption & Load-weighted price (\$/MWh) & 34 & 33 & 32 & 64 & 57 & 103\\
 & Compliance-adjusted load-weighted price (\$/MWh) & 43 & 41 & 42 & 64 & 57 & 103\\
\addlinespace
Capacity & Adequacy shadow price (\$ thousand/MW-yr) & -- & 13 & 105 & -- & 0 & 92\\
 & Carbon shadow price (\$/tCO\textsubscript{2}) & -- & -- & -- & -- & 10 & 160\\
\bottomrule
\end{tabular}
\endgroup
\end{table}

\begin{figure*}[!htbp]
\centering
\includegraphics[width=\linewidth]{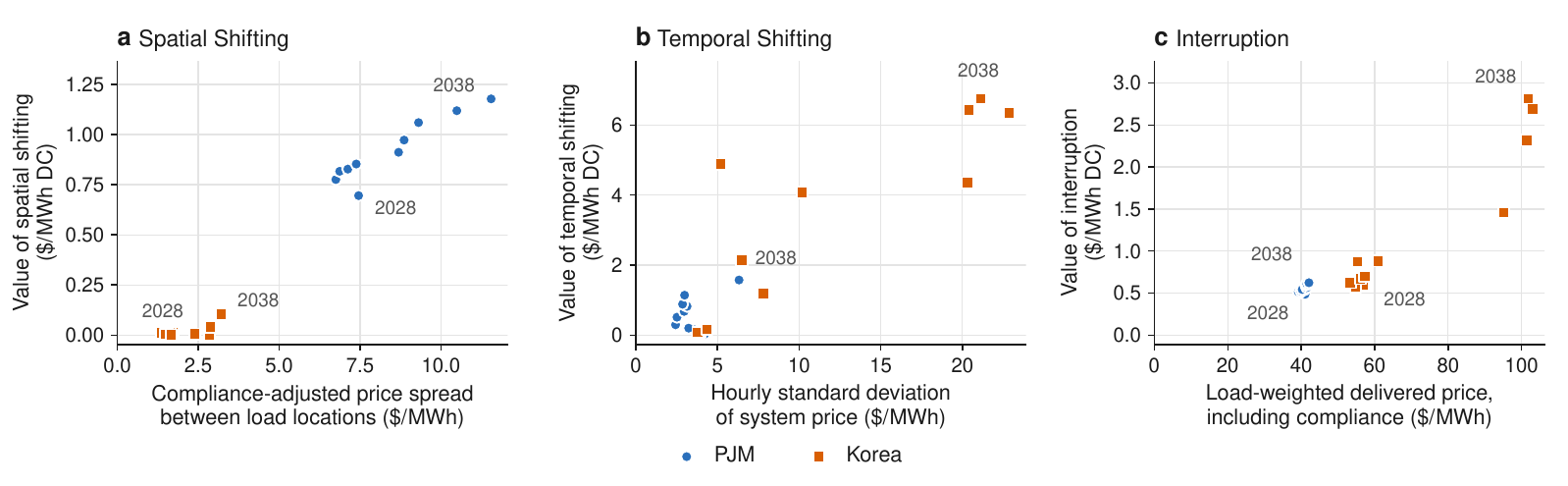}
\caption{\rev{\textbf{Indicators computed without flexibility against the value of each mechanism, 2028--2038.} Each point represents one target year on one grid. Horizontal-axis price indicators are computed by fixing investments at the all-firm CEM solution and reoptimizing dispatch. The vertical axis is the increase in total annual system cost when spatial shifting (a), temporal shifting (b), or interruption (c) is disabled in the reference tier mix, normalized by original data-center demand. Investments and operations are reoptimized in both the reference and leave-one-out cases, with policy constraints and performance-based capacity accreditation retained.
}}
\label{fig:indicators}
\end{figure*}

\rev{Table~\ref{tab:leave-one-out} reports results for two target years, so we ask whether quantities a planner can compute without modeling data-center flexibility help explain each mechanism's value across a broader set of conditions. For each target year from 2028 to 2038, we solve an all-firm CEM to establish the benchmark system and derive its price indicators. Planning-constraint shadow prices are taken from the all-firm CEM. We compare these indicators with leave-one-out values in the reference tier mix of the same year, independently reoptimizing generation, storage and transmission investments together with operations in both the reference and leave-one-out cases (Table~\ref{tab:indicators}, Fig.~\ref{fig:indicators} and Sec.~\ref{sec:leave-one-out-method}). Each leave-one-out value is the resulting increase in optimized total cost per MWh of data-center demand, including changes in operating cost, investment and policy compliance.}

\rev{\emph{Spatial shifting} remains closely associated with the spread in delivered price between load locations, provided compliance is included (Fig.~\ref{fig:indicators}a). The plain price spread does not track PJM's spatial value across the 11 years (\res{$r=-0.02$}), and neither does the congestion markup on line flows (\res{$r=-0.23$}), but the compliance-adjusted spread does (\res{$r=0.94$}). PJM's adjusted spread of \res{\$6.8--11.6/MWh} accompanies spatial values of \res{\$0.70--1.18/MWh} of data-center demand. Korea has no RPS constraints, smaller spreads of \res{\$1.4--3.2/MWh}, and spatial values of at most \res{\$0.10/MWh}, with a weaker within-grid association (\res{$r=0.67$}). The pooled correlation is high (\res{$r=0.98$}), although it also reflects the separation between the two systems and does not establish a common linear calibration.}

\rev{\emph{Temporal shifting} depends on both hourly operating opportunities and the capacity that shifting can replace (Fig.~\ref{fig:indicators}b). Hourly price variability is associated with its reoptimized value in Korea (\res{$r=0.85$}), but the relationship is weaker in PJM (\res{$r=0.39$}). Low price variability does not necessarily imply low planning value: in Korea in 2031, the hourly standard deviation is only \res{\$5.21/MWh}, while temporal value reaches \res{\$4.88/MWh} of data-center demand. Disabling temporal shifting also removes the shiftable tier's performance-based capacity credit and changes the optimal generation and storage portfolio. In Korea in 2038, removing temporal shifting raises battery additions from \res{1.18 to 5.22~GW}, giving a temporal value of \res{\$6.35/MWh}. In PJM, temporal value is more closely associated with the all-firm adequacy shadow price (\res{$r=0.98$}) than with hourly price variability.}

\rev{Wind and solar PV capacity relative to peak provides a complementary structural indicator of temporal value. Korea's ratio rises from \res{41\%} to \res{88\%}, and its correlation with reoptimized temporal value is \res{$r=0.90$}. However, the relationship is not monotonic: annual changes in available candidates and adequacy needs alter the portfolio that flexibility replaces. PJM's ratio remains within \res{34--46\%} and has little association with temporal value (\res{$r=0.06$}). The ratio thus helps describe Korea's increasing need for balancing resources, but does not provide a common predictor across both systems. Here it includes existing and newly built wind and solar PV capacity in the all-firm solution; it is available without modeling flexible data centers, but still requires an expansion-model solution.}

\rev{\emph{Interruption} remains associated with the compliance-adjusted level of the delivered price (Fig.~\ref{fig:indicators}c), with \res{$r=0.77$} in PJM and \res{$r=0.95$} in Korea. Without the compliance adjustment, the correlation in PJM is negative (\res{$r=-0.83$}), consistent with the importance of avoided RPS obligations in addition to energy costs. The reference tier mix may interrupt 1\% of total data-center energy, making 1\% of the average delivered price a useful operating-value benchmark. In Korea in 2038, the benchmark is \res{\$1.03/MWh}, while the reoptimized value is \res{\$2.69/MWh}. Moreover, interruption is valued conditional on spatial and temporal flexibility remaining available, so its incremental benefit need not equal an estimate based on all-firm prices.}

\rev{Together, these results suggest using the indicators to identify the means through which flexibility creates value. Compliance-adjusted spatial price differences reveal opportunities for relocating workloads, while hourly price variation and renewable capacity relative to peak help characterize opportunities for temporal shifting. Adequacy shadow prices provide additional information when flexibility can replace accredited capacity, and delivered prices indicate the operating benefit of interruption. Their usefulness depends on the mechanism and grid: operating-price indicators alone do not determine the full value when investments respond. These are descriptive relationships from two systems, over annual scenarios that share partly interpolated inputs and common trends. They do not establish causal effects, universal thresholds, or predictive performance on an additional grid.}

\subsection{ The value of the interruptible tier depends on event-shape limits} \label{sec:inter-event-shape}

\begin{figure*}[!htbp]
\centering
\centering
\includegraphics[width=\linewidth]{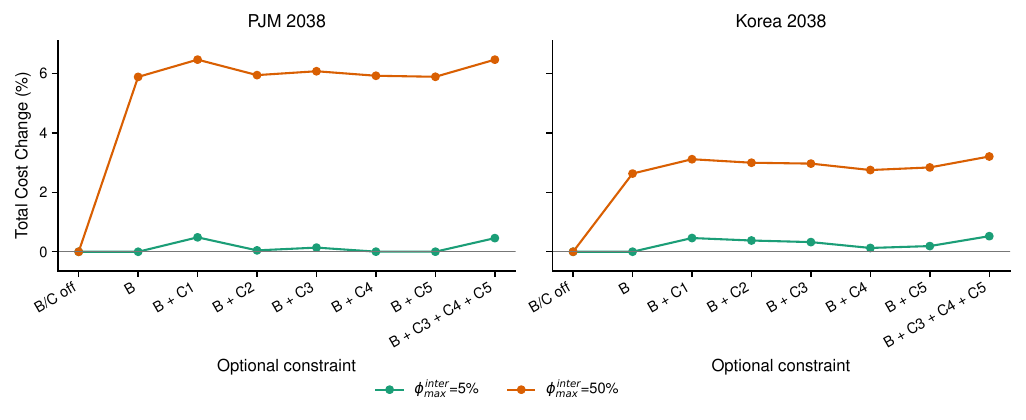}
\caption{\textbf{Effect of interruption limits on total cost in 2038.} Each point adds one optional constraint on the interruptible tier to the \rev{annual interruptible-energy cap} ($\phi^{\mathrm{inter}}_{\max}$), and the rightmost point combines three of them. Layer~B caps the number of interruption hours per year. C1 caps how deeply power withdrawal may be cut in any hour, C2 restricts interruption to a fixed time-of-day window, C3 caps the hour-to-hour ramp, C4 caps consecutive interruption hours, and C5 sets a minimum \rev{space between events (recovery) starts (Sec.~\ref{sec:flexmodel}). Cost changes are relative to the reference tier mix at the same annual interruptible-energy cap with no optional constraint (B/C off). At a \res{5\%} annual interruptible-energy cap, the hour cap alone has negligible effect, but depth and combined shape limits can raise cost by about \res{0.5\%}. At \res{50\%}, the hour cap accounts for most of the increase, with shape limits adding further cost.}}
\label{fig:eventshapesweep}
\end{figure*}

The all-interruptible case is consistently  the cheapest among all simulations (Tables~\ref{tab:s1_2028} and~\ref{tab:s2_2038}). It reduces total cost by \rev{\res{24\%}} and  \rev{\res{57\%}} in PJM in 2028 and 2038 and by  \rev{\res{13\%}} and \rev{\res{25\%}} in  Korea. However, with the \rev{annual interruptible-energy cap} set to one and no event-shape limits, a fully interruptible tier is equivalent to removing the data-center load. \rev{It is an unconstrained upper bound. Under the 5\% cap an all-interruptible tier saves \res{1.5\%} and \res{7.2\%} in PJM and \res{0.7\%} and \res{3.3\%} in Korea.} The \rev{reference tier} mix uses far less \rev{interruption}, at \rev{\res{1.4~TWh} and \res{6.6~TWh}} in PJM in 2028 and 2038 and \rev{\res{0.43~TWh} and \res{0.78~TWh}} in  Korea in 2028 and 2038, because the 5\%  \rev{cap} binds in every case.

This gap in outcomes with different assumptions on interruptible data-center flexibility   shows why this flexibility tier  must be modeled with realistic service-quality limits rather than as an unconstrained (on/off) block (as in \cite{norris2025rethinking}). To test the effects of the event-shape limits for the interruptible tier \rev{we raise the annual interruptible-energy cap of the reference tier mix to 50\%, add an annual cap of 480 interruption hours (Layer~B), and then add the event-shape limits one at a time (Layer~C, Methods \ref{sec:flexmodel}).}

\rev{At the 5\% annual interruptible-energy  cap, the cap of 480 interruption hours per year alone has negligible cost, but event-shape limits can still increase cost. In PJM in 2028, adding the depth cap increases cost by \res{0.2\%}, while the time-of-day window, duration limit and spacing between event starts each add approximately \res{0.01\%} or less, within the 0.1\% optimality gap to which these mixed-integer program runs are solved. In 2038, the depth cap increases cost by about \res{0.5\%} in both systems. At looser annual interruptible-energy caps, the annual interruptible-hour cap becomes more restrictive. In PJM in 2028, it alone raises cost by about \res{2.6\%} at a 50\% cap and \res{4.3\%} at 80\%, and with the depth cap by \res{2.8\%} and \res{4.5\%}. Since 480 hours are 5.5\% of the year, the hour cap reduces interrupted energy from \res{14.2} to \res{1.4~TWh} at the 50\% cap, removing almost all of the benefit of the looser annual interruptible-energy cap. An event-shape limit can also increase cost through capacity accreditation. In Korea in 2038, restricting interruption to a time-of-day window from 16:00 to 21:00 increases cost by \res{0.4\%} even at the 5\% cap, because some stress hours fall outside the window. Thus, the interruptible tier then earns no capacity credit, and \res{1.4~GW} more batteries are built.} Ignoring these limits entirely, as the all-interruptible upper-bound case does, overstates how much flexibility is actually available.

\subsection{ The least-cost \rev{shiftable}/interruptible composition is system- and policy-dependent} \label{sec:sensitivity}

\begin{figure*}[!htbp]
\centering
\begin{subfigure}[t]{0.91\textwidth}
\centering
\includegraphics[width=0.9\linewidth]{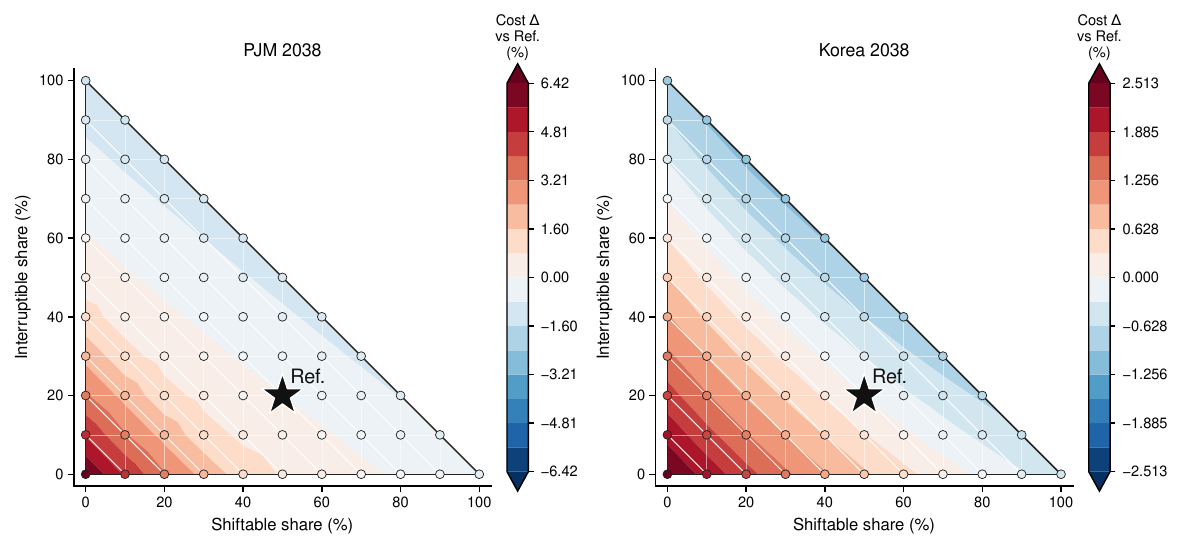}
\caption{Total cost across tier shares (2038\rev{, annual interruptible-energy cap 5\%}).}
\end{subfigure}
\vspace{2mm}
\begin{subfigure}[t]{0.91\textwidth}
\centering
\includegraphics[width=0.9\linewidth]{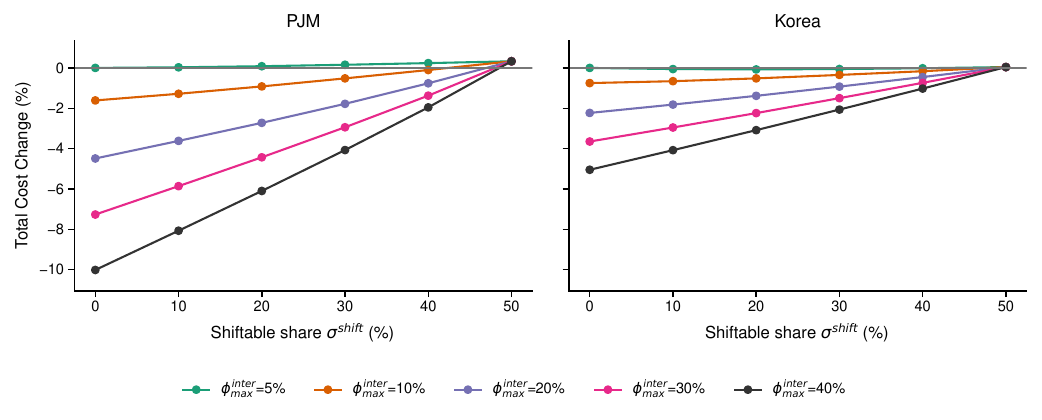}
\caption{Total cost across \rev{shiftable shares with the firm share fixed at 50\%, at five annual interruptible-energy caps (2038).}}
\end{subfigure}
\caption{\textbf{Tier-share sensitivities in 2038.} (a) \rev{Shiftable} and
interruptible shares varied at 10\% granularity, with the firm share as the
residual. Color is the interpolated cost change relative to the \rev{reference tier} mix
(firm: 30\%, \rev{shiftable}: 50\%, interruptible: 20\%), marked by a star. \rev{This panel keeps the annual interruptible-energy cap at 5\%. It therefore differs from the all-interruptible case of Tables~\ref{tab:s1_2028} and~\ref{tab:s2_2038}, which relaxes the cap to 100\% and is an unconstrained upper bound.} %In PJM in 2038, all-interruptible under the 5\% cap saves \res{\$4.3B}, against \res{\$3.9B} for all-shiftable and \res{\$33.9B} for the upper bound. The two panels use different color scales: the range is \res{$-1.3\%$ to $+6.4\%$} in PJM and \res{$-1.0\%$ to $+2.5\%$} in Korea.} 
(b) The \rev{shiftable} share varied from 0 to 50\% with the firm share fixed at 50\%, at five values of the \rev{annual interruptible-energy cap. Cost is shown relative to the composition with 50\% firm and 50\% interruptible load at a 5\% cap.} All five curves meet at a 50\% \rev{shiftable} share, where no interruptible tier remains and the cap has no effect.}
\label{fig:sensitivity_updates}
\end{figure*}

\rev{Fig.~\ref{fig:sensitivity_updates}a shows how total cost changes as the three tier shares vary, at the base annual interruptible-energy cap of 5\%. In both systems the firm share matters most. Total cost decreases steadily as the firm share decreases, and the contours run close to lines of constant firm share. In PJM, cost ranges from \res{6.4\%} above the reference tier mix with all load firm to \res{1.3\%} below it with all load interruptible. In Korea the range is from \res{2.5\%} above to \res{1.0\%} below. At a given firm share, the cheapest split between the other two tiers differs between the systems. In PJM, cost decreases as interruptible load replaces shiftable load. With no firm load, it is \res{0.7\%} below the reference tier mix when all load is shiftable and \res{1.3\%} below when all is interruptible. In Korea, a mix is cheapest, which with no firm load is 20\% shiftable and 80\% interruptible, because temporal shifting has an operating value under the carbon cap that interruption under the 5\% cap cannot supply.}

\rev{Within a fixed firm share, the least-cost split between the shiftable and interruptible tiers depends mainly on the annual interruptible-energy cap $\phi^{\mathrm{inter}}_{\max}$, not on the share of load assigned to the interruptible tier by itself} (Fig.~\ref{fig:sensitivity_updates}b). \rev{With the firm share fixed at 50\% and a 5\% cap, total cost in 2038 varies across the six compositions by only \res{0.3\%} in PJM, where the composition 50/0/50 is cheapest, and by \res{0.1\%} in Korea, where 50/20/30 is cheapest (compositions are written as firm/shiftable/interruptible, in percent). Relaxing the annual interruptible-energy cap makes 50/0/50 the cheapest composition in both systems from a 10\% cap upward, and it widens the gap: at a 40\% cap that composition costs \res{10\%} less than at 5\% in PJM and \res{5\%} less in Korea, while 50/50/0 is unaffected, since a composition with no interruptible tier does not depend on the cap.}

\rev{This does not mean that the shiftable tier has no value.} When a large interruptible share is allowed and interruption carries little penalty, the cap is what sets the cheapest composition. Neither endpoint is a portfolio a data-center operator could actually offer, since the share of workload that tolerates being switched off is bounded well below one. Planners should therefore value the two tiers separately and design connection rules around the composition a data-center operator can credibly supply.

\subsection{\rev{Sensitivities rank what drives the value of flexibility}}\label{sec:robust}

\begin{table}[!htbp]
\centering

\caption{\rev{\textbf{Sensitivity of the value of flexibility.} Saving of the reference tier mix against all-firm (\% of all-firm cost) under each change, ranked by the largest departure of the 2038 saving from its base value on either grid. Each case with a changed system or policy has its own all-firm run. Type marks a system characteristic (S), a policy, market or accreditation rule (P), a flexibility assumption (F) or a modeling choice (M). A dash means the factor does not apply to that grid or year.}}
\label{tab:ranking}
\begingroup\rescolor
\scriptsize
\setlength{\tabcolsep}{2.5pt}
\begin{tabular}{@{}>{\raggedright\arraybackslash}p{0.22\linewidth}c>{\raggedright\arraybackslash}p{0.29\linewidth}rrrr@{}}
\toprule
& & & \multicolumn{2}{c}{PJM} & \multicolumn{2}{c}{Korea}\\
\cmidrule(lr){4-5}\cmidrule(lr){6-7}
Factor & Type & Range tested & 2028 & 2038 & 2028 & 2038\\
\midrule
\emph{Base case} & & & 0.83 & 6.03 & 0.16 & 2.45\\
\addlinespace
Gas candidates in the queue & S & Cycle~1 removed (12.1~GW of gas left in 2038) & 1.15 & 25.4 & -- & --\\
Annual interruptible-energy cap & F & 1\%, 10\%, 20\%, 100\% (base 5\%) & 0.59--5.71 & 5.49--16.5 & 0.06--2.60 & 1.97--7.26\\
Data-center capacity credit & P & None; fixed at a class rating & 0.75--0.83 & 2.01--6.05 & 0.16 & 0.85--1.77\\
Stress hours for the credit & P & 5, 25, 100 (base 10) & 0.77--0.83 & 2.47--6.04 & 0.16 & 1.10--2.45\\
Mechanism disabled & F & Spatial; temporal; both & 0.37--0.82 & 2.90--4.72 & 0.13--0.16 & 0.79--2.43\\
Battery cost and duration & S & ATB~2024, ATB~2025; 2--10~h & 0.83 & 6.03 & 0.16 & 1.85--5.35\\
Data-center load & S & 25\%, 50\% of the forecast & 0.15--0.43 & 3.15--4.82 & 0.04--0.08 & 0.76--1.38\\
Temporal and spatial costs & F
& \$0.6--2,000/MWh each (49 cases)
& 0.37--1.19 & 2.90--6.58 & 0.13--0.19 & 0.79--2.81\\
RPS regime & P & None; hard, with generic resources & 0.37--0.39 & 5.26--7.54 & -- & --\\
RPS obligation basis & P & Nominal load & 0.43 & 4.64 & -- & --\\
Shift cap $\sigma^{\mathrm{sh}}_{\max}$ & F & 0.01--0.5 (base 0.3) & 0.41--1.12 & 4.77--6.89 & 0.16 & 2.43--2.45\\
Tier shares & F & All shiftable; all interruptible at a 5\% cap & 0.97--1.54 & 6.64--7.23 & 0.05--0.67 & 2.91--3.34\\
ATB cost case & S & Advanced; Conservative & 0.83--0.84 & 5.90--6.06 & 0.16--0.17 & 1.87--3.15\\
Storage mandate & P & State targets as hard limits & 0.75 & 5.33 & -- & --\\
Forecast pool requirement & P & 0.9232 (base 0.9401) & 0.76 & 5.36 & -- & --\\
Make-up horizon & F & 4, 12~h (base 24~h) & 0.82--0.83 & 5.92--6.01 & 0.16 & 1.79--2.40\\
ACP level & P & $\times 0.5$; $\times 2$ & 0.61--1.26 & 5.67--6.64 & -- & --\\
Usable line rating $K$ & M & 0.6, 0.8 (base 0.7) & 0.83 & 5.96--6.16 & 0.16--0.17 & 1.99--2.74\\
Representative days & M & Other capacity bases; added stress days & 0.81--0.85 & 6.04--6.09 & 0.16 & 2.44--2.45\\
Value of lost load & P & \$4,000--10,000/MWh nominal & 0.83 & 6.05--6.07 & 0.16 & 2.45\\
Investment tax credit & P & Batteries and nuclear, 2028 & 0.83 & -- & 0.16 & --\\
\bottomrule
\end{tabular}
\endgroup
\end{table}

\rev{Table~\ref{tab:ranking} ranks the factors tested, by how far they move the 2038 saving on either grid. The largest effects come from the system and from its accreditation rules, together with the annual interruptible-energy cap. The candidate set is the largest: without PJM's Cycle~1 interconnection queue projects only 12.1~GW of gas candidates remain, and the 2038 saving rises to \res{\$22B} (\res{25\%}), of which \res{\$14B} is avoided load shedding and \res{\$5B} avoided resource-adequacy shortfall, so that case bounds queue attrition and does not estimate the value of flexibility. The capacity credit and the number of stress hours over which it must be earned come next. With no credit the 2038 saving is \res{2.0\%} in PJM and \res{0.9\%} in Korea. Earning the credit over 5 hours instead of 10 changes nothing. Over 100 hours, which stand for \res{2,500} hours of the year in PJM, the saving falls to \res{2.5\%} in PJM and \res{1.1\%} in Korea, because deferred work must be made up within the day and the interruptible tier's annual budget is spread thin. The credit is therefore valuable when system stress is concentrated in a few hours, as in the net-load profiles used here. With 25 stress hours the saving is still \res{5.9\%} in PJM and \res{2.2\%} in Korea. In Korea the battery assumptions matter as much as the credit: across ATB~2024 and ATB~2025 costs and durations of 2 to 10~h, the 2038 saving ranges from \res{1.8\%} to \res{5.4\%}, since the credit replaces batteries and, under the Korean adequacy rules, a longer battery costs more for the same credit. The all-firm case builds \res{6.7--8.7~GW} of batteries in all of these cases and the reference tier mix \res{0.6--3.8~GW}. In PJM no battery is built in any of them, with or without flexibility, and the saving does not move.}

\rev{Data-center penetration changes the total saving but has little effect on the saving per MWh of data-center demand. Scaling PJM's data-center load to 25\% and 50\% of the forecast lowers the 2038 saving from \res{6.0\%} to \res{3.2\%} and \res{4.8\%} of system cost, but per MWh of data-center demand it is \res{\$6.4} and \res{\$6.2/MWh}, against \res{\$5.4/MWh} in the base case. In Korea the same scaling gives \res{\$10.0} and \res{\$9.7/MWh}, against \res{\$10.0/MWh}. At 25\% of its forecast, PJM's data-center load is about 10\% of its peak, close to Korea's 7\%, and its saving per MWh is still below Korea's, so the difference between the systems is not a matter of penetration.}

\begin{figure*}[!htbp]
\centering
\begin{subfigure}[t]{0.96\textwidth}
\centering
\includegraphics[width=\linewidth]{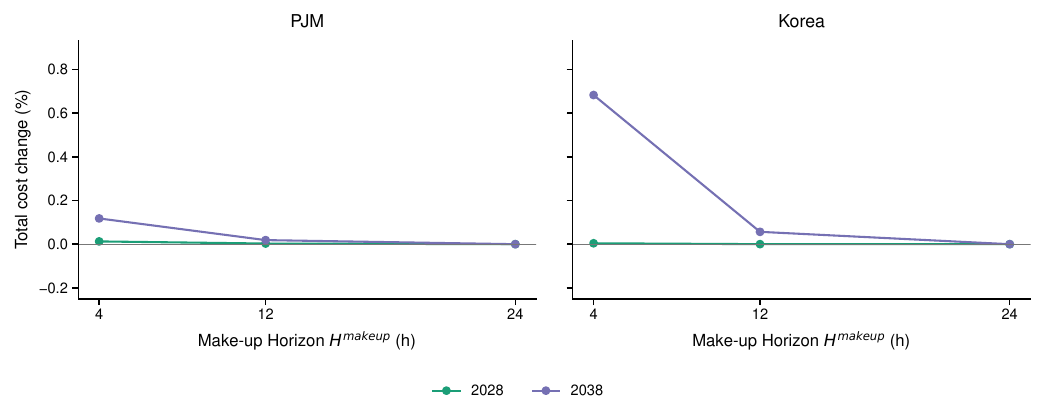}
\caption{Make-up horizon sensitivities}
\end{subfigure}
\vspace{2mm}
\begin{subfigure}[t]{0.96\textwidth}
\centering
\includegraphics[width=\linewidth]{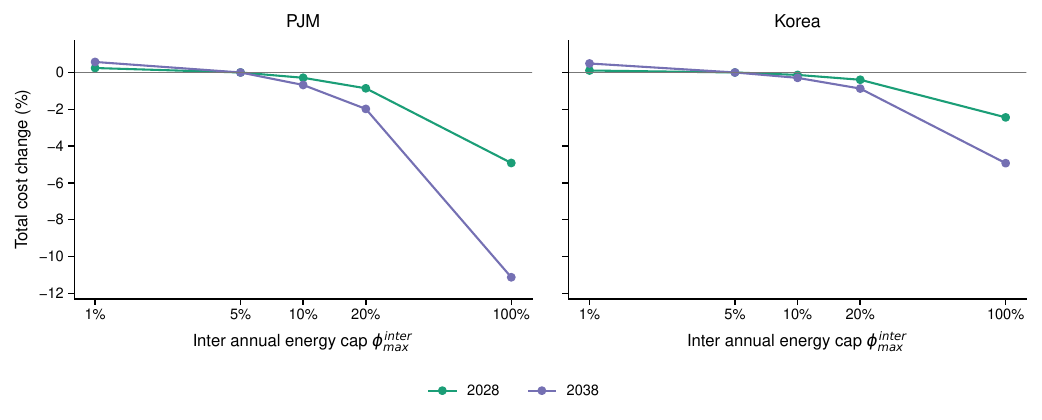}
\caption{\rev{Annual interruptible-energy} cap sensitivities}
\end{subfigure}
\caption{\textbf{Make-up horizon and \rev{annual interruptible-energy cap} sensitivities.} \rev{Change in the total cost of the reference tier mix when} the make-up horizon ($H^{\mathrm{makeup}}$) and the \rev{annual interruptible-energy cap} 
($\phi^{\mathrm{inter}}_{\max}$) \rev{depart from their base values of 24~h and 5\%.} 
Korean 2038 cost rises by \rev{\res{0.7\%}} as $H^{\mathrm{makeup}}$ falls from 24 to 4~h,
consistent with its temporal (storage-like) value being largely intraday,
whereas PJM's cost \rev{rises by \res{0.1\%}}.}
\label{fig:hdef_phi_sweeps}
\end{figure*}

\rev{The flexibility assumptions other than the interruptible-energy cap matter less. Across deferral and relocation costs from \$0.6 to \$100/MWh each, a range of more than 150 times, the 2038 saving stays between \res{4.3\%} and \res{6.6\%} in PJM and between \res{1.5\%} and \res{2.8\%} in Korea. Even at a deferral cost of \$1,000--2,000/MWh, at which no load is deferred, it is \res{3.7\%} and \res{0.9\%}. The volumes change far more than the saving does. PJM's temporal shifting (deferred volume) decreases from \res{58} to \res{2.4~TWh} and its spatial shifting (relocated volume) from \res{48~TWh} to zero. The contrast between the grids does not depend on the lower cost given to spatial shifting in the base case. With both costs at \$10/MWh, PJM still relocates \res{41~TWh} and defers \res{12~TWh}, and with relocation at \$0.6/MWh Korea relocates only \res{1--2~TWh}. The total saving also depends little on whether data centers can relocate load at all. At a shift cap of 1\% of the shiftable tier, which allows almost no relocation, PJM's 2038 saving is \res{4.8\%}, and at 50\% it is \res{6.9\%}. PJM's relocated volume changes in proportion to the cap, from \res{1.6} to \res{79~TWh}, while Korea's saving stays at \res{2.4--2.5\%} throughout. This matters because evidence for relocating workloads between sites is limited to inference, and because the mix of tiers is a scenario. The interruptible-energy cap is the exception. Raising it from 5\% to 20\% raises the 2038 saving from \res{6.0\%} to \res{7.9\%} in PJM and from \res{2.5\%} to \res{3.3\%} in Korea, and removing it gives \res{16\%} and \res{7.3\%} (Fig.~\ref{fig:hdef_phi_sweeps}).}

\rev{The RPS policy sensitivity results answer whether PJM's value depends on allowing RPS shortfalls. Doubling or halving every ACP moves the 2038 saving to \res{\$4.5B} or \res{\$3.1B}, against \res{\$3.6B}. Removing the RPS gives \res{\$2.7B}. A hard RPS with unlimited generic solar PV and wind in each state, under which no shortfall occurs, gives \res{\$3.5B} in 2038, close to the base case, but only \res{\$0.08B} in 2028, against \res{\$0.19B}. In that case the 2038 reference tier mix builds no gas generation capacity at all and the all-firm case builds \res{30~GW}, so a strict policy does not by itself make data-center flexibility less valuable in PJM by 2038. With generic resources available the hard requirement and its control, which allows shortfalls at the ACP, give identical results, because building is cheaper than paying the ACP. A lower forecast pool requirement and a state storage mandate imposed as a hard constraint lower the 2038 saving only slightly, to \res{\$3.1B} and \res{\$3.2B}.}

\rev{Korea is more sensitive than PJM to the network and to the make-up horizon of the shiftable tier. Tightening the usable line rating from 70\% to 60\% raises Korea's reference tier mix cost by \res{1.9\%} in 2028 and \res{2.6\%} in 2038, and relaxing it to 80\% lowers it by \res{0.5\%} and \res{1.5\%}; PJM changes by at most \res{0.2\%} (supplementary Fig.~\ref{fig:tx_sweep}). Reducing the make-up horizon from 24 to 4~h raises Korea's 2038 cost by \res{0.7\%}: temporal shifting falls from \res{9.0} to \res{4.4~TWh}, the credit from \res{6.2} to \res{4.3~GW}, and \res{1.5~GW} more batteries are built (Fig.~\ref{fig:hdef_phi_sweeps}). In PJM the same change raises 2038 cost by \res{0.1\%}, as the shiftable tier's credit falls by \res{1.4~GW} and \res{1.9~GW} more gas is built. Neither grid responds to the value of lost load, since no firm load is shed in the reference tier mix, or to the choice of representative days: selecting them on another capacity basis or adding the year's highest net-load days changes the 2038 saving by at most \res{\$0.04B} in PJM and \res{\$0.005B} in Korea.}

\rev{Full-year runs over all 8,760 hours give 2038 savings of \res{6.03\%} in PJM and \res{2.33\%} in Korea, close to the \res{6.0\%} and \res{2.5\%} from the representative days (Sec.~\ref{sec:repdays}), and show how often each mechanism is used (Supplementary Fig.~\ref{fig:duration_curves}). In PJM, spatial shifting runs at a nearly constant \res{1.1~GW} in 2028, and in 2038 at \res{4.3--5.5~GW} in every hour of the year and at \res{5.5~GW} in about \res{6,300~h}, while temporal shifting and interruption reach their largest values in only a few hundred hours. Spatial shifting in PJM therefore acts like siting the load in another zone rather than responding to conditions in particular hours, and its value comes from RPS obligations that apply in every hour. Raising the cost of spatial shifting from \$2 to \$10/MWh decreases relocated energy by only \res{18\%} in 2028 and \res{15\%} in 2038. In 2038 almost all of this decrease is load relocated out of ComEd, which decreases from about \res{1.2~GW} to nearly zero in most hours, while load relocated from other zones stays near \res{4.1~GW}. The saving from relocating ComEd load is therefore below \$10/MWh in most hours, whereas relocating load out of other zones remains worthwhile at the higher cost. In Korea, spatial shifting is small, about \res{0.1~GW}, and at the higher cost it decreases by \res{94\%} in 2028 and, with deferral at \$2/MWh, by \res{77\%} in 2038, which agrees with the small price spreads between load locations in Korea. In Korea in 2038, temporal shifting uses the full shiftable tier for about \res{900~h} a year, deferring load into midday solar hours each day, whereas in PJM it acts mainly in the highest net-load hours, as a capacity resource would. Lowering the deferral cost from \$10 to \$2/MWh raises PJM's deferred energy from \res{12.9} to \res{45.8~TWh} and Korea's from \res{8.4} to \res{12.5~TWh}. At \$100/MWh for both shifting mechanisms, PJM relocates no load and defers in only \res{10}~h, as many as its stress hours, where deferral still earns capacity credit, and Korea defers its full shiftable tier in only \res{57}~h. Interruption uses its full annual cap in every case in both systems.}

\section{Discussion}\label{sec3}

\rev{Accounting for data-center flexibility in CEM shows that the flexibility mechanisms lower different components of system cost. Temporal shifting (or deferral) and interruption during stress hours reduce the required accredited capacity that must be built. Spatial shifting (or relocation) and interruption reduce RPS compliance cost in PJM, and temporal shifting in Korea defers load into midday solar hours, where the carbon cap drives solar PV capacity deployment. Which of these mechanisms matters most differs between grids and changes over time within each system. In 2028, when PJM's existing supply already comes close to meeting its resource-adequacy requirement and Korea's fully meets it, flexibility is worth less than 1\% of system cost. In PJM most of that comes from spatial shifting, which lowers compliance cost by relocating RPS obligations between states, while interruption also lowers RPS obligations by reducing the load served. By 2038 the resource-adequacy constraints become active in both systems. In Korea the adequacy constraint has a shadow price of \res{\$92,000/MW-yr}, while PJM has its shortfall price of \res{\$105,000/MW-yr}. In both systems, removing the capacity credit that the flexible tiers earn in stress hours removes about two thirds of the  cost saving. What that credit replaces depends on each system's own accreditation rules. It is natural gas generation capacity in PJM, where a 4-hour battery is rated at only 22\% of its installed capacity, and batteries in Korea, where storage is credited at its full installed capacity. In PJM in 2038, spatial shifting relocates nearly four times as much energy as temporal shifting defers (\res{47.6} against \res{12.2~TWh}) but contributes less (\res{\$1.18} against \res{\$1.57/MWh}), so the volume of flexible energy is a poor guide to its value. With investment held fixed at the all-firm solution, spatial shifting retains the most operating and compliance value in PJM in both years, whereas in Korea interruption does in 2028 and temporal shifting in 2038 (Table~\ref{tab:leave-one-out}). These results indicate that each mechanism should be valued against the constraint it relieves on a given grid and in a given year, not through one product that pays every mechanism alike, such as a single capacity payment per MW of flexible load.}

\rev{These results inform how data-center flexibility should be procured and accredited. First, spatial, temporal and interruptible value can be supported by different instruments. Capacity accreditation based on the contributions during stress hours would capture most of the temporal and interruptible value in both systems, without paying for flexibility the system does not use. A product that rewards intraday load shaping would enhance Korea's operating value. Second, accreditation rules decide what flexible load displaces, so the rules for flexible load and for storage should be set together. Flexible load is worth more per MWh in Korea than in PJM largely because the resource it displaces there, a battery credited at its full capacity, is the cheapest source of accredited capacity. Under duration-based ratings such as in PJM, if applied in Korea, batteries would earn less credit and flexible load could displace a different resource. Third, regulators should decide whether RPS obligations follow relocated load. Most of PJM's relocated volume and nearly all of its spatial value depend on an obligation leaving with the load, from Virginia and Illinois toward Ohio and Pennsylvania. This lowers compliance cost for data centers and their suppliers, but it is largely a transfer, and it leaves renewable builds unchanged in our runs. A state seeking to keep the obligation would remove most of that incentive. A recent field demonstration relocated inference workload out of Dominion during a winter peak, which shows that such relocation can be operated \cite{williams2026power}. Our model also relocates load out of Dominion, but mainly to AEP and to lower compliance cost rather than to relieve a peak. Fourth, valuing flexibility against capacity alone can raise emissions. Without a carbon limit, the gas combined-cycle plants that flexibility avoids would have displaced older coal-fired plants, and PJM's 2038 emissions rise by a third. Flexibility rules should therefore be interpreted in the context of carbon policy and generation retirement plans \cite{DOE2025GridReliability}.} \rev{Fifth}, interruptible load must be accredited with realistic event-shape assumptions. At high \rev{annual interruptible-energy caps}, omitting the annual-hour, depth, ramp and recovery limits materially overstates its value\rev{, and a time-of-day window that excludes stress hours removes the tier's capacity credit}.

\rev{Finally, the differences between the data-center flexibility mechanisms affect how the model's transmission results should be interpreted. Spatial shifting could substitute for transmission only where it relieves congestion. In PJM spatial shifting does not relieve congestion, and the model adds almost no transmission in any case, so data-center flexibility neither replaces nor requires new transmission lines there. In Korea, transmission line additions reach \res{58--88~GW} by 2038, and they are somewhat larger with flexibility than without (\res{85} against \res{76~GW} in the reference tier mix), consistent with transmission lines partly substituting for the \res{5.8~GW} of batteries that the reference tier mix no longer builds. These transmission line additions} should be interpreted as a signal of where the current capacity is worth upgrading, not as a construction schedule. Near-term transmission is constrained by grid-equipment supply chains \cite{yao2026grid}, so a \rev{transmission line} appearing in the 2028 expansion cannot realistically be energized by then. Raising the capacity of existing \rev{transmission lines} to accommodate data-center load growth is  faster, and the June 2026 FERC orders  direct  ISOs to consider advanced transmission technologies (grid-enhancing technologies like dynamic line rating or advanced conductors) in their large-load studies \cite{ferc2026largeload}.

\rev{Our results do not point to a single best composition of tiers. At the base annual interruptible-energy cap of 5\%, compositions with the same firm share differ in cost by only \res{0.1--0.3\%}, and the cheapest split differs between the grids. Once the cap reaches about 10\% the composition with the largest interruptible share is the cheapest on both.} Neither extreme is a portfolio \rev{a data-center} operator could realistically offer, since the share of workload that tolerates being switched off is limited. The value of a \rev{shiftable} or interruptible megawatt is therefore  both system- and policy-dependent.

\begin{table}[!htbp]
\centering
\caption{\rev{\textbf{Transferability of the framework.} Which parts apply to another power system as they stand, and which require new inputs or a different formulation.}}
\label{tab:transfer}
\begingroup\revcolor
\scriptsize
\setlength{\tabcolsep}{4pt}
\renewcommand{\arraystretch}{1.15}
\begin{tabular}{@{}>{\raggedright\arraybackslash}p{0.17\linewidth}>{\raggedright\arraybackslash}p{0.34\linewidth}>{\raggedright\arraybackslash}p{0.42\linewidth}@{}}
\toprule
Component & Transfers as it stands & Needs recalibration or reformulation\\
\midrule
Flexibility tiers (Layers A--C) & Tier definitions, backlog, make-up and shifting constraints, event-shape constraints: all are stated at the point of interconnection & Tier shares, make-up horizons, shift caps and event-shape limits, from local tariffs and the local data-center mix\\
Investment and operation & Linear CEM on representative days; storage and transmission variables & Candidates (a queue in markets, a plan under central planning); costs and financing\\
Adequacy & Accredited-capacity constraint; data-center credit derived from the reduction delivered in stress hours & Requirement, ELCC class ratings and shortfall price. Energy-only markets such as ERCOT have no capacity requirement to anchor the credit, and European capacity mechanisms differ by country\\
Policy & A requirement as a hard constraint, or with a shortfall priced at its statutory alternative & State RPS (other US markets); provincial renewable obligations (China); a carbon cap (Korea); a carbon price in the objective (EU, Singapore)\\
Network & Line flow limits, contingency derating, linear losses & Spatial resolution. Bidding zones in Europe and inter-provincial transmission lines in China suit a zonal model; in a small system such as Singapore, spatial shifting has little domestic role\\
Leave-one-out tests and indicators & Procedures and formulas, computed from an all-firm solution & Include whichever compliance price applies; use the adequacy price where capacity is priced; thresholds must be re-estimated\\
\bottomrule
\end{tabular}
\endgroup
\end{table}

\rev{The framework can be applied to other systems, and Table~\ref{tab:transfer} summarizes transferable insights and aspects that require further detailed consideration. The tiered data-center flexibility formulation, the leave-one-out tests and indicators transfer directly, since they are stated in terms of power withdrawal and do not depend on market rules. Adequacy and policy are the components most specific to a system, and they are also the ones that decide much of the value, so they should be considered for each setting.}

Several limitations apply to these conclusions. We analyze two planning target years in detail, 2028 for the short term and 2038 for the long term,  and solve the 11 target years from 2028 to 2038 only to test the transferable indicators.  The model does not resolve the sequencing of investments between the target years or the option value of staging them. \rev{ELCC class ratings of generation, storage and demand response resources are inputs, i.e., we use PJM's ratings for each delivery year and the Korean rules for Korea. Only the data-center credit is derived within the model, because no published rating exists for flexible data-center loads. The derivation assumes \textit{a priori known} stress hours and that the committed flexibility is always delivered. In practice flexible load, in particular temporally-shiftable load, that flattens the net peak may also change the accreditation of other resources, notably solar PV and storage. The credit assumes that stress is concentrated in a few hours, as the sensitivity to stress hours shows. Each system follows its own resource-adequacy rules, and the storage results depend on them. The reduction in PJM's 4-hour battery ELCC rating from 0.59 to 0.22 is the main reason PJM builds no batteries, and full credit for storage is the reason Korea's all-firm case builds them. 
The 21-zone PJM network does not account for intra-zonal congestion. 
Because the total savings are attained with almost no spatial shifting, this mostly affects shifted volumes and destinations, which follow the assumed shift cap and should be read as zone-level signals. The data-center load tiers are represented in aggregate. The model does not simulate individual workloads, job schedulers or site-level limits, so tier shares and shifting capability are scenario assumptions. %The candidates assume that the whole queue is available at its requested date, and no retirements are applied. The tiers are represented in aggregate: the model does not simulate individual workloads, job schedulers or site-level limits, so tier shares and shifting capability are scenario assumptions. Evidence for relocating workloads between sites comes from inference, not from large synchronous training. Korean fuel and transmission coefficients have no documented price year and are taken as 2023 dollars.}
}
Because individual hyperscale data centers are large enough to affect grid-wide investment decisions, designing flexibility products for them directly is a natural lever, and quantifying how those products interact with on-site \rev{generation and storage} resources is left to future work,   \cite{kim2026flexibility}. 

\section{Methods}\label{sec:methods}

\subsection{Model overview}

We formulate a deterministic CEM that co-optimizes investment in candidate generation, storage and transmission for a target year against the operation of the resulting power grid. We solve the model for two target years 2028 and 2038. Future-year operation is  evaluated over  a set of \rev{representative days (Sec.~\ref{sec:repdays})}, each weighted by the number of days it represents, $w_o$, and resolved at $24$ cyclic hours, with an hourly resolution. The model is implemented in Julia/JuMP and solved with Gurobi. \rev{It is a linear program unless the optional Layer~B or~C constraints are active, which make it a mixed-integer program (MIP).} The same objective and constraint set apply to both host power grids, and only the input data and policy constraints differ as applicable for PJM and Korea. \rev{Resource adequacy uses the same accredited-capacity structure in both systems (Sec.~\ref{sec:ra}).}

\paragraph{Sets and indices.} Buses $n\in\mathcal{N}$; existing generators $g\in\mathcal{G}$ and candidate generators $k\in\mathcal{K}$; existing and candidate storage $s\in\mathcal{S}=\mathcal{S}^{\mathrm{ext}}\cup\mathcal{S}^{\mathrm{cand}}$;  \rev{transmission lines} $l\in\mathcal{L}=\mathcal{L}^{\mathrm{ext}}\cup\mathcal{L}^{\mathrm{cand}}$; demands $d\in\mathcal{D}$ with data-center eligibility for \rev{spatial shifting} $\mathcal{D}^{\mathrm{sh}}\subseteq\mathcal{D}$; representative scenarios $o\in\mathcal{O}$ with intra-day hours $t\in\mathcal{T}$, $|\mathcal{T}|=24$, where $t^-$ denotes the preceding hour and wraps around the representative day, $t^-=t-1$ for $t>1$ and $1^-=24$, so that intra-day storage and backlog dynamics are cyclic; data-center tiers $\{\mathrm{firm,flex,inter}\}$; reserve-providing resources $\mathcal{R}$; policy areas (PJM states) $a\in\mathcal{A}$, each covering the load zones $\mathcal{Z}_a$.

\subsection{Objective}
The  CEM objective is to minimize the total system cost:
\begin{equation}
\min\; I^{\mathrm{gen}} + I^{\mathrm{line}} + I^{\mathrm{sto}} + O + C^{\mathrm{ls,firm}}
+ C^{\mathrm{ls,dc}} + \Pi
\end{equation}
with the annualized investment costs in generation (gen), transmission (line) and storage (sto):
\begin{equation}
I^{\mathrm{gen}}=\sum_{k\in\mathcal K} c^{\mathrm{ann}}_k x^{\mathrm{gen}}_k,
\quad
I^{\mathrm{line}}=\rev{\mathrm{FCR}^{\mathrm{tx}}}\sum_{l\in\mathcal L^{\mathrm{cand}}}
c^{\mathrm{inv}}_\ell\,x^{\mathrm{line}}_\ell,\quad
I^{\mathrm{sto}}=\sum_{s\in\mathcal S^{\mathrm{cand}}}
\left(c^{\mathrm{ann,p}}_s x^{\mathrm{p}}_s+c^{\mathrm{ann,e}}_s x^{\mathrm{e}}_s\right),
\end{equation}
where generation and storage candidate costs are already annualized as
\rev{$c^{\mathrm{ann}}=\mathrm{FCR}_k \cdot \mathrm{OCC}_k+\mathrm{FOM}_k$, with $\mathrm{FCR}_k$ the fixed charge rate of technology $k$ (Sec.~\ref{sec:costs}).} Battery Energy Storage System (BESS) assumes a 15-year life, most generation candidates use 30 years, and transmission uses a 40-year life. The transmission term annualizes a distance-scaled per-MW capital cost, $c^{\mathrm{inv}}_\ell=c^{\mathrm{tx}}L_\ell$,
where \rev{$c^{\mathrm{tx}}=\$2{,}500$/MW-mile (2023 dollars) for PJM \cite{epri2025regen}}, $c^{\mathrm{tx}}=\$1{,}710$/MW-mile for Korea \cite{kang2024transmission} and $L_\ell$ is the \rev{transmission line}
length, so that $c^{\mathrm{inv}}_\ell\,x^{\mathrm{line}}_\ell$ is the overnight cost of
building $x^{\mathrm{line}}_\ell$~MW along \rev{transmission line} $l$. %Korea uses year-specific 2028 and 2038 cost projections before annualization. 
The expected operation cost is expressed as:
\begin{equation}
O=\sum_{o\in\mathcal O} w_o \sum_{t\in\mathcal T}
\Big(\sum_{g\in\mathcal G} c^{\mathrm{op}}_g\, p_{g,o,t}
+\sum_{k\in\mathcal K} c^{\mathrm{op}}_k\, p^{\mathrm{cand}}_{k,o,t}\Big),
\end{equation}
where $p_{g,o,t}$ and $p^{\mathrm{cand}}_{k,o,t}$ are the hourly power outputs (MW) of existing generator $g$ and candidate generator $k$ in scenario $o$ and hour $t$, and $c^{\mathrm{op}}_g$ and $c^{\mathrm{op}}_k$ are their variable operating costs (\$/MWh), i.e., fuel plus variable O\&M.
Firm load-shedding and data-center flexibility costs are expressed as:
\begin{align}
C^{\mathrm{ls,firm}}&=\sum_{o} w_o\sum_{t}\sum_{d}
C^{\mathrm{ls}}\, p^{\mathrm{ls}}_{d,o,t},\\
C^{\mathrm{ls,dc}}&=\sum_{o} w_o\sum_{t}\Big[\sum_{d}
\big(C^{\mathrm{temporal}} p^{\rev{\mathrm{temporal}}}_{d,o,t}
+C^{\mathrm{ls,inter}} p^{\mathrm{inter}}_{d,o,t}\big)
+\sum_{d}\sum_{d'\ne d} C^{\mathrm{spatial}}\, p^{spatial}_{d,d',o,t}\Big].
\end{align}
The \rev{shiftable tier} carries an adder for compute shifted in time and a smaller one for compute  shifted in space. Interruption carries none, since the interruptible tier is constrained instead by its annual \rev{interruptible-energy cap}. Involuntary firm shedding is priced at the value of lost load \rev{(VoLL)}. \rev{In both systems, storage charging and discharging also carry a cost of \$0.01/MWh. This is a numerical tie-breaker that prevents simultaneous charging and discharging, not an economic cost, consistent with the ReEDS, which  uses the same value for its minimum storage variable cost \cite{avraam2025regional}.}

\begin{revblock}
Policy costs in PJM are alternative compliance payments (ACPs) on RPS shortfalls and the price of any resource-adequacy shortfall:
\begin{equation}
\Pi=\sum_{a\in\mathcal A}\sum_{c\in\mathcal C}\pi_{a,c}\,\xi_{a,c}+\pi^{\mathrm{ra}}\xi^{\mathrm{ra}},
\end{equation}
where $\mathcal C$ holds the requirement classes (Tier~1, solar carve-out, Tier~2). An ACP is the statutory alternative to retiring a renewable energy certificate (REC), so a shortfall is a lawful compliance choice priced at the state's own value. Table~\ref{tab:cost_params} lists the prices, flexibility adders and the value of lost load. This term applies only to PJM. Korea carries no such terms ($\Pi=0$) and instead enforces its climate policy as a hard cap on total CO\textsubscript{2}e emissions (Eq.~\ref{eq:co2cap}), so the different policy regimes in PJM and Korea enter the same objective through different modeling mechanisms. To test this difference, the PJM runs include hard-RPS, no-RPS and scaled-ACP sensitivities (Sec.~\ref{sec:robust}).
\end{revblock}

\begin{table}[!htbp]
\centering
\scriptsize
\caption{\rev{\textbf{Policy, flexibility and shedding cost parameters, in 2023 dollars}. Nominal values are converted with the GDP chain-type price index, which is 1.1253 in 2028 and 1.3850 in 2038 with 2023 = 1. ACP ranges cover the states with a listed ACP; requirements without one (IL, NC, MI, PA solar) use the sales-weighted average of the listed ACPs of that class and year.}}
\label{tab:cost_params}
\setlength{\tabcolsep}{4pt}
\renewcommand{\arraystretch}{1.15}
\begingroup\revcolor
\begin{tabular}{@{}>{\raggedright\arraybackslash}p{0.27\linewidth}l>{\raggedright\arraybackslash}p{0.21\linewidth}>{\raggedright\arraybackslash}p{0.36\linewidth}@{}}
\toprule
Parameter & Symbol & Value & Basis\\
\midrule
\multicolumn{4}{@{}l}{\emph{Policy (PJM)}}\\
Tier~1 ACP & $\pi_{a,\mathrm{T1}}$ & \$21.8--44.4/MWh (2028); \$16.1--38.5 (2038) & Statutory, by state and year \cite{pjmeis2025rps}\\
Solar carve-out ACP & $\pi_{a,\mathrm{S}}$ & \$28.9--355.5/MWh (2028); \$16.2--216.6 (2038) & As above\\
Tier~2 ACP & $\pi_{a,\mathrm{T2}}$ & \$13.3--44.4/MWh (2028); \$10.8--36.1 (2038) & As above\\
Resource-adequacy shortfall & $\pi^{\mathrm{ra}}$ & \$105,417/MW-yr & \$325/MW-day UCAP price cap of the 2028/2029 auction, also its clearing price \cite{pjm2028brarep}, held constant in real terms\\
\addlinespace
\multicolumn{4}{@{}l}{\emph{Data-center flexibility adders}}\\
Temporal shifting & $C^{\mathrm{temporal}}$ & \$10/MWh & Assumption. Set above the marginal cost of wind and solar generation, so that load is not deferred when nothing is gained, and below that of gas-fired generation, so that deferring load away from hours when gas is on the margin lowers cost. Varied \$0.6--2,000/MWh, close to the value of lost load, at which no load is deferred\\
Spatial shifting & $C^{\mathrm{spatial}}$ & \$2/MWh & Assumption, on the same basis. Set below the temporal adder because relocated work is served in the same hour. Varied \$0.6--100/MWh\\
Interruption & $C^{\mathrm{ls,inter}}$ & \$0/MWh & By definition: interruption is contracted, and it is limited by the annual interruptible-energy cap, not by a price\\
\addlinespace
\multicolumn{4}{@{}l}{\emph{Involuntary shedding}}\\
Value of lost load & $C^{\mathrm{ls}}$ & \$1,777/MWh (2028); \$1,444/MWh (2038) & \$2,000/MWh nominal offer cap of FERC Order No.~831 \cite{ferc2016order831}; varied to \$10,000/MWh\\
\bottomrule
\end{tabular}
\endgroup
\end{table}

\subsection{Network, generation and storage operation}
Power balance is enforced at every node, scenario and hour using a net-injection \rev{transport} network model, in which \rev{transmission line} flows are limited  by their contingency-derated thermal ratings. The power balance augments conventional supply/demand with the data-center flexibility terms:
\begin{align}
&\sum_{g\in\mathcal G_n} p_{g,o,t}+\sum_{k\in\mathcal K_n} p^{\mathrm{cand}}_{k,o,t}
+\sum_{s\in\mathcal S_n}\!\big(p^{\mathrm{dis}}_{s,o,t}-p^{\mathrm{ch}}_{s,o,t}\big)
+\!\!\sum_{\ell:j_\ell=n}\!\! f_{\ell,o,t}-\!\!\sum_{\ell:i_\ell=n}\!\! f_{\ell,o,t} \nonumber\\
&\quad=\mu_{n,o,t}D_n + D^{\mathrm{dc}}_n + p^\ell_{n,o,t} - p^{\mathrm{ls}}_{n,o,t}
- p^{\mathrm{temporal}}_{n,o,t} - p^{\mathrm{inter}}_{n,o,t} + m_{n,o,t}
+ \!\!\sum_{\substack{d'\in\mathcal D^{\mathrm{sh}}\\ d'\ne n}}\!\! p^{spatial}_{d',n,o,t}
- \!\!\sum_{\substack{d'\in\mathcal D^{\mathrm{sh}}\\ d'\ne n}}\!\! p^{spatial}_{n,d',o,t},
\end{align}
where $m$ is \rev{shiftable} make-up consumption, the last two terms on the right-hand side are the spatial-shifted inflows  and outflows associated with node $n$, and $p^\ell_{n,o,t}$ is the physical transmission-loss withdrawal assigned to node $n$. Losses follow a ReEDS-style linear model \cite{ho2021regional}: each line loses 1\% of absolute flow per 100 miles, divided evenly between its two endpoint buses. Thus $p^\ell_{n,o,t}=\frac{1}{2}\sum_{\ell\in\mathcal L(n)}\lambda_\ell |f_{\ell,o,t}|$, where $\lambda_\ell=0.01 L_\ell/100$ when $L_\ell$ is in miles. 

\paragraph{Candidate Generation, Storage and Transmission}
The investment in generation, storage and transmission capacity is constrained by the candidate set as follows:
\begin{align}
&0 \le x_k^{\mathrm{gen}}\le \bar P_k
&& \forall k\in\mathcal{K},\\
&0\le x_\ell^{\mathrm{line}}\le \bar{F}_\ell && \forall l\in\mathcal{L}^{\mathrm{cand}},\\
& 0\le x^p_s \le \overline{P}_s,
&& \forall s \in \mathcal{S}^{\mathrm{cand}}, \\
& 0\le x^e_s \le \overline{E}_s,
&& \forall s \in \mathcal{S}^{\mathrm{cand}}, \\
& x^e_s = \tau x^p_s,
&& \forall s \in \mathcal{S}^{\mathrm{cand}}, \\
\end{align}
Storage investment is constrained by both energy and capacity, having a 4 hour duration ($\tau =4$) as a baseline. \rev{Sensitivities use 6, 8 and 10~h, each with its own ATB cost and PJM ELCC class rating.}

\paragraph{Generator Operation}

Generation operation is constrained as below, including operating reserve constraints where a system-wide minimum can be  optionally included.  Generators are limited by capacity (dispatchable) or by an availability factor $\kappa_{\cdot,o,t}$ times capacity (renewables); candidate generators are limited by their built capacity $x^{\mathrm{gen}}_k$. Reserve provision is limited by remaining headroom and by ramp capability over a $\Delta t=\tfrac{1}{6}$~h (10-minute) deliverability window.
\begin{align}
& 0\le p_{g,o,t}\le \bar P_g
&& \forall g\in\mathcal{G}^{\mathrm{disp}},o,t,\\
&0\le p_{g,o,t}\le \kappa_{g,o,t}\bar P_g
&& \forall g\in\mathcal{G}^{\mathrm{ren}},o,t,\\
&0\le p^{\mathrm{cand}}_{k,o,t}\le x_k^{\mathrm{gen}}
&& \forall k\in\mathcal{K}^{\mathrm{disp}},o,t,\\
&0\le p^{\mathrm{cand}}_{k,o,t}\le \kappa_{k,o,t}x_k^{\mathrm{gen}}
&& \forall k\in\mathcal{K}^{\mathrm{ren}},o,t,\\
&0\le r_{g,o,t}\le \bar P_g-p_{g,o,t}
&& \forall g\in\mathcal{G}^{\mathrm{disp}},o,t,\\
&0\le r_{g,o,t}\le \Delta t\,\rho_g
&& \forall g\in\mathcal{G}^{\mathrm{disp}},o,t,\\
&0\le r_{k,o,t}\le x_k^{\mathrm{gen}}-p^{\mathrm{cand}}_{k,o,t}
&& \forall k\in\mathcal{K}^{\mathrm{disp}},o,t,\\
&0\le r_{k,o,t}\le \Delta t\,\rho_k x_k^{\mathrm{gen}}
&& \forall k\in\mathcal{K}^{\mathrm{disp}},o,t,\\
&p_{g,o,t}+r_{g,o,t}-p_{g,o,t^-}\le \rho_g
&& \forall g\in\mathcal{G}^{\mathrm{disp}},o,t \in \mathcal{T}\setminus\{1\},\\
&p_{g,o,t}-p_{g,o,t^-}-r_{g,o,t^-}\ge -\rho_g
&& \forall g\in\mathcal{G}^{\mathrm{disp}},o,t \in \mathcal{T}\setminus\{1\},\\
&p^{\mathrm{cand}}_{k,o,t}+r_{k,o,t}-p^{\mathrm{cand}}_{k,o,t^-}\le \rho_k x_k^{\mathrm{gen}}
&& \forall k\in\mathcal{K}^{\mathrm{disp}},o,t \in \mathcal{T}\setminus\{1\},\\
&p^{\mathrm{cand}}_{k,o,t}-p^{\mathrm{cand}}_{k,o,t^-}-r_{k,o,t^-}\ge -\rho_k x_k^{\mathrm{gen}}
&& \forall k\in\mathcal{K}^{\mathrm{disp}},o,t \in \mathcal{T}\setminus\{1\},
\end{align}

\paragraph{Network Operation}

We approximate $N\!-\!1$ security with a contingency factor $K$ that derates every
\rev{transmission line (interface)}, rather than enumerating an explicit contingency set, following values used in large nodal CEMs \cite{lee2025canopi, frysztacki2021strong, horsch2018linear, horsch2018pypsa}. We set $K=0.7$ in all nominal cases and vary it from 0.6 to 0.8 for sensitivity analysis (Fig.~\ref{fig:tx_sweep}).
\begin{align}
&-K x_\ell^{\mathrm{line}}\le f_{\ell,o,t}\le K x_\ell^{\mathrm{line}} && \forall l\in\mathcal{L}^{\mathrm{cand}},o,t,\\
&-K F_\ell^{\mathrm{rev}}\le f_{\ell,o,t}\le K F_\ell^{\mathrm{fwd}} && \forall l\in\mathcal{L}^{\mathrm{ext}},o,t.
\end{align}
Candidate \rev{transmission lines} are bounded by their built capacity $x_\ell^{\mathrm{line}}$, capped at the rated $\bar F_\ell$, and existing \rev{transmission lines} are constrained by their fixed forward/reverse ratings. The same multiplier $K$ derates both ratings. 

\paragraph{Storage Operation}
Storage is modeled using  standard state-of-charge dynamics with charging/discharging efficiencies, a 4-hour duration link $x^e_s=\tau x^p_s$ ($\tau=4$), and \rev{a daily boundary at half charge, so that every day starts and ends at $e_{s,o,T}=0.5\,x^e_s$}:
\begin{align}
& 0\le p^{\mathrm{ch}}_{s,o,t} \le \overline{P}_s,
&& \forall s \in \mathcal{S}^{\mathrm{ext}},\ o,\ t, \\
& 0\le p^{\mathrm{dis}}_{s,o,t} \le \overline{P}_s,
&& \forall s \in \mathcal{S}^{\mathrm{ext}},\ o,\ t, \\
& 0\le e_{s,o,t} \le \overline{E}_s,
&& \forall s \in \mathcal{S}^{\mathrm{ext}},\ o,\ t, \\
& 0\le p^{\mathrm{ch}}_{s,o,t} \le x^p_s,
&& \forall s \in \mathcal{S}^{\mathrm{cand}},\ o,\ t, \\
& 0\le p^{\mathrm{dis}}_{s,o,t} \le x^p_s,
&& \forall s \in \mathcal{S}^{\mathrm{cand}},\ o,\ t, \\
& 0\le e_{s,o,t} \le x^e_s,
&& \forall s \in \mathcal{S}^{\mathrm{cand}},\ o,\ t, \\
& e_{s,o,t}
= e_{s,o,t-1}
+ \eta^{\mathrm{ch}} p^{\mathrm{ch}}_{s,o,t}
- \frac{1}{\eta^{\mathrm{dis}}} p^{\mathrm{dis}}_{s,o,t},
&& \forall s \in \mathcal{S},\ o,\ t \in \mathcal{T}\setminus\{1\}, \\
& e_{s,o,1}
= 0.5 x^e_s
+ \eta^{\mathrm{ch}} p^{\mathrm{ch}}_{s,o,1}
- \frac{1}{\eta^{\mathrm{dis}}} p^{\mathrm{dis}}_{s,o,1},
&& \forall s \in \mathcal{S},\ o, \\
& \rev{e_{s,o,T} = 0.5\, x^e_s},
&& \forall s \in \mathcal{S},\ o, \\
& r_{s,o,t} + p^{\mathrm{dis}}_{s,o,t} - p^{\mathrm{ch}}_{s,o,t} \le x^p_s,
&& \forall s \in \mathcal{S},\ o,\ t, \\
& r_{s,o,t} \Delta t \le \eta^{\mathrm{dis}} e_{s,o,t},
&& \forall s \in \mathcal{S},\ o,\ t,\\
\end{align}

\paragraph{Firm Load Shedding}

Firm load shedding from the base load and firm data center load is constrained as follows:

\begin{align}
& 0 \le p_{d,o,t}^{\mathrm{ls}}
\le
\mu_{d,o,t}D_d+\sigma_{\mathrm{firm}}D_d^{\mathrm{dc}}
&& \forall d,o,t .
\end{align}

\subsection{Tiered data-center flexibility}
 \label{sec:flexmodel}
Data-center load at node $d$ is divided into firm, \rev{shiftable} and interruptible shares, and the shares are constrained as $\sigma^{\mathrm{firm}}+\sigma^{\rev{\mathrm{shift}}}+\sigma^{\mathrm{inter}}=1$. They correspond to the EPRI DCFlex  definitions as classified in Table~1 of Sec.~\ref{sec1}~\cite{epri2025gridflexibility}. The firm share enters the power balance as load and may only be involuntarily shed ($p^{\mathrm{ls}}\le \mu D + \sigma^{\mathrm{firm}}D^{\mathrm{dc}}$). The \rev{shiftable} and interruptible shares are governed by the three layers of constraints, which we label Layers~A, B and~C and state formally below. Layer~A is always active and sets the structural limits on \rev{temporal shifting}, \rev{spatial shifting} and interruption.  Layers~B and~C are optional and restrict how often and how sharply the interruptible share may be curtailed. Assigning distinct \rev{interruption} terms to distinct service classes follows the priority-service principle~\cite{chao1987priority}.
 
\paragraph{Layer~A: structural flexibility (always active).}
\emph{Per-hour tier bounds.} \rev{Temporal shifting} plus spatial shift-out cannot exceed
the \rev{shiftable} allotment $\sigma^{\rev{\mathrm{shift}}}D^{\mathrm{dc}}_d$, and interruption cannot exceed the interruptible allotment $\sigma^{\mathrm{inter}}D^{\mathrm{dc}}_d$:
\begin{equation}
p^{\mathrm{temporal}}_{d,o,t}+\!\!\sum_{d'\ne d}\!\! p^{spatial}_{d,d',o,t}
\le \sigma^{\rev{\mathrm{shift}}} D^{\mathrm{dc}}_d,\qquad
p^{\mathrm{inter}}_{d,o,t}\le \sigma^{\mathrm{inter}} D^{\mathrm{dc}}_d.
\end{equation}

\emph{Temporal backlog (cyclic).} Following the treatment of deferrable
and shiftable demand as a virtual energy buffer~\cite{papavasiliou2013large,
qureshi2009cutting, liu2013data}, \rev{deferred} workload is conserved rather than
lost: it accumulates in a buffer, which we call the temporal backlog, and must be made up within a horizon
$H^{\mathrm{makeup}}$. \rev{Let $q_{d,o,t,a}\ge 0$ be work deferred in hour $t$ and made up $a$ hours later, for $a=1,\dots,H^{\mathrm{makeup}}$, with hour indices taken cyclically within the day. Then}
\begin{revblock}
\begin{equation}
p^{\mathrm{temporal}}_{d,o,t}=\sum_{a} q_{d,o,t,a},\quad
m_{d,o,t}=\sum_{a} q_{d,o,t-a,a},\quad
b_{d,o,t}=\sum_{a}\sum_{j=0}^{a-1} q_{d,o,t-j,a},
\end{equation}
so every deferred task finishes within $H^{\mathrm{makeup}}$ hours, not only within the size of the buffer. Make-up and work shifted in from other sites share the receiving site's shiftable-tier capacity, which limits the workload a site can accept:
\begin{equation}
m_{d,o,t}+\sum_{d'\ne d} p^{spatial}_{d',d,o,t}\le \sigma^{\rev{\mathrm{shift}}}D^{\mathrm{dc}}_d.
\end{equation}
\end{revblock}

\emph{\rev{Annual interruptible-energy cap.}} The interruptible tier carries an
availability-style limit on the total curtailed energy:
\begin{equation}
\sum_{o}\sum_{t} w_o\, p^{\mathrm{inter}}_{d,o,t}
\le \phi^{\mathrm{inter}}_{\max}\,\sigma^{\mathrm{inter}} D^{\mathrm{dc}}_d
\sum_{o}\sum_{t} w_o. \label{eq:a3}
\end{equation}

\emph{Inter-zonal workload shifting.} Latency-tolerant workloads may migrate
across zones, following geographical load balancing~\cite{liu2011greening,
qureshi2009cutting} and realizing the spatial-shift flexibility~\cite{epri2025gridflexibility}.
There is no self-shifting ($p^{spatial}_{d,d}=0$), and \rev{both outflow from and inflow to each node are capped at
a fraction of its shiftable allotment}:
\begin{equation}
\sum_{d'\ne d} p^{spatial}_{d,d',o,t}\le \sigma^{\mathrm{sh}}_{\max}\,\sigma^{\rev{\mathrm{shift}}}
D^{\mathrm{dc}}_d,\qquad \rev{\sum_{d'\ne d} p^{spatial}_{d',d,o,t}\le \sigma^{\mathrm{sh}}_{\max}\,\sigma^{\rev{\mathrm{shift}}}
D^{\mathrm{dc}}_d.}
\end{equation}

\paragraph{Layer~B: annual interruption-hour cap (optional).} Real
interruptible tariffs cap the number of interruption \emph{hours} rather than energy alone. A binary indicator $z_{d,o,t}\in\{0,1\}$ flags each
interruption hour; interruption may occur only in flagged hours, and the
representative-day-weighted annual count is capped at
$H^{\mathrm{inter}}_{\max}$:
\begin{align}
p^{\mathrm{inter}}_{d,o,t} &\le \sigma^{\mathrm{inter}} D^{\mathrm{dc}}_d\, z_{d,o,t},\label{eq:b1}
&& \forall d,o,t,\\
\sum_{o}\sum_{t} w_o\, z_{d,o,t} &\le H^{\mathrm{inter}}_{\max}, && \forall d.
\end{align}

\paragraph{Layer~C: event-shape restrictions (optional).} Five further
restrictions shape the interruptible tier's \rev{interruption} profile to reflect a feasible interruption schedule. C1--C2 follow
standard demand-response program design (depth-limited and peak-window
\rev{interruption}), while C3--C5 adapt the classical unit-commitment ramping and
minimum-up/-down-time formulations~\cite{carrion2006computationally, rajan2005minimum} to the
interruption indicator. They are disabled in the base case and enabled
individually in the optional-constraint sensitivity; we let $t\oplus k$ denote
the cyclic forward hour index within a representative day.

\emph{(C1) Per-hour depth cap.} Interruption in any hour is bounded by a
fraction $\delta_{\max}$ of the interruptible allotment:
\begin{equation}\label{eq:c1}
p^{\mathrm{inter}}_{d,o,t}\le \delta_{\max}\,\sigma^{\mathrm{inter}} D^{\mathrm{dc}}_d,
\qquad \forall d,o,t.
\end{equation}
 
\emph{(C2) Time-of-day window.} Interruption is permitted only during an
allowed set of hours $\mathcal H^{\mathrm{allow}}$:
\begin{equation}\label{eq:c2}
p^{\mathrm{inter}}_{d,o,t}=0,\qquad \forall d,o,\; t\notin\mathcal H^{\mathrm{allow}}.
\end{equation}

\emph{(C3) Notification / ramp limit.} The hour-to-hour change in
interruption is bounded two-sidedly by $\rho^{\mathrm{dc}}$ of the
interruptible allotment:
\begin{equation}\label{eq:c3}
\big|\,p^{\mathrm{inter}}_{d,o,t}-p^{\mathrm{inter}}_{d,o,t^-}\,\big|
\le \rho^{\mathrm{dc}}\,\sigma^{\mathrm{inter}} D^{\mathrm{dc}}_d,
\qquad \forall d,o,t.
\end{equation}

\emph{(C4) Maximum consecutive interruption hours (duration limit).} Over every
cyclic window of $H^{\mathrm{consec}}+1$ hours, at most $H^{\mathrm{consec}}$ may be
interruption hours, which forbids runs longer than $H^{\mathrm{consec}}$:
\begin{equation}\label{eq:c4}
\sum_{k=0}^{H^{\mathrm{consec}}} z_{d,o,\,t\oplus k}\le H^{\mathrm{consec}},
\qquad \forall d,o,t.
\end{equation}

\emph{(C5) Minimum spacing between event starts.} An event-start indicator
$u_{d,o,t}\in\{0,1\}$ marks the onset of an interruption, and at most one onset may occur within any cyclic window of $R^{\mathrm{rec}}+1$ hours, with
$R^{\mathrm{rec}}=12$:
\begin{align}\label{eq:c5}
u_{d,o,t} &\ge z_{d,o,t}-z_{d,o,t^-}, && \forall d,o,t,\\
\sum_{k=0}^{R^{\mathrm{rec}}} u_{d,o,\,t\oplus k} &\le 1, && \forall d,o,t.
\end{align}
Because C4 and C5 are expressed in terms of the interruption-hour indicator
$z_{d,o,t}$, enabling either requires Layer~B's linking constraint
$p^{\mathrm{inter}}_{d,o,t}\le\sigma^{\mathrm{inter}}D^{\mathrm{dc}}_d z_{d,o,t}$
to be active as well, so that $z$ is tied to actual interruption rather than
left free; the annual-hour cap itself (the second Layer~B inequality) may remain
relaxed if only the event-shape behavior is desired.

%\emph{Novelty.} The individual formulations above are drawn from established demand-response, geographical-load-balancing, and unit-commitment models. Our contribution is their assembly into a single tiered, three-layer data-center flexibility module embedded in a network-constrained CEM, and its consistent application across the PJM and South Korean systems.
 
\subsection{Resource adequacy}\label{sec:ra}
\begin{revblock}
Resource adequacy is enforced as an accredited-capacity constraint. Each resource contributes its installed capacity times its effective load-carrying capability (ELCC) class rating, and the shiftable and interruptible data-center load tiers are credited with the load they reduce during stress hours. Their credits, $C^{\mathrm{shift}}$ and $C^{\mathrm{inter}}$, are decision variables bounded by the reduction each tier delivers in every stress hour:
\begin{align}
&\sum_{g\in\mathcal G}\gamma_g \overline P_g + \sum_{k\in\mathcal K}\gamma_k
x^{\mathrm{gen}}_k + \gamma^{\mathrm s}\!\Big(\sum_{s\in\mathcal S^{\mathrm{ext}}}\!
\overline P_s + \sum_{s\in\mathcal S^{\mathrm{cand}}}\! x^p_s\Big)
+ \gamma^{\mathrm{DR}}P^{\mathrm{DR}} \nonumber\\
&\qquad + \kappa^{\mathrm{shift}}C^{\mathrm{shift}}+\kappa^{\mathrm{inter}}C^{\mathrm{inter}}+\xi^{\mathrm{ra}}
\;\ge\; \mathrm{FPR}\cdot D^{\mathrm{peak}},\label{eq:ra_adequacy}\\
&C^{\mathrm{shift}}\le \sum_{d}\big(p^{\mathrm{temporal}}_{d,o,t}-m_{d,o,t}\big),\qquad
C^{\mathrm{inter}}\le \sum_{d}p^{\mathrm{inter}}_{d,o,t},\qquad \forall (o,t)\in\mathcal H^{\mathrm{stress}}.\label{eq:ra_credit}
\end{align}
The model decides together how much load to defer and interrupt during the stress hours and how much capacity to build, and each tier is credited only with the reduction its dispatch delivers in every one of those hours. The credit thus works as a performance obligation, as under non-firm service, because the planner can count on a tier only for the reduction it is committed to deliver when the system is stressed. We derive the credit in this way only for the data-center tiers because other resources have published ELCC class ratings and no similar rating exists yet for shiftable or interruptible data-center load.

We also tested a two-step variant, in which the credit is fixed as accredited capacity and the model is solved again without the bounds~\eqref{eq:ra_credit}. Total cost changes by at most \res{0.03\%} across the 16 base cases, but the tiers then no longer deliver what they are credited with. In the PJM reference tier mix in 2038 the reduction delivered in the stress hours decreases from \res{36.5~GW} to zero, and in Korea from \res{6.2} to \res{1.7~GW}. We therefore keep the bounds in the reported results. Two sensitivities replace the earned credit (Sec.~\ref{sec:robust}). One gives no credit. The other gives a fixed credit per MW of data-center load that does not depend on dispatch, equal to the demand-response ELCC class rating for the shiftable tier in PJM (0.91 in 2028 and 0.77 in 2038), 1 for the interruptible tier in PJM, and 0.50 for both tiers in Korea. The factors $\kappa^{\mathrm{shift}}$ and $\kappa^{\mathrm{inter}}$ scale the credit. They equal 1 in all reported cases, so the delivered reduction counts in full, and the no-credit sensitivity sets them to 0.

The stress hours $\mathcal H^{\mathrm{stress}}$ are the ten hours of the modeled days with the highest net load. They are selected before the model is solved, so they do not depend on its solution. Net load is conventional plus data-center load minus wind and solar output, with renewable capacity on the same basis as the representative days, namely the existing fleet plus the queued wind and solar projects at full size, which the model builds in every case. In PJM the stress hours fall between 17:00 and 23:00, after solar output has dropped. Most are on the peak net-demand day (12 July in 2028 and 3 August in 2038), the others on summer evenings and, in 2028, one December evening. They represent \res{207} and \res{130} hours of the year after applying the representative-day weights. In Korea the stress hours are selected in the same way from the BPESD capacities of the target year. They are hours 16--22 on 6 July and hours 9, 18 and 19 on 4 January in 2028, and hours 17--23 on 5 July and hours 9, 18 and 19 on 4 January in 2038, and they represent \res{271} and \res{268} hours of the year. Because the bounds use the same dispatch variables as the power balance, the credit reflects load that is actually reduced, at its cost and within every limit on the tiers. Spatial shifting relocates load between zones without lowering the system peak, so it earns no credit.

For PJM, ELCC class ratings follow PJM's published ratings for each delivery year, using the 2028/29 values for the 2028 runs and PJM's projected 2038/39 values for the 2038 runs \cite{pjm2026elcc}. Ratings fall sharply for storage and wind as their penetration grows, for example from 0.59 to 0.22 for a 4-hour battery. Existing units are credited by plant type, and candidate technologies at the ELCC values in Table~\ref{tab:capital_cost_comparison}. A fixed demand-response portfolio of 8{,}396~MW is credited at the demand-response ELCC class rating, 0.91 in 2028 and 0.77 in 2038. The requirement uses the forecast pool requirement of the 2028/2029 auction, $\mathrm{FPR}=0.9401$ \cite{pjm2028brarep}, in both target years, and PJM's latest published value of 0.9232 is a sensitivity. A shortfall $\xi^{\mathrm{ra}}$ is priced at the auction's price cap, which approximates the steep part of PJM's Variable Resource Requirement curve.

For Korea, the adequacy constraint has the same structure, with $(1+\mathrm{IRM})$ in place of the FPR:
\begin{align}
&\sum_{g\in\mathcal G}\gamma_g \overline P_g
+ \sum_{k\in\mathcal K}\gamma_k x^{\mathrm{gen}}_k
+ \sum_{s\in\mathcal S^{\mathrm{ext}}}\overline P_s
+ \sum_{s\in\mathcal S^{\mathrm{cand}}}x^p_s
+ P^{\mathrm{DR}} \nonumber\\
&\qquad
+ \kappa^{\mathrm{shift}}C^{\mathrm{shift}}
+ \kappa^{\mathrm{inter}}C^{\mathrm{inter}}
\;\ge\; (1+\mathrm{IRM})D^{\mathrm{peak}}.
\label{eq:ra_adequacy_korea}
\end{align}
The installed reserve margin (IRM) is 20\% in 2028 and 22\% in 2038, following the 11th BPESD \cite{MOTIE202511th}. Conventional generators, storage and demand response receive full capacity credit, and adequacy is enforced as a hard constraint with no shortfall variable, also following the 11th BPESD. Wind and solar PV capacity credits are adapted from \cite{paik2024elcc} using their shares of total generation capacity in the 11th BPESD, and they are held fixed across the flexibility cases.
\end{revblock}

\subsection{Policy Constraints}

Grid-specific policy constraints reflect the present  \textbf{PJM} and \textbf{Korea} status.%Togglable optional constraints exist to see the effect of reflecting realistic constraints. 

\paragraph{PJM Policy Constraints}
\begin{revblock}
PJM policy is enforced at the state level through each state's statutory RPS. Each state $a$ has up to three requirements: a main tier including solar (Tier~1), a solar carve-out, and a secondary tier (Tier~2), with percentages and ACPs from the PJM Environmental Information Services comparison of state programs \cite{pjmeis2025rps}. Virginia's schedules follow the current statute \cite{vcea2020}. A requirement applies to the state's retail sales, including all data-center load as it is actually served. Annual data-center energy of zone $z$ is
\begin{equation}
E^{\mathrm{dc}}_z=\sum_{o,t}w_o\Big(D^{\mathrm{dc}}_z-p^{\mathrm{temporal}}_{z,o,t}+m_{z,o,t}-p^{\mathrm{inter}}_{z,o,t}+\sum_{z'}\big(p^{spatial}_{z',z,o,t}-p^{spatial}_{z,z',o,t}\big)\Big),
\end{equation}
so interrupted energy carries no obligation, deferred workload counts when it is made up, and relocated workload counts where it is consumed. With $E^{\mathrm{conv}}_z$ the conventional load served, $\alpha_{z,a}$ the share of zone $z$'s conventional load in state $a$ from EIA-861 retail sales \cite{eia2025861}, $h(z)$ the state hosting zone $z$'s data-center load, $\kappa_a$ the share of a state's sales its RPS covers, and $\rho_{a,c,z}$ the requirement of class $c$, the obligation is
\begin{equation}
Q_{a,c}=\kappa_a\sum_{z}\rho_{a,c,z}\big(\alpha_{z,a}E^{\mathrm{conv}}_z+ \mathbf{1} [h(z)=a]E^{\mathrm{dc}}_z\big).
\end{equation}
It is met with RECs $x_{r,a,g,p}$ of technology group $g$ from state $r$, retired toward pool $p$ of state $a$, or by paying the ACP:
\begin{equation}
\sum_{(r,g)\in\mathcal E_{a,c}}x_{r,a,g,p(c)}+\xi_{a,c}\ge Q_{a,c}\quad\forall a,c,\qquad
\sum_{a,p}x_{r,a,g,p}\le\sum_{i\in\mathcal G_{r,g}}\sum_{o,t}w_o\,p_{i,o,t}\quad\forall r,g.
\end{equation}
Set $\mathcal E_{a,c}$ holds the pairs of supply state and technology group whose RECs are eligible for class $c$ under the law of state $a$. In most states these include supply from anywhere in PJM, and each MWh is retired only once. A pool $p$ is a group of requirement classes that the same retired REC counts toward. The solar carve-out and Tier~1 share one pool, so a solar REC retired for the carve-out also counts toward Tier~1, while Tier~2 has its own pool. Virginia and North Carolina also require a minimum share of their RECs from in-state supply. State capacity targets are not imposed. Offshore-wind targets are authorizations or procurement goals, the one committed offshore wind project is already in the existing fleet, solar generation capacity policy enters through the carve-outs, and state storage targets have no statutory shortfall price. A storage-mandate sensitivity imposes these storage targets as hard constraints.
\end{revblock}

\paragraph{Korean Policy Constraints}
The CO\textsubscript{2}e emission goal derived from the Nationally Determined Contribution is the strongest planning pressure for the Korean power grid. The 11th Basic Plan for Long-Term Electricity Supply and Demand (BPESD) provides a CO\textsubscript{2}e emission goal for each year until 2038. The CO\textsubscript{2}e emission constraint is enforced as a hard constraint ($\bar E^{\mathrm{CO}_2\mathrm{e}} = 189.9 $ and $83.1$~Mt for 2028 and 2038, respectively):
\begin{align}
&\sum_{o\in\mathcal{O}}w_o
\sum_{t\in\mathcal{T}}
\left(
\sum_{g\in\mathcal{G}}\varepsilon_g p_{g,o,t}
+\sum_{k\in\mathcal{K}}\varepsilon_k p^{\mathrm{cand}}_{k,o,t}
\right)
\le \bar E^{\mathrm{CO\textsubscript{2}e}}
\label{eq:co2cap}
\end{align}
where $\varepsilon_g$ is the CO\textsubscript{2}-equivalent emission rate (tCO\textsubscript{2}/MWh) of unit $g$ as reported in the 11th BPESD.

\subsection{Representative days, systems, and scenarios}\label{sec:repdays}

\begin{figure*}[!htbp]
\centering
\centering
\includegraphics[width=\linewidth]{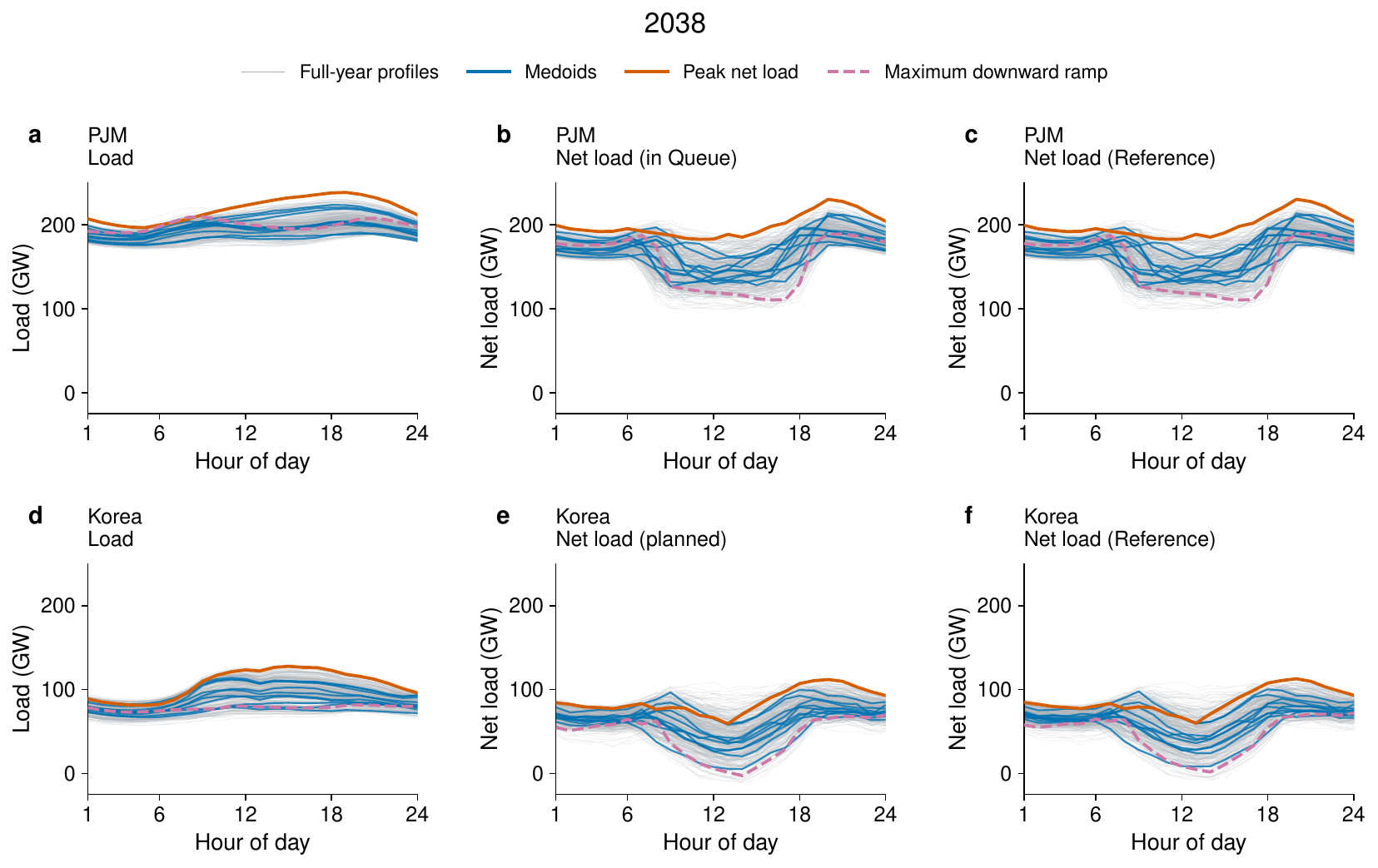}
\caption{\textbf{Representative days in 2038.} \rev{(a,d) Gross load. (b,e) Net load using existing plus queued wind and solar capacity in PJM, or planned capacity in Korea. (c,f) Net load using existing capacity plus optimized Reference case buildouts. Gray lines show all 365 days and blue lines show  medoids, orange the peak-net-load day, and dashed magenta the maximum downward-ramp day. Both extreme days are selected using the capacity in queue (PJM) or planned (Korea).}}
\label{fig:repdays}
\end{figure*}

\begin{revblock}
Operating conditions are modeled with 12 representative days of 24 hours each. For each grid we build 8,760-hour series of system demand including data centers, available wind and available solar (installed capacity times capacity factor), standardize each by its mean and standard deviation, and describe each day by its 72 values. Distance between days is the squared Euclidean distance divided by 72. The day holding the highest hourly net demand and the day holding the largest one-hour net-demand decrease are fixed as medoids, and the other ten are chosen by $k$-medoids from 10,000 starts \cite{scott2019clustering}. Weights start from cluster sizes and are adjusted by a quadratic program to preserve 365 days and annual gross demand, with each weight at least one day. Sampling days may affect planning results \cite{hilbers2019importance}, and this procedure includes both a peak net-demand day and a ramp extreme. Selection precedes the CEM and does not depend on its solution. We experimented with  different number of representative days, and 12 days deliver numerically stable outcomes.  The renewable capacity used for selection is the existing fleet plus the queue of that year for PJM and the BPESD totals of that year for Korea; it does not change installed capacity in the CEM. Table~\ref{tab:repdays} lists the details of the selected representative days. Supplementary Fig.~\ref{fig:repdays_2028} and Fig.~\ref{fig:repdays} illustrate them against all days of 2028 and 2038, respectively. In PJM, selecting days on another capacity basis changes the saving by at most \res{\$0.04B}, and adding the year's highest net-load days changes it by \res{\$0.01B}; in Korea both change it by at most \res{\$0.005B}. As a further check, the four base cases of each grid and year were solved over all 365 days, each day keeping its own storage and backlog cycle. Total cost then differs from the 12-day result by at most \res{1.2\%}, and the saving of the reference tier mix is \res{0.9\%} and \res{6.0\%} in PJM and \res{0.2\%} and \res{2.3\%} in Korea, in 2028 and 2038, against \res{0.8\%}, \res{6.0\%}, \res{0.2\%} and \res{2.5\%} with 12 days (Sec.~\ref{sec:robust}). Using 12 representative days therefore reproduces the full-year results closely while keeping the model tractable.
\end{revblock}

\begin{table}[!htbp]
\centering
\scriptsize
\caption{\rev{\textbf{Representative days and weights.} Day of the year, with its weight in days in parentheses.}}
\label{tab:repdays}
\begingroup\revcolor
\setlength{\tabcolsep}{4pt}
\setlength{\tabcolsep}{3pt}\begin{tabular}{@{}>{\raggedright\arraybackslash}p{0.11\linewidth}>{\raggedright\arraybackslash}p{0.15\linewidth}>{\raggedright\arraybackslash}p{0.15\linewidth}>{\raggedright\arraybackslash}p{0.51\linewidth}@{}}
\toprule
Grid, year & Peak net demand & Maximum ramp-down & Other medoids\\
\midrule
PJM 2028 & 193, 12 July (7.85) & 84, 25 March (5.04) & 129 (32.0), 45 (32.9), 273 (31.1), 249 (35.3), 220 (41.4), 312 (36.8), 143 (22.5), 290 (31.6), 289 (52.4), 353 (36.2)\\
PJM 2038 & 215, 3 August (4.08) & 84, 25 March (11.15) & 358 (40.3), 235 (34.3), 209 (18.6), 128 (36.2), 7 (42.6), 327 (48.3), 204 (18.9), 323 (35.6), 289 (24.6), 133 (50.3)\\
Korea 2028 & 187, 6 July (19.20) & 58, 27 February (15.92) & 4 (45.6), 302 (25.7), 170 (20.9), 345 (22.8), 264 (48.0), 46 (20.1), 246 (32.9), 95 (33.0), 315 (41.7), 339 (39.3)\\
Korea 2038 & 186, 5 July (16.39) & 58, 27 February (15.75) & 4 (51.1), 302 (25.1), 170 (20.6), 345 (22.5), 264 (46.9), 46 (20.3), 246 (32.5), 95 (33.0), 315 (41.1), 339 (39.8)\\
\bottomrule
\end{tabular}
\endgroup
\end{table}

\begin{table}[!htbp]
\centering
\scriptsize
\caption{\rev{\textbf{Demand and renewables data.} Hourly statistics over 8,760~h. Demand includes data-center load, which is flat in each hour before flexibility. Capacity factors are capacity-weighted means of output over installed capacity in the selection basis of Sec.~\ref{sec:repdays}. Sources: PJM demand from the 2026 Load Forecast \cite{pjm2026load}, wind and solar profiles from the ICARUS PJM Dataset \cite{khanal2025icaruspjm}; Korean demand from the 11th BPESD \cite{MOTIE202511th} and KPG~193 \cite{Song2025KPG}, profiles from KPG~193 \cite{Song2025KPG}.}}
\label{tab:demand_stats}
\begingroup\revcolor
\setlength{\tabcolsep}{3.5pt}
\begin{tabular}{@{}lrrrr@{}}
\toprule
& \multicolumn{2}{c}{PJM} & \multicolumn{2}{c}{Korea}\\
\cmidrule(lr){2-3}\cmidrule(lr){4-5}
& 2028 & 2038 & 2028 & 2038\\
\midrule
Annual demand (TWh)                 & 1,022 & 1,707 & 646 & 775\\
of which data centers (TWh, \%)     & 142 (14\%) & 655 (38\%) & 43 (6.6\%) & 78 (10\%)\\
Mean demand (GW)                    & 117 & 195 & 74 & 88\\
Standard deviation (GW)             & 13.8 & 11.1 & 11.3 & 13.6\\
Minimum, peak (GW)                  & 87, 172 & 170, 239 & 49, 108 & 58, 129\\
Data-center capacity (GW, \% of peak) & 16.3 (9.4\%) & 74.8 (31\%) & 4.9 (4.5\%) & 8.9 (6.9\%)\\
Peak net demand (GW)                & 157 & 230 & 98 & 112\\
Solar capacity factor, mean         & 0.25 & 0.25 & 0.18 & 0.18\\
Wind capacity factor, mean          & 0.44 & 0.43 & 0.26 & 0.26\\
\bottomrule
\end{tabular}
\endgroup
\end{table}

\begin{table}[!htbp]
\centering
\scriptsize
\caption{\rev{\textbf{Sources for the case study design for PJM and Korea.}}}
\label{tab:case_study_design}
\setlength{\tabcolsep}{2pt}
\renewcommand{\arraystretch}{1.15}
\begin{tabular}{L{3.6cm} C{2.2cm} C{2.2cm} C{2.2cm} C{2.2cm}}
\toprule
& \multicolumn{2}{c}{\textbf{PJM}} 
& \multicolumn{2}{c}{\textbf{Korea}}\\
\cmidrule(lr){2-3}\cmidrule(lr){4-5}
\centering Item 
& 2028 & 2038 
& 2028 & 2038\\
\midrule

Generator Candidates
& \rev{PJM cycle and serial queues to 2028}
& \rev{PJM cycle and serial queues to 2038}
& BPESD until 2028
& BPESD until 2038 \\

Transmission Candidates
& In-Corridor Expansion (+100\%)
& In-Corridor Expansion (+100\%)
& BPESD until 2028
& BPESD until 2038 \\

Storage Candidates
& \rev{PJM queue to 2028}
& \rev{PJM queue to 2038}
& \rev{Central BESS Market Tender}
& \rev{Central BESS Market Tender} \\

\rev{Candidates (GW)}
& \rev{36.7}
& \rev{226.5}
& --
& -- \\

Data center capacity (GW)
& 16.3
& 74.8
& 4.9\textsuperscript{*}
& 8.9\textsuperscript{**} \\

Emissions target (MtCO$_2$eq)
& --
& --
& 189.9
& 83.1 \\

Retrofits 
& -- 
& -- 
& -- 
& Coal \& LNG \\

Operating Reserve (GW, optional) 
& -- 
& -- 
& 4.5 
& 4.5 \\

\bottomrule
\end{tabular}

\vspace{1mm}
\begin{minipage}{0.96\linewidth}
\scriptsize
\textsuperscript{*} Extrapolated from data-center power use permits reported in the BPESD.\\
\textsuperscript{**} Post-calculated from the summer peak load in the 11th BPESD using the peak contract-capacity utilization rate \cite{MOTIE202310th}.\\
\end{minipage}
\end{table}

\begin{revblock}
\textbf{PJM} is represented as a 21-zone network, consistent with PJM's zones, taken from the ICARUS PJM Dataset \cite{khanal2025icaruspjm}, which provides the resource fleet, transmission network and renewable profiles. The fleet starts from EPA's NEEDS database \cite{epa2025needs} and was updated with EIA-860M (August 2026) \cite{eia2026860m} and PJM's capacity-auction resource lists \cite{pjm2028brarep}. Retirements are not applied \cite{DOE2025GridReliability}. Candidates are live projects in PJM's Serial and Cycle queues (October 2026) \cite{pjm2026queueserial, pjm2026queuecycle} requesting service by the target year, each sized by the MW it adds, so that uprates and later positions at one site are not double counted. Projects due by 2026 that are under construction or in engineering and procurement join the existing resource fleet (3.0~GW). The candidates are 36.7~GW by 2028 and 226.5~GW by 2038, of which 10.1 and 93.2~GW are gas and 4.1 and 58.6~GW are storage. Most early-stage projects historically withdraw, so a sensitivity removes Cycle~1, which leaves 12.1~GW of gas in 2038. Solar-plus-storage hybrids are modeled as solar.
\end{revblock}

\textbf{Korea} is represented on the KPG 193 system with generator, topology and capacity-factor data from KPG 193 (nuclear ramp capped at 3\,\%/h) \cite{Song2025KPG}. Policy constraints follow the 11th BPESD: CO\textsubscript{2}e emission caps (189.9~Mt in 2028, 83.1 in 2038) and the 2038 case allows co-firing retrofits (green-hydrogen LNG and 20\%-ammonia coal). For  Korea, co-firing retrofit candidates are modeled as continuous investment options at eligible buses, and unidentified BPESD candidates including solar PV, wind, LNG and nuclear are optimized within eligible buses with capacity caps. \rev{Candidate
sites for battery energy storage systems (BESS) are mapped to KPG193 buses based on the target substations from the
Korean ESS central contract market tender \cite{kpx2025essTender}, capping total capacities to 1.5 GW and 22.5 GW for 2028 and 2038, respectively. 
The national solar PV and wind caps equal the BPESD 2038 totals of 77.2 and 40.7~GW.}

\paragraph{Scenarios.} Two static generation-interconnection scenarios are each evaluated under four flexibility cases (100\% firm; 100\% \rev{shiftable}; 100\% interruptible; \rev{reference tier mix} $0.3/0.5/0.2$): target years 2028 and 2038 include candidate resources available through these years. These two snapshots are chosen deliberately to bracket the planning problem---an immediate, queue-constrained horizon and a longer-term one---rather than to approximate a continuous annual build, which would conflate the flexibility mechanisms with assumptions about the precise investment trajectory. 

\rev{Sensitivity analyses are listed in Table~\ref{tab:ranking}. In both systems they vary the tier shares, the flexibility costs, the shift cap, the make-up horizon, the annual interruptible-energy cap, the capacity-credit settings, the number of stress hours, data-center penetration, battery cost and duration, the ATB cost case, the usable line rating, the value of lost load and the representative-day basis, together with the optional Layer~B/C constraints. In PJM they also vary the RPS regime, the ACP level, the obligation basis, the forecast pool requirement, the queue and the state storage mandate.} Reported costs are annualized; energy quantities are annualized by the representative-day weights.

\subsection{Cost Assumptions}\label{sec:costs}

\begin{revblock}
All costs in both systems are in real 2023 dollars, the dollar year of the NLR ATB 2025 \cite{nlr2025atb}. For PJM, overnight costs and fixed O\&M follow the ATB 2025 Moderate case, with the Advanced and Conservative cases as sensitivities; fuel prices follow the EIA Annual Energy Outlook 2026 \cite{eia2026aeo}. Values stated in other dollars are converted with the GDP chain-type price index \cite{bea2026gdp}. Each target year is solved as a single year, so nothing is discounted between 2028 and 2038. Multiplying the reported costs by 1.1253 (2028) or 1.3850 (2038) gives nominal dollars of the target year.

Capital is annualized at ATB's fixed charge rate, as in the ReEDS model \cite{avraam2025regional}:
\begin{equation}
\mathrm{FCR}_k=\mathrm{CRF}(r_k,T_k)\,\mathrm{PFF}_k,\quad
\mathrm{CRF}(r,T)=\frac{r}{1-(1+r)^{-T}},\quad
\mathrm{PFF}_k=\frac{1-\tau\,\mathrm{PVD}_k}{1-\tau},
\end{equation}
where $r_k$ is the technology's real after-tax weighted average cost of capital from ATB, $T_k$ the recovery period (30 years for generation, 15 for batteries, 40 for transmission), $\tau=25.74\%$ the combined tax rate and $\mathrm{PVD}_k$ the present value of depreciation. Real rates range from 3.1\% (solar) to 4.3\% (gas, nuclear), which gives the rates listed under Table~\ref{tab:capital_cost_comparison}. The base case has no tax credits, as in ATB's Without Tax Credits case. A 2028 sensitivity gives batteries and nuclear ATB's 40\% investment tax credit, net of a 4\% transfer discount. It changes no decision: a 4-hour battery then costs \$186,000 per accredited MW-yr, against \$157,000 for gas.
\end{revblock}
\rev{Korea uses the same fixed charge rates, the same ATB 2025 Moderate battery costs, and ATB 2025 variable O\&M mapped by technology. Its other overnight and fixed costs follow published Korean data~\cite{Moon2025LCOE}, already in 2023 dollars, and its transmission keeps the Korean construction rate with the same financing. The value of lost load is deflated as in PJM. Sensitivities use ATB 2024 \cite{NREL2024ATB} and ATB 2025 \cite{nlr2025atb} battery costs at durations of 2 to 10~h.} Co-firing retrofits for green-hydrogen LNG and 20\%-ammonia coal are available with the following costs with 2.0 and 1.5 times multipliers for fuel and variable O\&M costs~\cite{BNEF2022Retro,Oberg2022Retro}. Dashes denote technologies that are existing-only (non-investable) or absent in that system.

\begin{table}[!htbp]
\centering
\scriptsize
\caption{\rev{\textbf{Capital cost and capacity-credit assumptions for PJM and Korea candidate technologies}}. Overnight capital cost (OCC) is in \$/kW, fixed O\&M (FOM) in \$/kW-yr; ELCC is the dimensionless capacity credit and Ann.\ is the annualized fixed charge (\rev{$\mathrm{FCR}\cdot\mathrm{OCC}+\mathrm{FOM}$}, \$/kW-yr). \textit{Notes}: OCC includes grid-connection costs, incorporates the construction finance factor, and corresponds to the CAPEX reported in \rev{\cite{nlr2025atb}} for PJM and the OCC reported in \cite{Moon2025LCOE} for  Korea. The ELCC values of solar PV and onshore wind in Korea are reported in  \cite{paik2024elcc} \rev{Fixed charge rates, the same in both systems: gas and coal 0.0692, nuclear 0.0633, solar 0.0539, onshore wind 0.0546, offshore wind 0.0556, batteries 0.0890, transmission 0.0609. PJM ELCCs are class ratings for 2028/29 and 2038/39. Under the Korean adequacy rules, conventional generation and storage count at their full capacity (Sec.~\ref{sec:ra}). All dollars are 2023.}}
\label{tab:capital_cost_comparison}
\setlength{\tabcolsep}{3pt}
\renewcommand{\arraystretch}{1.10}
\begin{tabular}{l c rrrr rrrr}
\toprule
& & \multicolumn{4}{c}{\textbf{PJM}}
& \multicolumn{4}{c}{\textbf{Korea}}\\
\cmidrule(lr){3-6}\cmidrule(lr){7-10}
Technology & Year
& OCC & FOM & ELCC & Ann.
& OCC & FOM & ELCC & Ann.\\
&
& \multicolumn{1}{c}{\$/kW}
& \multicolumn{1}{c}{\$/kW-yr}
&
& \multicolumn{1}{c}{\$/kW-yr}
& \multicolumn{1}{c}{\$/kW}
& \multicolumn{1}{c}{\$/kW-yr}
&
& \multicolumn{1}{c}{\$/kW-yr}\\
\midrule
LNG / NGCC
& 2028 & \rev{1,336} & \rev{30} & \rev{0.78} & \rev{122.1}
       & \rev{1,251} & \rev{42} & \rev{1.00} & \rev{128.5}\\
& 2038 & \rev{1,275} & \rev{28} & \rev{0.75} & \rev{116.4}
       & \rev{1,469} & \rev{42} & \rev{1.00} & \rev{143.6}\\
Coal
& 2028 & \rev{3,955} & \rev{87} & \rev{0.85} & \rev{360.6}
       & \rev{1,670} & \rev{66} & \rev{1.00} & \rev{181.5}\\
& 2038 & \rev{3,753} & \rev{83} & \rev{0.77} & \rev{342.7}
       & \rev{1,961} & \rev{66} & \rev{1.00} & \rev{201.7}\\
Nuclear
& 2028 & \rev{7,871} & \rev{144} & \rev{0.96} & \rev{641.9}
       & \rev{3,459} & \rev{139} & \rev{1.00} & \rev{357.9}\\
& 2038 & \rev{6,930} & \rev{144} & \rev{0.93} & \rev{582.4}
       & \rev{4,450} & \rev{139} & \rev{1.00} & \rev{420.5}\\
Solar PV
& 2028 & \rev{1,285} & \rev{22} & \rev{0.098} & \rev{91.0}
       & \rev{1,008} & \rev{24} & \rev{0.074} & \rev{78.3}\\
& 2038 & \rev{904} & \rev{18} & \rev{0.081} & \rev{66.8}
       & \rev{759} & \rev{24} & \rev{0.063} & \rev{64.9}\\
Onshore wind
& 2028 & \rev{1,727} & \rev{37} & \rev{0.34} & \rev{131.7}
       & \rev{1,875} & \rev{46} & \rev{0.16} & \rev{148.4}\\
& 2038 & \rev{1,340} & \rev{29} & \rev{0.21} & \rev{101.9}
       & \rev{1,563} & \rev{46} & \rev{0.11} & \rev{131.4}\\
Offshore wind, fixed
& 2028 & \rev{5,925} & \rev{84} & \rev{0.6} & \rev{412.8}
       & -- & -- & -- & --\\
& 2038 & \rev{3,760} & \rev{76} & \rev{0.32} & \rev{285.1}
       & -- & -- & -- & --\\
4h BESS
& 2028 & \rev{1,287} & \rev{46} & \rev{0.59} & \rev{160.2}
       & \rev{1,287} & \rev{46} & \rev{1.00} & \rev{160.2}\\
& 2038 & \rev{1,032} & \rev{36} & \rev{0.22} & \rev{127.7}
       & \rev{1,032} & \rev{36} & \rev{1.00} & \rev{127.7}\\
Coal--ammonia retrofit
& 2028 & -- & -- & -- & --
       & \rev{184} & \rev{0} & -- & \rev{12.7}\\
& 2038 & -- & -- & -- & --
       & \rev{216} & \rev{0} & -- & \rev{14.9}\\
LNG--hydrogen retrofit
& 2028 & -- & -- & -- & --
       & \rev{313} & \rev{0} & -- & \rev{21.6}\\
& 2038 & -- & -- & -- & --
       & \rev{367} & \rev{0} & -- & \rev{25.4}\\
\bottomrule
\end{tabular}
\end{table}

\begin{table}[!htbp]
\centering
\scriptsize
\caption{\textbf{Data-center flexibility parameters.} Baseline values applied
identically to PJM and  Korea. The basis for each value is given per row.
Values marked as assumptions are adopted in this work and examined through the
sensitivity analyses of Sec.~\ref{sec:sensitivity} and Sec.~\ref{sec:robust}. \rev{The tier shares and shifting capability are scenarios: the evidence behind them comes from a mix of workloads and from inference traffic, and it does not establish them for large synchronous training.}}
\label{tab:dcflex_params}
\setlength{\tabcolsep}{4pt}
\renewcommand{\arraystretch}{1.15}
\begin{tabular}{
>{\raggedright\arraybackslash}p{0.26\linewidth}
>{\raggedright\arraybackslash}p{0.12\linewidth}
>{\raggedright\arraybackslash}p{0.10\linewidth}
>{\raggedright\arraybackslash}p{0.43\linewidth}
}
\toprule
Parameter & Symbol & Baseline & Basis\\
\midrule
\multicolumn{4}{@{}l}{\emph{Tier composition}}\\
\cmidrule(r){1-4}
Firm share & $\sigma^{\mathrm{firm}}$ & 0.30 & EPRI DCFlex workload classes~\cite{epri2025gridflexibility}; varied in Sec.~\ref{sec:sensitivity}\\
\rev{Shiftable} share & $\sigma^{\rev{\mathrm{shift}}}$ & 0.50 & As above\\
Interruptible share & $\sigma^{\mathrm{inter}}$ & 0.20 & As above\\
\addlinespace
\multicolumn{4}{@{}l}{\emph{\rev{Shiftable} tier (Layer~A)}}\\
\cmidrule(r){1-4}
Make-up horizon & $H^{\mathrm{makeup}}$ & 24~h & Assumption; varied 4--24~h in Sec.~\ref{sec:robust}\\
Shift cap, out of and into a site & $\sigma^{\mathrm{sh}}_{\max}$ & 0.30 & \rev{Assumption; varied 0.01--0.5 in Sec.~\ref{sec:robust}. Evidence for shifting is from inference \cite{williams2026power}}\\
\addlinespace
\multicolumn{4}{@{}l}{\emph{Interruptible tier (Layer~A)}}\\
\cmidrule(r){1-4}
\rev{Annual interruptible-energy cap} & $\phi^{\mathrm{inter}}_{\max}$ & 0.05 & Assumption, consistent with EPRI interruptible-tariff practice~\cite{epri2025gridflexibility}. Equals 5\% of interruptible-tier energy, or 1\% of data-center energy in the \rev{reference tier} mix, at the top of the 0.25--1\% curtailment-headroom range of Norris et al.~\cite{norris2025rethinking}. \rev{Varied 0.01--1 in Sec.~\ref{sec:robust}}\\
\addlinespace
\multicolumn{4}{@{}l}{\emph{\rev{Capacity credit}}}\\
\cmidrule(r){1-4}
\rev{Stress hours} & \rev{$|\mathcal H^{\mathrm{stress}}|$} & \rev{10} & \rev{Assumption; varied 5--100 in Sec.~\ref{sec:robust}}\\
\rev{Credit derate} & \rev{$\kappa^{\rev{\mathrm{shift}}},\kappa^{\mathrm{inter}}$} & \rev{1} & \rev{Delivered reduction counted in full; no credit is tested}\\
\addlinespace
\multicolumn{4}{@{}l}{\emph{Interruptible tier (Layer~B, off by default)}}\\
\cmidrule(r){1-4}
Annual hour cap & $H^{\mathrm{inter}}_{\max}$ & 480~h & Assumption, within the 180--720~h/yr interruptible-tariff range~\cite{epri2025gridflexibility}\\
\addlinespace
\multicolumn{4}{@{}l}{\emph{Event-shape restrictions (Layer~C, off by default)}}\\
\cmidrule(r){1-4}
Per-hour depth cap & $\delta_{\max}$ & 0.30 & Assumption, within the 25--50\% turndown range~\cite{epri2025gridflexibility}, corroborated by measured 18--55\% workload flexibility~\cite{acun2026investigating} and sustained 10--40\% field reductions~\cite{williams2026power}\\
Allowed hours & $\mathcal{H}^{\mathrm{allow}}$ & 16--21 & Assumption\\
Ramp limit & $\rho^{\mathrm{dc}}$ & 0.25/h & Assumption\\
Max consecutive & $H^{\mathrm{consec}}$ & 4~h & Assumption\\
Min recovery & $R^{\mathrm{rec}}$ & 12~h & Assumption\\
\bottomrule
\end{tabular}
\end{table}

\subsection{\rev{Leave-one-out tests and transferable indicators}} \label{sec:leave-one-out-method}

\begin{revblock}
\emph{Leave-one-out tests.} The contribution of mechanism $f$, which is temporal shifting, spatial shifting or interruption, is
\begin{equation}
V_f=\frac{C_{\mathrm{without}\,f}-C_{\mathrm{reference}}}{E^{\mathrm{orig}}_{\mathrm{DC}}}\quad[\$/\mathrm{MWh}],
\end{equation}
where $C_{\mathrm{without}\,f}$ is the optimal cost of the reference tier mix with the variables of $f$ fixed to zero and investment reoptimized, and $E^{\mathrm{orig}}_{\mathrm{DC}}$ is annual data-center energy before flexibility. These conditional effects need not add to the total saving against all-firm. A second set of tests holds investment at the all-firm solution and reoptimizes operation only, with and without $f$; the difference is the operating and compliance value of the mechanism. The value of the capacity credit is obtained in the same way, by removing the credit of both tiers from the adequacy constraint while all three mechanisms stay available. For PJM a further test keeps the RPS obligation of every spatially shifted MWh in its home state.

\emph{Indicators.} Indicators are computed from the dual of the power balance in an all-firm run, with investments fixed for economic dispatch. With $D_{nt}$ load at node $n$, $q_{nt}=D_{nt}/\sum_m D_{mt}$, $\lambda_{nt}$ the LMP and $w_t$ the hour's weight, the load-weighted spatial spread is
\begin{equation}
G_{\mathrm{spatial}}=\frac{\sum_t w_t\sum_i\sum_j q_{it}q_{jt}\,|\lambda_{it}-\lambda_{jt}|}{\sum_t w_t}.
\end{equation}
The RPS-adjusted measures replace $\lambda_{nt}$ by $p_{nt}=\lambda_{nt}+\rho_n$, where $\rho_n=\sum_r a_{nr}\nu^{\mathrm{RPS}}_r$, $a_{nr}$ is the incremental RPS obligation per MWh of data-center demand at $n$ and $\nu^{\mathrm{RPS}}_r$ the obligation's shadow price. For Korea $\rho_n=0$. With $\lambda^{\mathrm{sys}}_t=\sum_n q_{nt}\lambda_{nt}$ and $\bar\lambda^{\mathrm{sys}}$ its weighted mean, the hourly variability of the system price is its weighted standard deviation,
\begin{equation}
\sigma_{\mathrm{LMP}}=\sqrt{\frac{\sum_t w_t(\lambda^{\mathrm{sys}}_t-\bar\lambda^{\mathrm{sys}})^2}{\sum_t w_t}}\quad[\$/\mathrm{MWh}],
\end{equation}
and its normalized form is $100\,\sigma_{\mathrm{LMP}}/\bar\lambda^{\mathrm{sys}}$.
The renewable indicator is installed wind and solar generation capacity over peak load. The interruption indicator is the load-weighted mean of $\lambda_{nt}$ or $p_{nt}$. The congestion indicator is the mean of the line-limit shadow prices weighted by absolute flow. The shadow prices of the adequacy constraint and of the Korean carbon cap are taken from the all-firm CEM, because the adequacy dual is not unique once investments are fixed. For the 2028--2038 comparison, annual CEMs are solved independently from a common 2026 base grid with year-specific inputs, interpolated between target years where a source gives no annual value. For each indicator, we compute the Pearson correlation with the corresponding mechanism's contribution over the 11 target years from 2028 to 2038, separately for each grid (11 points) and, where stated, for both systems together (22 points).
\end{revblock}

\subsection{Data Center Data Collection}
Nodal data-center load forecasts in Korea were obtained by scaling the nation-wide data-center load to a given spatial distribution. The spatial distribution of data centers was collected from Data Center Map \cite{DataCenterMap2026Korea} as of May 14, 2026. Missing capacities were imputed using Bayesian ridge regression based on observed capacity, whitespace, market, and development-year features following the method of the PJM ICARUS data-center dataset \cite{Reed2025ICARUS}. Each data center is mapped to the nearest KPG 193 bus using the geographic distance between facility and bus coordinates.

\section*{Data availability}
The processed model inputs and the result tables underlying all figures and
tables are available at Zenodo (\url{https://doi.org/10.5281/zenodo.22016953}).
The underlying system datasets are third-party and subject to their respective
use terms: the PJM 21-zone reduction and ICARUS data-center dataset
\cite{khanal2025icaruspjm, Reed2025ICARUS}, KPG 193 v2.0 \cite{Song2025KPG},
the 11th BPESD \cite{MOTIE202511th}, and the 2027/28 PJM Base Residual Auction
\cite{pjm2027bra}.

\section*{Code availability}
The capacity expansion model, including the tiered data-center flexibility module, and the preprocessing and figure-generation scripts are available alongside the data at the same Zenodo record
(\url{https://doi.org/10.5281/zenodo.22016953}) under an MIT license.

\section*{Author contributions}
Y.D.\ and J.K.\ conceived and designed the study. S.K., J.K. and Y.D. \ developed the
methodology,  G.R.\ and B.Y. provided extensive feedback. S.K., G.R. and B.Y. performed the
simulations for the PJM system; G.R.\ adapted the formulation to the
Korean system and performed the simulations for it. S.K. and G.R. conducted the initial 
analysis and result synthesis; J.K. and Y.D. provided feedback. B.Y.\ contributed to the PJM data pipeline
and the transmission representation. A.S.\ contributed the regulatory and
policy framing. D.G.\ and C.K.\ advised on the modeling and analysis.
S.K.\ and G.R. drafted the paper; Y.D. edited the manuscript.
Y.D.\ and J.K.\ supervised the project and secured funding. All authors
revised and approved the paper.

\section*{Competing interests}
The authors declare no competing interests.

\section*{Additional information}
\textbf{Supplementary information} is available for this paper.\\
\textbf{Correspondence and requests for materials} should be addressed
to Y.D.


\bibliography{sn-bibliography}% common bib file

@article{zuluaga2027nodal,
  title={Nodal capacity expansion planning with flexible large-scale load siting},
  author={Zuluaga, Tomas Valencia and Pang, Simon and Watson, Jean-Paul},
  journal={Electric Power Systems Research},
  volume={263},
  pages={113786},
  year={2027},
  publisher={Elsevier}
}

@article{boyu_npj_supply_chain,
  author = {Yao, Boyu and Jeong, Hannah and Mehrtash, Mahdi and Allan, Bentley and Ockerman, Daniel and Dvorkin, Yury},
  title = {Understanding supply chain constraints for the US clean energy transition},
  journal = {npj Clean Energy},
  year = {2025},
  volume = {1},
  number = {1},
  pages = {9},
  doi = {10.1038/s44406-025-00009-1},
  url = {https://doi.org/10.1038/s44406-025-00009-1}
}

@article{shehabi20242024,
  title={2024 united states data center energy usage report},
  author={Shehabi, Arman and Newkirk, Alex and Smith, Sarah J and Hubbard, Alex and Lei, Nuoa and Siddik, Md Abu Bakar and Holecek, Billie and Koomey, Jonathan and Masanet, Eric and Sartor, Dale},
  year={2024}
}

@techreport{nerc2026assessment,
  author      = {{NERC Large Loads Working Group}},
  title       = {Assessment of Gaps in Existing Practices, Requirements, and Reliability Standards for Emerging Large Loads},
  institution = {North American Electric Reliability Corporation (NERC)},
  year        = {2026},
  month       = {March},
  type        = {White Paper},
  url         = {https://www.nerc.com/globalassets/our-work/guidelines/reliability/white-paper---assessment-of-gaps.pdf}
}

@misc{brancucci2025flexible,
  title={Flexible Data Centers: A Faster, More Affordable Path to Power},
  author={Brancucci, Cinzia and Cutler, D and Jenkins, J},
  year={2025},
  publisher={Camus Energy, Encoord, and Princeton ZERO Lab San Francisco, CA, USA}
}

@techreport{epri2026powering,
  author      = {{Electric Power Research Institute (EPRI)}},
  title       = {Powering Intelligence 2026: Updated Scenarios of U.S. Data Center Electricity Use and Power Strategies},
  institution = {Electric Power Research Institute (EPRI)},
  year        = {2026},
  month       = {February},
  number      = {3002034696},
  type        = {Technical Report},
  url         = {https://www.epri.com/research/products/000000003002034696}
}

@article{lee2025canopi,
  title={CANOPI: Contingency-Aware Nodal Optimal Power Investments with High Temporal Resolution},
  author={Lee, Thomas and Sun, Andy},
  journal={arXiv preprint arXiv:2510.03484},
  year={2025}
}

@article{frysztacki2021strong,
  title={The strong effect of network resolution on electricity system models with high shares of wind and solar},
  author={Frysztacki, Martha Maria and H{\"o}rsch, Jonas and Hagenmeyer, Veit and Brown, Tom},
  journal={Applied Energy},
  volume={291},
  pages={116726},
  year={2021},
  publisher={Elsevier}
}

@article{horsch2018linear,
  title={Linear optimal power flow using cycle flows},
  author={H{\"o}rsch, Jonas and Ronellenfitsch, Henrik and Witthaut, Dirk and Brown, Tom},
  journal={Electric Power Systems Research},
  volume={158},
  pages={126--135},
  year={2018},
  publisher={Elsevier}
}

@article{horsch2018pypsa,
  title={PyPSA-Eur: An open optimisation model of the European transmission system},
  author={H{\"o}rsch, Jonas and Hofmann, Fabian and Schlachtberger, David and Brown, Tom},
  journal={Energy strategy reviews},
  volume={22},
  pages={207--215},
  year={2018},
  publisher={Elsevier}
}

@misc{pjm2027bra,
  author      = {{PJM Interconnection}},
  title       = {ELCC Class Ratings for the 2027/2028 Base Residual Auction},
  institution = {PJM Interconnection},
  year        = {2025}, 
  type        = {Technical Document},
  url         = {https://www.pjm.com/-/media/DotCom/planning/res-adeq/elcc/2027-28-bra-elcc-class-ratings.pdf}
}

@article{liu2026watts,
  title={Watts vs. Bytes: Turning Data Centers into Grid Assets via Storage Compute Co-Optimization},
  author={Liu, Shaohui and Shin, Sungho and Deka, Deepjyoti},
  journal={arXiv preprint arXiv:2605.16190},
  year={2026}
}

@article{Moon2025LCOE,
title = {Assessing the levelized cost of energy in South Korea},
journal = {Energy Strategy Reviews},
volume = {62},
pages = {101897},
year = {2025},
issn = {2211-467X},
doi = {https://doi.org/10.1016/j.esr.2025.101897},
author = {Hee Seung Moon and Sunhee Baik and Won Young Park},
}

@techreport{NREL2024ATB,
  author      = {Mirletz, Brian and others},
  title       = {Annual Technology Baseline: The 2024 Electricity Update},
  institution = {National Renewable Energy Laboratory (NREL)},
  year        = {2024},
  doi         = {10.25984/2377191}
}

@article{Oberg2022Retro,
title = {Exploring the competitiveness of hydrogen-fueled gas turbines in future energy systems},
journal = {International Journal of Hydrogen Energy},
volume = {47},
number = {1},
pages = {624-644},
year = {2022},
issn = {0360-3199},
doi = {https://doi.org/10.1016/j.ijhydene.2021.10.035},
author = {Simon Öberg and Mikael Odenberger and Filip Johnsson},
}

@techreport{BNEF2022Retro,
  author       = {{BloombergNEF}},
  title        = {Japan's Costly Ammonia Coal Co-Firing Strategy},
  institution  = {BloombergNEF},
  year         = {2022},
  month        = sep,
  day          = {28},
  url          = {https://assets.bbhub.io/professional/sites/24/BNEF-Japans-Costly-Ammonia-Coal-Co-Firing-Strategy_FINAL.pdf},
  note         = {Bloomberg Finance L.P.}
}

@article{kim2026flexibility,
  title={Flexibility-aware framework for efficient planner-initiated siting of data center},
  author={Kim, Dongjoo and Dong, Lin and Xie, Le},
  journal={Nature Communications},
  year={2026},
  publisher={Nature Publishing Group UK London}
}

@techreport{MOTIE202310th,
  author      = {{Ministry of Trade, Industry and Energy}},
  title       = {{Tenth Basic Plan for Long-Term Electricity Supply and Demand (2022--2036)}},
  institution = {{Ministry of Trade, Industry and Energy, Republic of Korea}},
  year        = {2023},
  month       = jan,
  number      = {MOTIE Notice No. 2023-036},
  urldate     = {2026-06-13}
}

@techreport{MOTIE202511th,
  author      = {{Ministry of Trade, Industry and Energy}},
  title = {{Eleventh Basic Plan for Long-Term Electricity Supply and Demand (2024--2038)}},
  institution = {{Ministry of Trade, Industry and Energy, Republic of Korea}},
  year        = {2025},
  month       = feb,
  number      = {MOTIE Notice No. 2025-169},
  url         = {https://www.motir.go.kr/kor/article/ATCLc01b2801b/70083/view},
  urldate     = {2026-06-13}
}

@misc{KEEI2026DomesticDC,
  author       = {{Korea Energy Economics Institute}},
  title        = {{Domestic Data Center Status}},
  year         = {2026},
  month        = apr,
  note         = {KESIS Monthly Energy Statistics, Issue 82; based on KDCC data},
  url          = {https://kesis.keei.re.kr/pdfOpenG.es?bid=0001&list_no=105&seq=2}
}

@misc{DistributedEnergyAct2024,
  author       = {{Republic of Korea}},
  title        = {{Distributed Energy Activation Act}},
  year         = {2024},
  note         = {Act establishing the power-grid impact assessment framework},
  url          = {https://www.law.go.kr/lsInfoP.do?lsiSeq=251685}
}

@misc{MOTIE2023DCDispersion,
  author       = {{Ministry of Trade, Industry and Energy}},
  title        = {{Measures to Mitigate Data Center Concentration in the Seoul Metropolitan Area}},
  year         = {2023},
  month        = mar,
  note         = {Policy measures for regional dispersion of data centers},
  url          = {https://eiec.kdi.re.kr/policy/callDownload.do?dtime=20230310162312&filenum=2&num=236270}
}

@incollection{chen2026defer,
  title={To Defer or To Shift? The Role of AI Data Center Flexibility on Grid Interconnection},
  author={Chen, Yize and Zheng, Xiaogui},
  booktitle={Proceedings of the 2026 ACM Sustainability Week},
  pages={322--327},
  year={2026}
}

@article{senga2026flexible,
  title = {Flexible data centers reduce power system costs but can increase emissions},
  author = {Senga, Juan Ramon L. and Wang, Shen and Knittel, Christopher R.},
  journal = {iScience},
  year = {2026},
  publisher = {Elsevier},
  doi = {10.1016/j.isci.2026.116497},
  url = {https://doi.org/10.1016/j.isci.2026.116497},
  issn = {2589-0042}
}

@dataset{Reed2025ICARUS,
  title        = {ICARUS Data Center Dataset (ICARUS-DC)},
  author       = {Reed, Owen and Khanal, Saroj and Yao, Boyu and Combariza, Dyer and Klemun, Magdalena and Dvorkin, Yury},
  year         = {2025},
  publisher    = {Johns Hopkins University, Ralph O'Connor Sustainable Energy Institute, NSF Global Center EPICS, and Johns Hopkins Applied Physics Laboratory},
  url          = {https://github.com/hopkinsicarus/ICARUS-DC-Dataset},
  note         = {Versioned GitHub repository. Data licensed under CC BY 4.0}
}

@dataset{khanal2025icaruspjm,
  title        = {ICARUS PJM Dataset},
  author       = {Khanal, Saroj and Yao, Boyu and Villafañe-Delgado, Marisel and Dvorkin, Yury},
  year         = {2025},
  publisher    = {Johns Hopkins University, Ralph O'Connor Sustainable Energy Institute, NSF Global Center EPICS, and Johns Hopkins Applied Physics Laboratory},
  url          = {https://github.com/HopkinsICARUS/ICARUS-PJM-Dataset},
  note         = {Versioned GitHub repository. Data licensed under CC BY 4.0}
}

@techreport{epri2025gridflexibility,
  author      = {{Electric Power Research Institute}},
  title       = {Grid Flexibility Needs and Data Center Characteristics},
  institution = {Electric Power Research Institute (EPRI)},
  year        = {2025},
  month       = {jun},
  address     = {Palo Alto, CA},
  number      = {3002031504},
  type        = {White Paper}
}

@article{norris2025rethinking,
  title={Rethinking load growth: Assessing the potential for integration of large flexible loads in us power systems},
  author={Norris, Tyler and Profeta, Timothy and Patino-Echeverri, Dalia and Cowie-Haskell, Adam},
  year={2025},
  publisher={Nicholas Institute for Energy, Environment \& Sustainability}
}

@legislation{TX-SB6-2025,
  title        = {Relating to the planning for, interconnection and operation of, and costs related to providing service for certain electrical loads and to the generation of electric power by a water supply or sewer service corporation},
  author       = {{Texas Legislature}},
  year         = {2025},
  number       = {S.B. No. 6},
  institution  = {Texas State Legislature},
  address      = {Austin, TX},
  note         = {89th Legislature, Regular Session},
  url          = {https://legiscan.com/TX/text/SB6/id/3248460}
}

@article{liu2011greening,
  title={Greening geographical load balancing},
  author={Liu, Zhenhua and Lin, Minghong and Wierman, Adam and Low, Steven H and Andrew, Lachlan LH},
  journal={ACM SIGMETRICS Performance Evaluation Review},
  volume={39},
  number={1},
  pages={193--204},
  year={2011},
  publisher={ACM New York, NY, USA}
}

@article{papavasiliou2013large,
  title={Large-scale integration of deferrable demand and renewable energy sources},
  author={Papavasiliou, Anthony and Oren, Shmuel S},
  journal={IEEE Transactions on Power Systems},
  volume={29},
  number={1},
  pages={489--499},
  year={2013},
  publisher={IEEE}
}

@inproceedings{qureshi2009cutting,
  title={Cutting the electric bill for internet-scale systems},
  author={Qureshi, Asfandyar and Weber, Rick and Balakrishnan, Hari and Guttag, John and Maggs, Bruce},
  booktitle={Proceedings of the ACM SIGCOMM 2009 conference on Data communication},
  pages={123--134},
  year={2009}
}

@article{carrion2006computationally,
  title={A computationally efficient mixed-integer linear formulation for the thermal unit commitment problem},
  author={Carri{\'o}n, Miguel and Arroyo, Jos{\'e} M},
  journal={IEEE Transactions on power systems},
  volume={21},
  number={3},
  pages={1371--1378},
  year={2006},
  publisher={IEEE}
}

@techreport{vaidhynathan2025vulcan,
  title       = {Vulcan Test Platform: Demonstrating the Data Center as a Flexible Grid Asset},
  author      = {Vaidhynathan, Deepthi and Prabakar, Kumaraguru and Martin, Gregory and Ramesh, Anand and Wheeler, Ben and Coco, Christopher and Clidaras, Jimmy and Monsch, Matthieu and Kim, Sangsun and Talukdar, Saurav and Marti, Shilpa},
  institution = {National Renewable Energy Laboratory},
  month       = jun,
  year        = {2025},
  doi         = {10.2172/2583521},
}

@article{rajan2005minimum,
  title={Minimum up/down polytopes of the unit commitment problem with start-up costs},
  author={Rajan, Deepak and Takriti, Samer and others},
  journal={IBM Res. Rep},
  volume={23628},
  pages={1--14},
  year={2005},
  publisher={Citeseer}
}

@article{chao1987priority,
  title={Priority service: Pricing, investment, and market organization},
  author={Chao, Hung-po and Wilson, Robert},
  journal={The American Economic Review},
  pages={899--916},
  year={1987},
  publisher={JSTOR}
}

@inproceedings{liu2013data,
  title={Data center demand response: Avoiding the coincident peak via workload shifting and local generation},
  author={Liu, Zhenhua and Wierman, Adam and Chen, Yuan and Razon, Benjamin and Chen, Niangjun},
  booktitle={Proceedings of the ACM SIGMETRICS/international conference on Measurement and modeling of computer systems},
  pages={341--342},
  year={2013}
}

@techreport{ho2021regional,
  title={Regional energy deployment system (reeds) model documentation (version 2020)},
  author={Ho, Jonathan and Becker, Jonathon and Brown, Maxwell and Brown, Patrick and Chernyakhovskiy, Ilya and Cohen, Stuart and Cole, Wesley and Corcoran, Sean and Eurek, Kelly and Frazier, Will and others},
  year={2021},
  institution={National Renewable Energy Laboratory (NREL), Golden, CO (United States)}
}

@article{colangelo2026ai,
  title={AI data centres as grid-interactive assets},
  author={Colangelo, Philip and Coskun, Ayse K and Megrue, Jack and Roberts, Ciaran and Sengupta, Shayan and Sivaram, Varun and Tiao, Ethan and Vijaykar, Aroon and Williams, Chris and Wilson, Daniel C and others},
  journal={Nature Energy},
  volume={11},
  number={2},
  pages={254--261},
  year={2026},
  publisher={Nature Publishing Group UK London}
}

@article{gu2025role,
  title={The Role of Flexible Connection in Accelerating Load Interconnection in Distribution Networks},
  author={Gu, Nan and Chen, Ge and Qin, Junjie},
  journal={arXiv preprint arXiv:2510.11476},
  year={2025}
}

@article{radovanovic2022carbon,
  title={Carbon-aware computing for datacenters},
  author={Radovanovi{\'c}, Ana and Koningstein, Ross and Schneider, Ian and Chen, Bokan and Duarte, Alexandre and Roy, Binz and Xiao, Diyue and Haridasan, Maya and Hung, Patrick and Care, Nick and others},
  journal={IEEE Transactions on Power Systems},
  volume={38},
  number={2},
  pages={1270--1280},
  year={2022},
  publisher={IEEE}
}

@misc{ferc2026largeload,
  author       = {{Federal Energy Regulatory Commission}},
  title        = {Show Cause Orders on the Integration of Large Loads
                  onto the Transmission System (Items E-7 through E-12)},
  howpublished = {Docket Nos. EL26-67-000 et al.},
  year         = {2026},
  month        = jun,
  note         = {Issued June 18, 2026},
  url          = {https://www.ferc.gov/news-events/news/fact-sheet-ferc-takes-action-supercharge-americas-grid-efficiency-reliability-and},
  urldate      = {2026-06-18}
}

@inproceedings{acun2026investigating,
  title={Investigating Power Consumption Flexibility of AI Data Centers for Demand Response Participation},
  author={Acun, Fatih and Hankendi, Can and Levine, Ethan and Reynolds, Hudson and Bardwick, Joshua and Coskun, Ayse K},
  booktitle={Proceedings of the 17th ACM International Conference on Future and Sustainable Energy Systems},
  pages={344--348},
  year={2026}
}

@article{williams2026power,
  title={Power-Flexible AI Data Centers: A New Paradigm for Grid-Responsive Compute},
  author={Williams, Chris and Colangelo, Philip and Coskun, Ayse and Levine, Ethan and Neale, Andy and Roberts, Ciaran and Sengupta, Shayan and Shirolkar, Nikhil and Sivaram, Varun and Soares, Sarah and others},
  journal={arXiv preprint arXiv:2606.25098},
  year={2026}
}

@article{shao2025stochastic,
  title={Stochastic long-term joint decarbonization planning for power systems and data centers: A case study in PJM},
  author={Shao, Zhentong and Yu, Nanpeng and Wong, Daniel},
  journal={International Journal of Electrical Power \& Energy Systems},
  volume={173},
  pages={111377},
  year={2025},
  publisher={Elsevier}
}

@misc{loher2026gemini,
  author       = {Loher, Lucia and Harrirou, Hussein Hassan},
  title        = {New ways to balance cost and reliability in the {Gemini API}},
  year         = {2026},
  month        = {apr},
  howpublished = {\url{https://blog.google/innovation-and-ai/technology/developers-tools/introducing-flex-and-priority-inference/}},
  note         = {Google The Keyword Blog. Accessed: 2026-06-29}
}

@article{yao2026grid,
  title={Grid-supporting equipment supply chains constrain the feasible pace of power system expansion},
  author={Yao, Boyu and Dvorkin, Yury},
  journal={Cell Reports Sustainability},
  year={2026},
  doi = {10.1016/j.crsus.2026.100923}
}

@INPROCEEDINGS{Song2025KPG,
  author={Song, Geonho and Kim, Jip},
  booktitle={2025 IEEE Power and Energy Society General Meeting (PESGM)}, 
  title={KPG 193: A Synthetic Korean Power Grid Test System for Decarbonization Studies}, 
  year={2025},
  volume={},
  number={},
  pages={1-5},
  doi={10.1109/PESGM52009.2025.11225382}}

@article{paik2024elcc,
  author  = {Paik, Chun-Hyun},
  title   = {Estimating the ELCC-based Marginal Capacity Credits of Renewable Energy in Korea},
  journal = {Journal of The Korean Solar Energy Society},
  year    = {2024},
  volume  = {44},
  number  = {4},
  pages   = {67--76},
  issn    = {1598-6411}
}

@misc{DataCenterMap2026Korea,
  author       = {{Data Center Map}},
  title        = {{South Korea Data Centers}},
  year         = {2026},
  howpublished = {\url{https://www.datacentermap.com/south-korea/}},
  note         = {Accessed: 2026-05-14}
}

@online{kang2024transmission,
  author       = {Kang, Da-eun},
  title        = {Building One Kilometer of Transmission Line Costs Up to KRW 25.8 Billion: Cost Is Also an Issue},
  organization = {The Chosun Daily},
  date         = {2024-09-13},
  url          = {https://www.chosun.com/economy/industry-company/2024/09/13/UHBU2RVL7VHMJNNEFPDSAHGYFA/},
  urldate      = {2026-08-04},
}

@article{rauh2026local,
  title={Local community opposition to data centers is clear and growing},
  author={Rauh, Adam and Reece, Mason and Simko, Tyler},
  year={2026}
}

@online{epri2025regen,
  title        = {{US-REGEN} Documentation v2025},
  organization = {{Electric Power Research Institute}},
  date         = {2025-05-23},
  url          ={https://us-regen-docs.epri.com/v2025/},
  urldate          ={2026-11-12}
}

@misc{pjmeis2025rps,
  title        = {Comparison of Renewable Portfolio Standards ({RPS}) Programs in {PJM} States},
  author       = {{PJM Environmental Information Services}},
  year         = {2025},
  note         = {15 April 2025},
  url          = {https://www.pjm-eis.com/-/media/DotCom/pjm-eis/documents/rps-comparison.pdf}
}

@techreport{pjm2028brarep,
  author       = {{PJM Interconnection}},
  title        = {{2028/2029 Base Residual Auction Report}},
  institution  = {PJM Interconnection},
  year         = {2026},
  month        = jul,
  day          = {14},
  url          = {https://www.pjm.com/-/media/DotCom/markets-ops/rpm/rpm-auction-info/2028-2029/2028-2029-bra-results-report.pdf},
}

@techreport{pjm2027brarep,
  author       = {{PJM Interconnection}},
  title        = {{2027/2028 Base Residual Auction Report}},
  institution  = {PJM Interconnection},
  year         = {2025},
  month        = dec,
  day          = {17},
  url          = {https://www.pjm.com/-/media/DotCom/markets-ops/rpm/rpm-auction-info/2027-2028/2027-2028-bra-report.pdf},
}

@misc{ferc2016order831,
  title        = {Offer Caps in Markets Operated by Regional Transmission Organizations and Independent System Operators, Order No. 831, 157 {FERC} \P~61,115},
  author       = {{Federal Energy Regulatory Commission}},
  year         = {2016}
}

@misc{nlr2025atb,
  title        = {2025 Electricity Annual Technology Baseline ({ATB})},
  author       = {{National Laboratory of the Rockies}},
  year         = {2026},
  note         = {v1 workbook, 27 August 2026},
  url          = {https://atb.nlr.gov/electricity/2025/index}
}

@misc{eia2026aeo,
  title        = {Annual Energy Outlook 2026},
  author       = {{U.S. Energy Information Administration}},
  year         = {2026},
  note         = {Supplement tables 54.10--54.13 and Table 20, 8 April 2026},
  url          = {https://www.eia.gov/outlooks/aeo/}
}

@misc{bea2026gdp,
  title        = {Gross Domestic Product (Chain-type Price Index), [A191RG3A086NBEA], retrieved from FRED},
  author       = {{U.S. Bureau of Economic Analysis}},
  year         = {2026},
  institution  = {Federal Reserve Bank of St. Louis},
  url          = {https://fred.stlouisfed.org/series/A191RG3A086NBEA}
}

@techreport{avraam2025regional,
  title={Regional Energy Deployment System (ReEDS) Model Documentation: 2025},
  author={Avraam, Charalampos and Ayad, Abdelrahman and Brown, Patrick and Carag, Vincent and Chen, Yunzhi and Chernyakhovskiy, Ilya and Cohen, Stuart and Cole, Wesley and Dhulipala, Surya and Duraes de Faria, Victor and others},
  year={2025},
  institution={National Renewable Energy Laboratory (NREL), Golden, CO (United States)}
}

@misc{pjm2026elcc,
  title        = {{Preliminary ELCC Class Ratings For Use in the Reliability Procurement Backstop (RBP)}},
  author       = {{PJM Interconnection}},
  year         = {2026},
  institution  = {PJM Interconnection},
  url = {https://www.pjm.com/-/media/DotCom/committees-groups/cifp-rbp/postings/preliminary-elcc-class-ratings-for-period-dy28-dy42.pdf}
}

@misc{eia2025861,
  title        = {Annual Electric Power Industry Report, Form {EIA-861}, detailed data files, 2025},
  author       = {{U.S. Energy Information Administration}},
  year         = {2026},
  url          = {https://www.eia.gov/electricity/data/eia861/}
}

@misc{vcea2020,
  title        = {Virginia Clean Economy Act, Va. Code \S~56-585.5, as amended through the 2026 session},
  author       = {{Commonwealth of Virginia}},
  year         = {2026},
  url          = {https://law.lis.virginia.gov/vacode/56-585.5}
}

@article{hilbers2019importance,
  title={Importance subsampling: improving power system planning under climate-based uncertainty},
  author={Hilbers, Adriaan P and Brayshaw, David J and Gandy, Axel},
  journal={Applied Energy},
  volume={251},
  pages={113114},
  year={2019},
  publisher={Elsevier}
}

@misc{epa2025needs,
  title        = {National Electric Energy Data System ({NEEDS}), rev. 11-28-2025},
  author       = {{U.S. Environmental Protection Agency}},
  year         = {2025},
  url          = {https://www.epa.gov/power-sector-modeling/national-electric-energy-data-system-needs}
}

@misc{eia2026860m,
  title        = {Preliminary Monthly Electric Generator Inventory (Form {EIA-860M}), August 2026},
  author       = {{U.S. Energy Information Administration}},
  year         = {2026},
  url          = {https://www.eia.gov/electricity/data/eia860m/}
}

@misc{pjm2026queuecycle,
  title        = {Cycle Service Request Status, exports of 23 October 2026},
  author       = {{PJM Interconnection}},
  year         = {2026},
  url          = {https://www.pjm.com/planning/m/cycle-service-request-status}
}

@misc{pjm2026queueserial,
  title        = {Serial Service Request Status, exports of 23 and 24 October 2026},
  author       = {{PJM Interconnection}},
  year         = {2026},
  url          = {https://www.pjm.com/planning/service-requests/serial-service-request-status}
}

@techreport{pjm2026load,
  title        = {2026 Load Forecast Report},
  author       = {{PJM Interconnection}},
  year         = {2026},
  institution  = {PJM Interconnection},
  url          = {https://www.pjm.com/-/media/DotCom/library/reports-notices/load-forecast/2026-load-report.pdf}
}

@article{scott2019clustering,
  title={Clustering representative days for power systems generation expansion planning: Capturing the effects of variable renewables and energy storage},
  author={Scott, Ian J and Carvalho, Pedro MS and Botterud, Audun and Silva, Carlos A},
  journal={Applied Energy},
  volume={253},
  pages={113603},
  year={2019},
  publisher={Elsevier}
}

@misc{kpx2025essTender,
  author = {{Korea Power Exchange}},
  title = {First Competitive Tender for the {ESS} Central Contract Market in 2025 (in {Korean})},
  year = {2025},
  month = may,
}

@inproceedings{choudhury2024mast,
  title={{MAST: Global Scheduling of ML Training across Geo-Distributed Datacenters at Hyperscale}},
  author={Choudhury, Arnab and Wang, Yang and Pelkonen, Tuomas and Srinivasan, Kutta and Jain, Abha and Lin, Shenghao and David, Delia and Soleimanifard, Siavash and Chen, Michael and Yadav, Abhishek and others},
  booktitle={18th USENIX Symposium on Operating Systems Design and Implementation (OSDI 24)},
  pages={563--580},
  year={2024}
}

@misc{pjm2026iras,
  author       = {{PJM Interconnection}},
  title        = {{Interim Resource Adequacy Service Proposal}},
  year         = {2026},
  month        = aug,
  howpublished = {Fact sheet},
  note         = {Proposal filed with FERC on 13 August 2026},
  url          = {https://www.pjm.com/-/media/DotCom/about-pjm/newsroom/fact-sheets/interim-resource-adequacy-service-proposal.pdf},
  urldate      = {2026-11-22}
}

@misc{ferc2026pjmlargeload,
  author       = {{Federal Energy Regulatory Commission}},
  title        = {{PJM Interconnection, L.L.C., Order Accepting Tariff Revisions in Part and Rejecting Tariff Revisions in Part}},
  howpublished = {Docket No. ER26-3515-000},
  year         = {2026},
  month        = oct,
  note         = {Issued 9 October 2026},
  url          = {https://www.pjm.com/pjmfiles/directory/etariff/FercOrders/9207/20261009-er26-3515-000.pdf}
}

@techreport{DOE2025GridReliability,
  author      = {{U.S. Department of Energy}},
  title       = {Evaluating the Reliability and Security of the United States Electric Grid},
  institution = {U.S. Department of Energy},
  year        = {2025},
  month       = {jul},
  type        = {Resource Adequacy Report},
  address     = {Washington, DC},
  note        = {Revised October 27, 2025; prepared in response to Executive Order 14262},
  url         = {https://www.energy.gov/sites/default/files/2025-11/DOE%20Final%20EO%20Report%20%28REVISED%20OCT%2027%29.pdf}
}

\clearpage

\paragraph{Supplementary Figures}

\begin{figure*}[!htbp]
\centering
\begin{subfigure}[t]{0.96\textwidth}
\centering
\includegraphics[width=\linewidth]{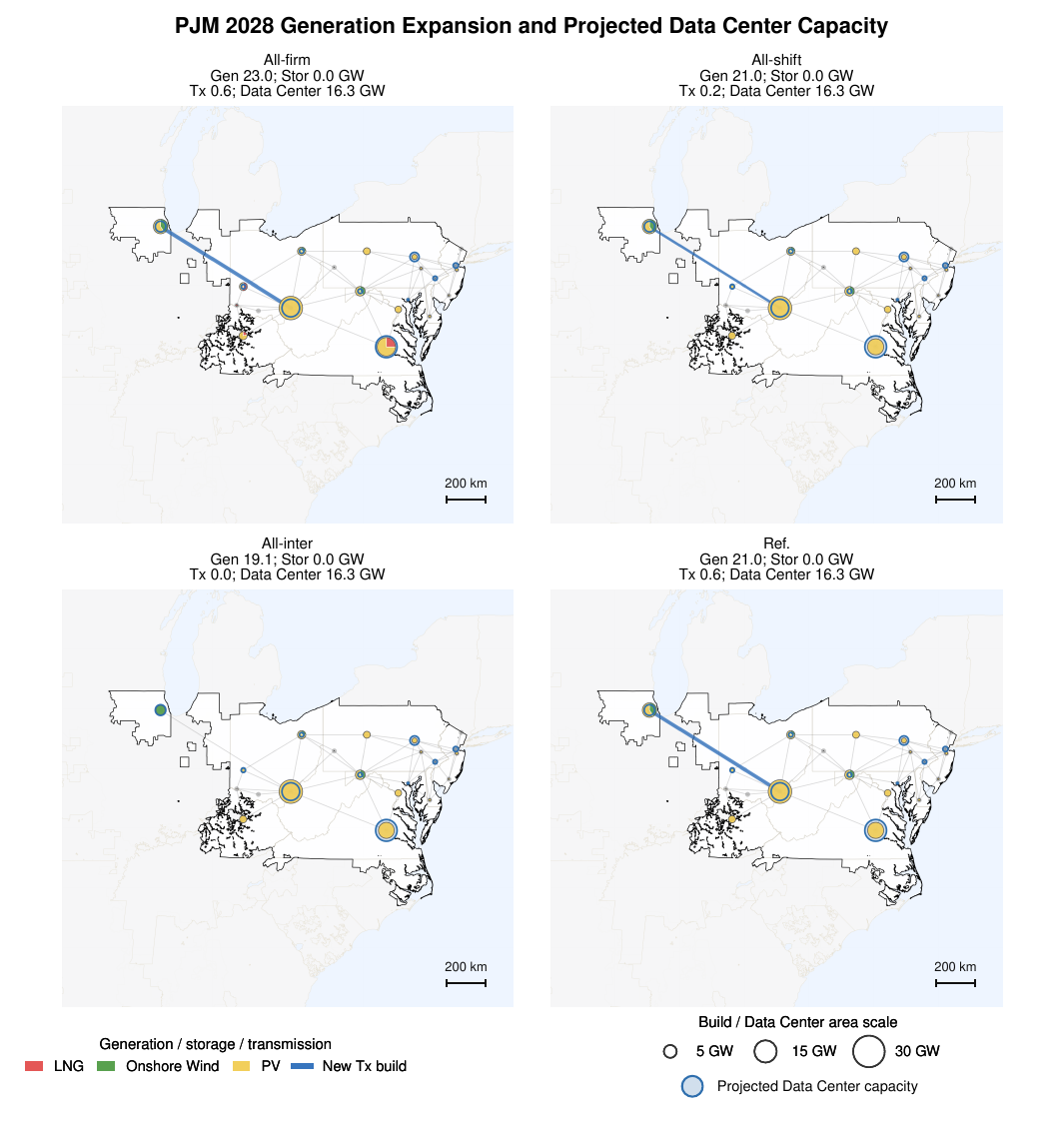}
\end{subfigure}
\caption{\textbf{PJM expansion, 2028.} Generation, storage and
transmission additions by zone under the all-firm baseline and the three
flexibility cases. \rev{(All-shift, All-inter and the reference tier mix, labelled Ref.). The all-firm case adds \res{2.0~GW} of gas and the other cases none. Solar and onshore wind are built to the limit of the queue except in the all-interruptible case, no storage is built, and the only new transmission, \res{0.6~GW} or less, is on the AEP--ComEd transmission line.}}
\label{fig:PJMexpS1}
\end{figure*}

\begin{figure*}[!htbp]
\centering
\begin{subfigure}[t]{0.96\textwidth}
\centering
\includegraphics[width=\linewidth]{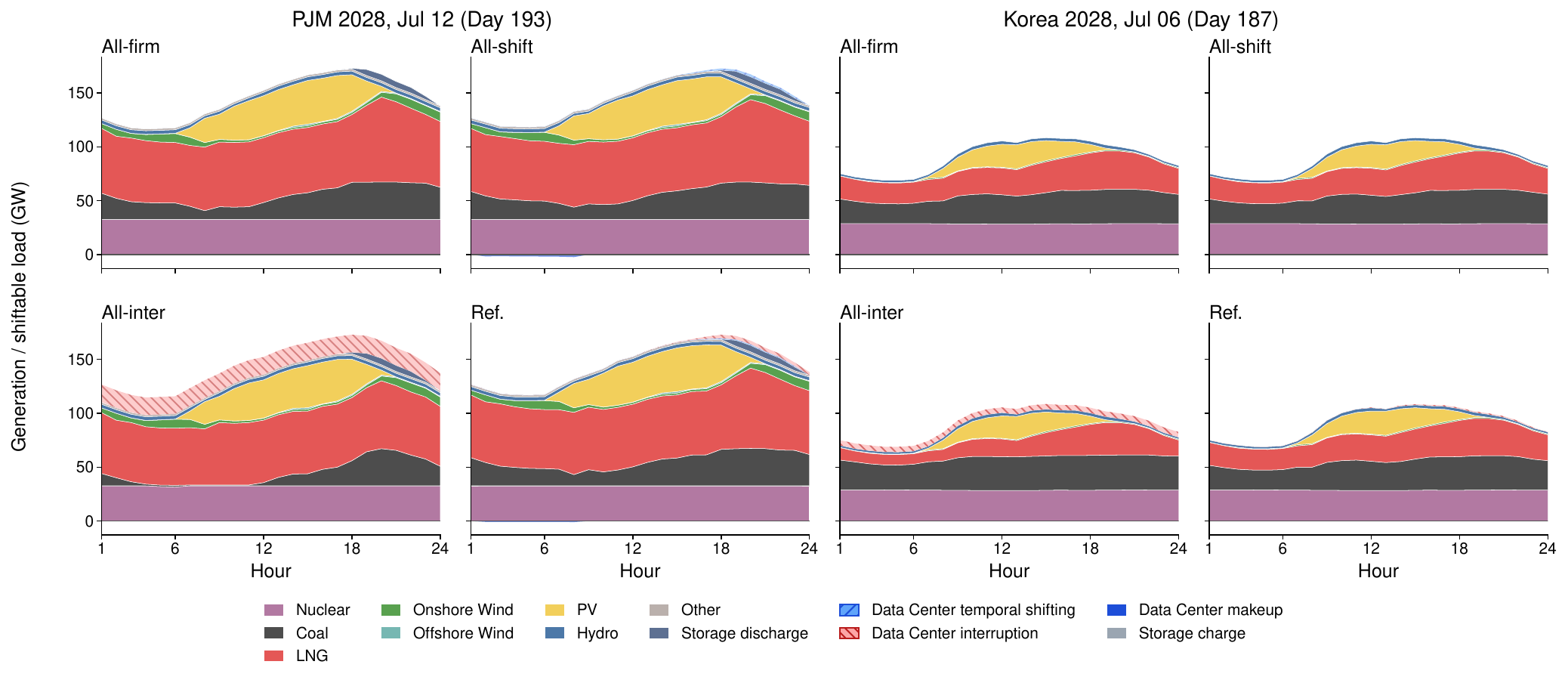}
\caption{Generation dispatch \rev{on the peak net-demand day} (2028) 
}
\end{subfigure}
\vspace{2mm}
\begin{subfigure}[t]{0.96\textwidth}
\centering
\includegraphics[width=\linewidth]{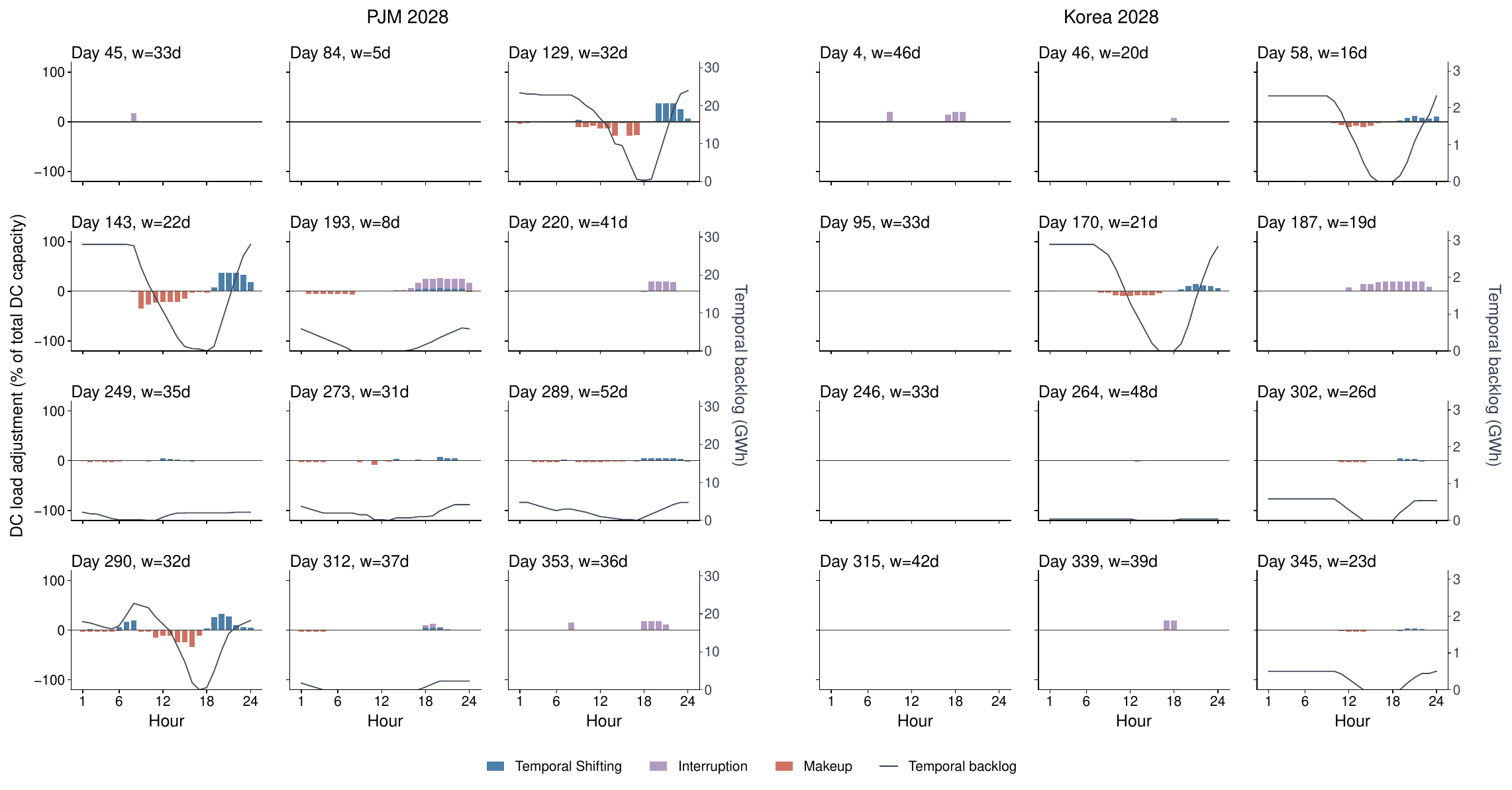}
\caption{\rev{Temporal shifting (deferral)}, make-up and interruption \rev{on the 12 representative days} (2028).}
\end{subfigure}
\caption{\textbf{Flexible \rev{data-center operation} in 2028.} \rev{(a) Generation stacks on each grid's peak net-demand day. (b) Temporal shifting (deferral), make-up and interruption in the reference tier mix on each representative day, as a percentage of total data-center (DC) load capacity, with the backlog in black (right axis, GWh); $w$ is the day's weight in days. The shiftable tier is used lightly in both systems in 2028 (\res{2.8~TWh} deferred in PJM and \res{0.13~TWh} in Korea, against \res{12.2} and \res{9.0~TWh} in 2038, Fig.~\ref{fig:flex_mechanisms_S2}), and interruption is concentrated in the evening hours of a few days.}}
\label{fig:flex_mechanisms_S1}
\end{figure*}

\begin{figure}[!htbp]
\centering
\includegraphics[width=\linewidth]{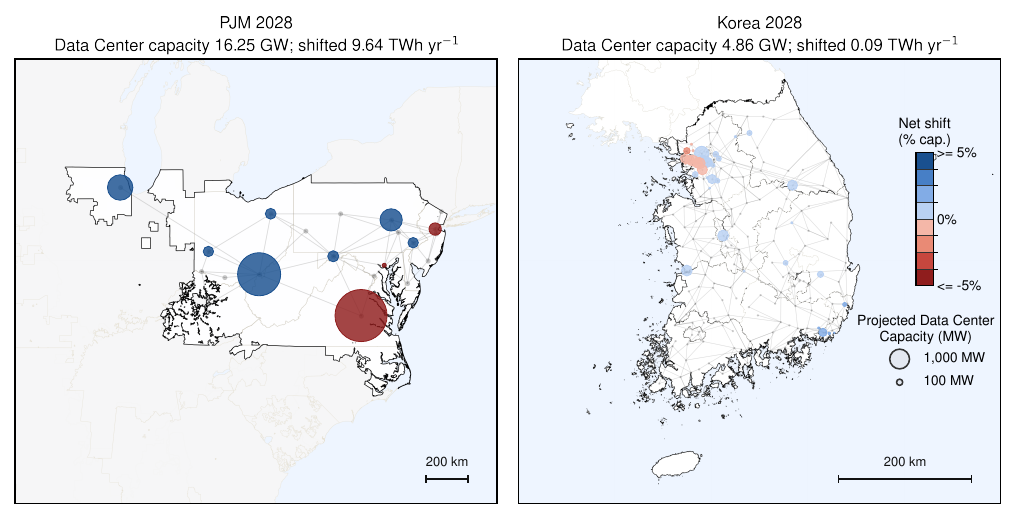}
\caption{\textbf{\rev{Spatial shifting} of data-center load with the 2028 \rev{reference}
mix.} Circle size is the projected data-center load at each node; color is the
net spatial shift relative to that node's capacity\rev{, blue for net inflow and red for net outflow. PJM relocates \res{9.6~TWh}, \res{92\%} of it out of Dominion and \res{65\%} of it into AEP, while Korea relocates \res{0.09~TWh}.}}
\label{fig:spatial_s1}
\end{figure}

\begin{figure*}[!htbp]
\centering
\begin{subfigure}[t]{0.96\textwidth}
\centering
\includegraphics[width=\linewidth]{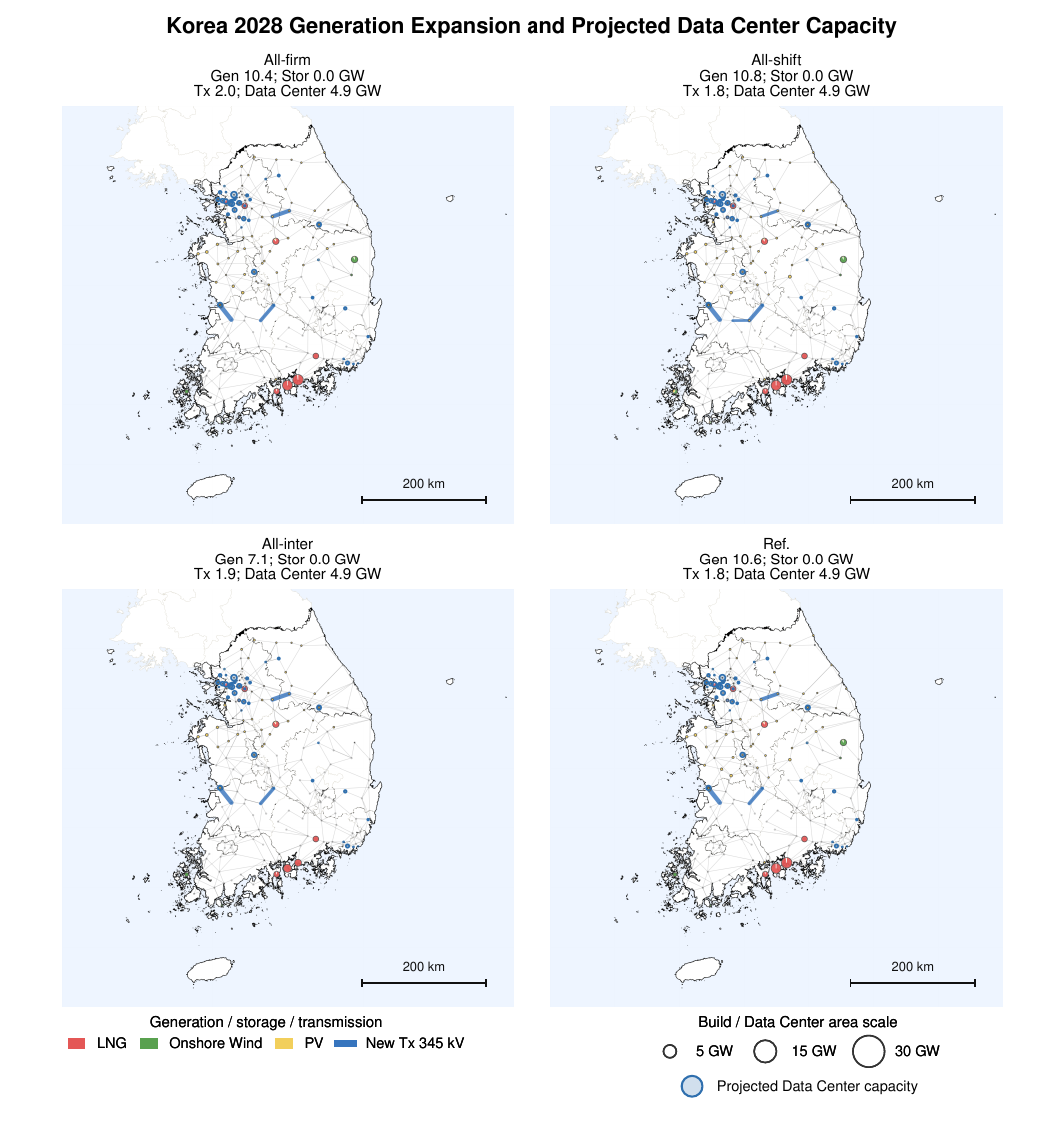}
\end{subfigure}
\caption{\textbf{Korea expansion, 2028.} Generation, storage and
transmission additions by zone under the All-firm baseline and the three
flexibility cases (\rev{All-shift}, All-inter, \rev{and the reference tier mix, labelled Ref.). New generation is \res{10.4--10.8~GW} except in the all-interruptible case (\res{7.1~GW}), no storage is built, and transmission line additions are about \res{2~GW} in every case.}.}
\label{fig:KPGexpS1}
\end{figure*}

\begin{figure*}[!htbp]
\centering
\begin{subfigure}[t]{0.96\textwidth}
\centering
    \includegraphics[width=\linewidth]{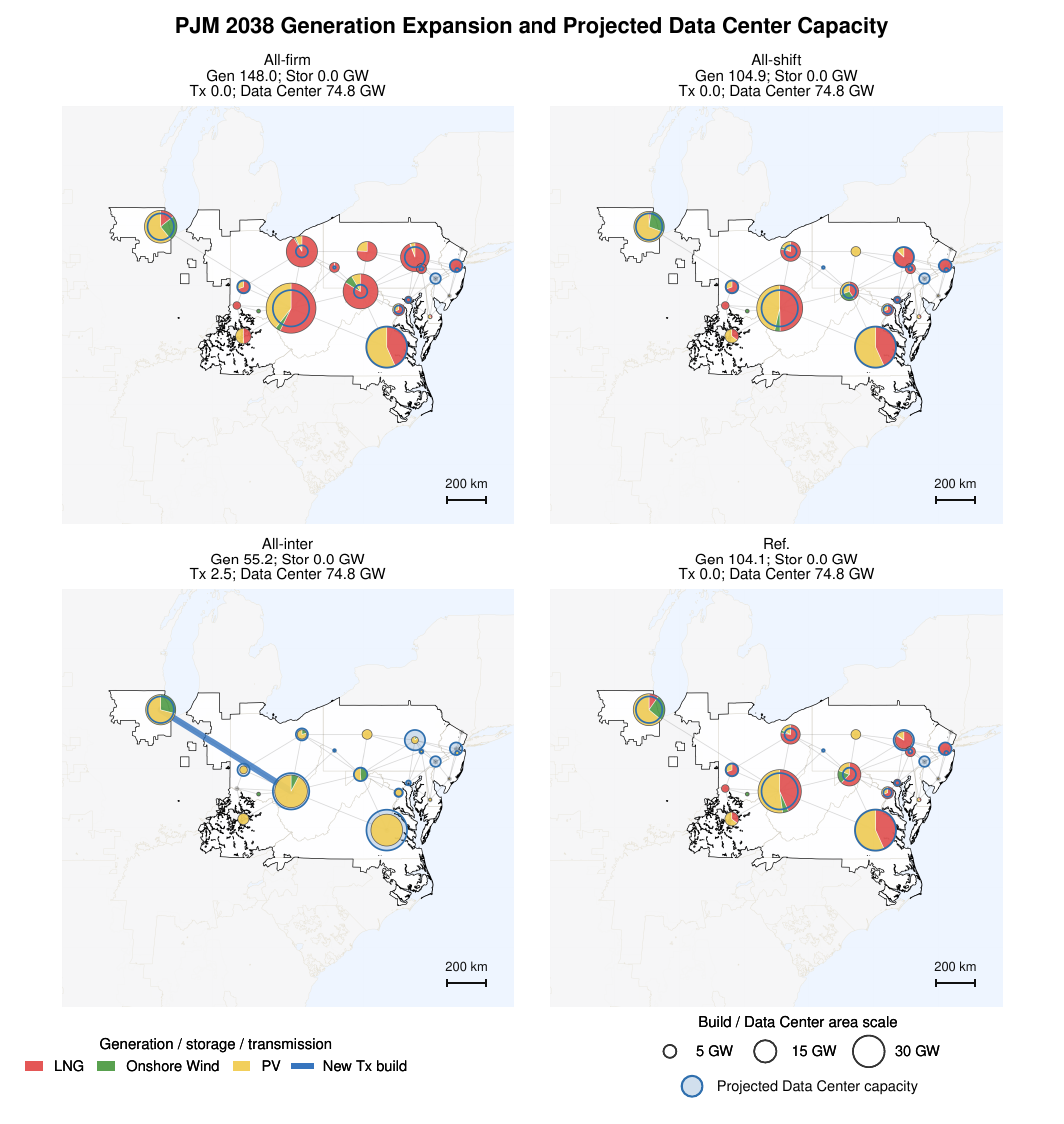}
\end{subfigure}
\caption{\textbf{PJM expansion, 2038.} Generation, storage and
transmission additions by zone under the All-firm baseline and the three
flexibility cases (\rev{All-shift}, All-inter, \rev{and the reference tier mix, labelled Ref.). New generation falls from \res{148~GW} in the All-firm case to \res{104~GW} in the reference tier mix, all of the difference gas. No storage is built, and transmission is added only in the all-interruptible case (\res{2.5~GW}).}}
\label{fig:PJMexpS2}
\end{figure*}

\begin{figure*}[!htbp]
\centering
\begin{subfigure}[t]{0.96\textwidth}
\centering
\includegraphics[width=\linewidth]{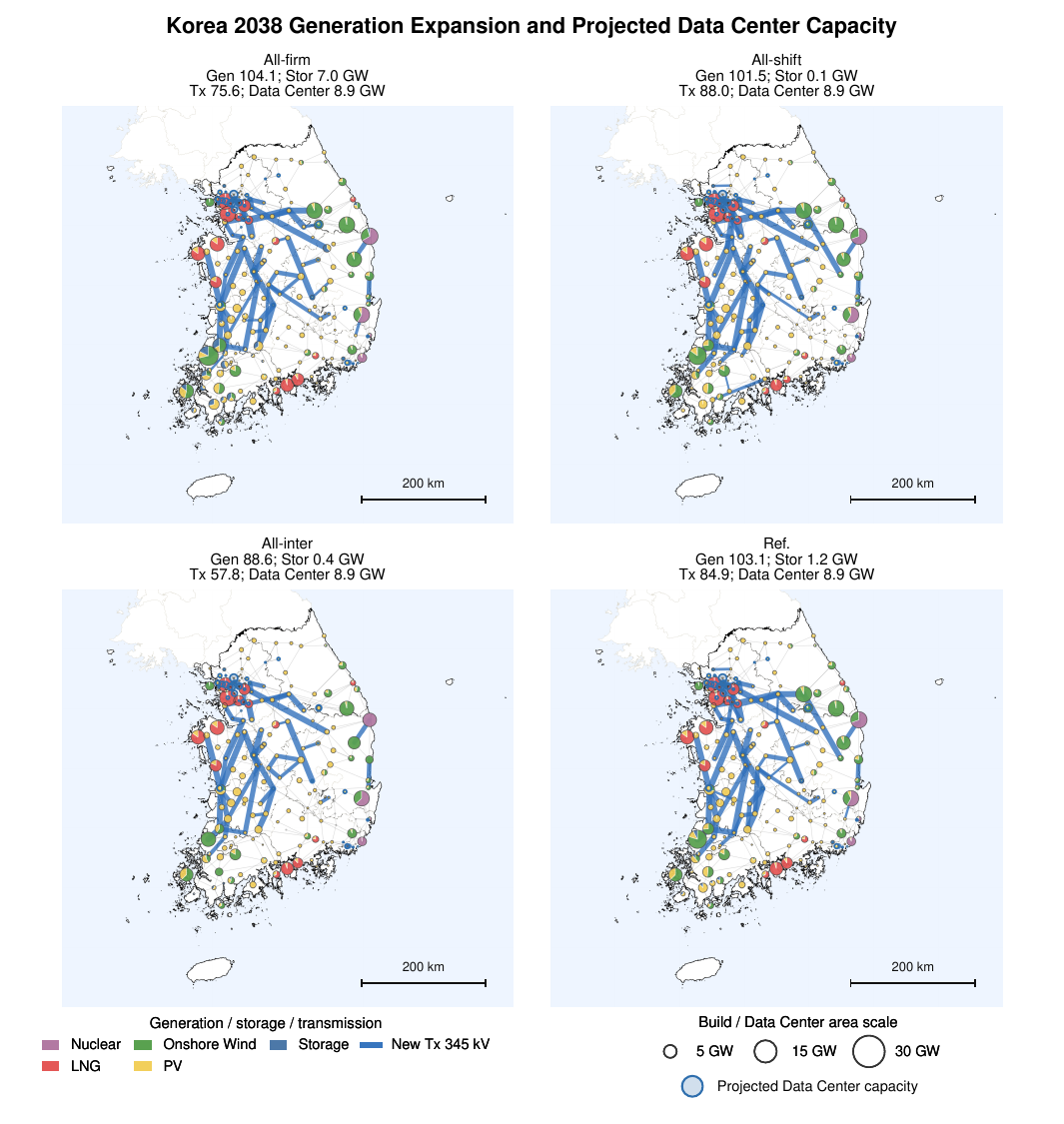}
\end{subfigure}
\caption{\textbf{Korea expansion, 2038.} Generation, storage and
transmission additions by zone under the All-firm baseline and the three
flexibility cases (\rev{All-shift}, All-inter, \rev{and the reference tier mix, labelled Ref.). Storage falls from \res{7.0~GW} in the All-firm case to \res{1.2~GW} in the reference tier mix, while transmission line additions rise from \res{76} to \res{85~GW}.}}
\label{fig:KPGexpS2}
\end{figure*}

\begin{figure*}[!htbp]
\centering
\begin{subfigure}[t]{0.96\textwidth}
\centering
\includegraphics[width=\linewidth]{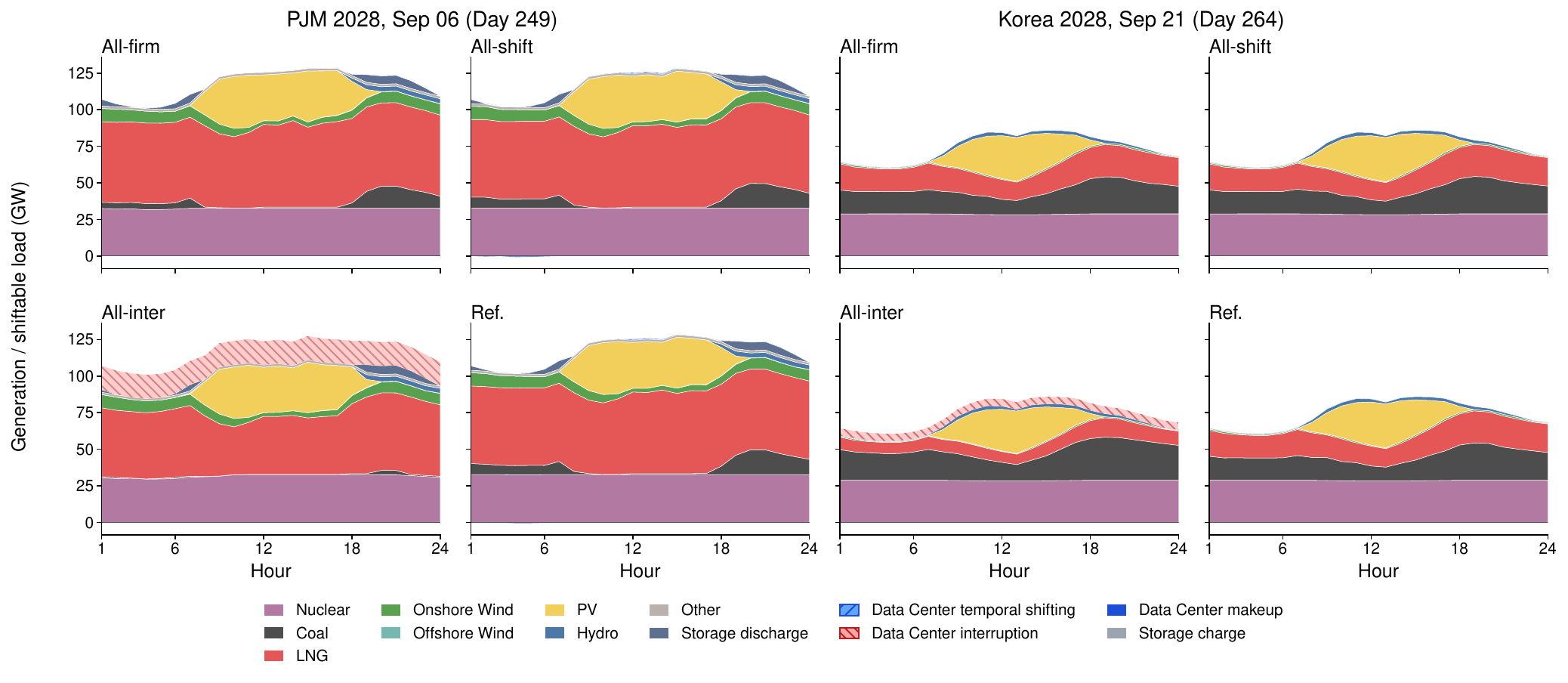}
\caption{Generation dispatch (2028)}
\end{subfigure}
\vspace{2mm}
\begin{subfigure}[t]{0.96\textwidth}
\centering
\includegraphics[width=\linewidth]{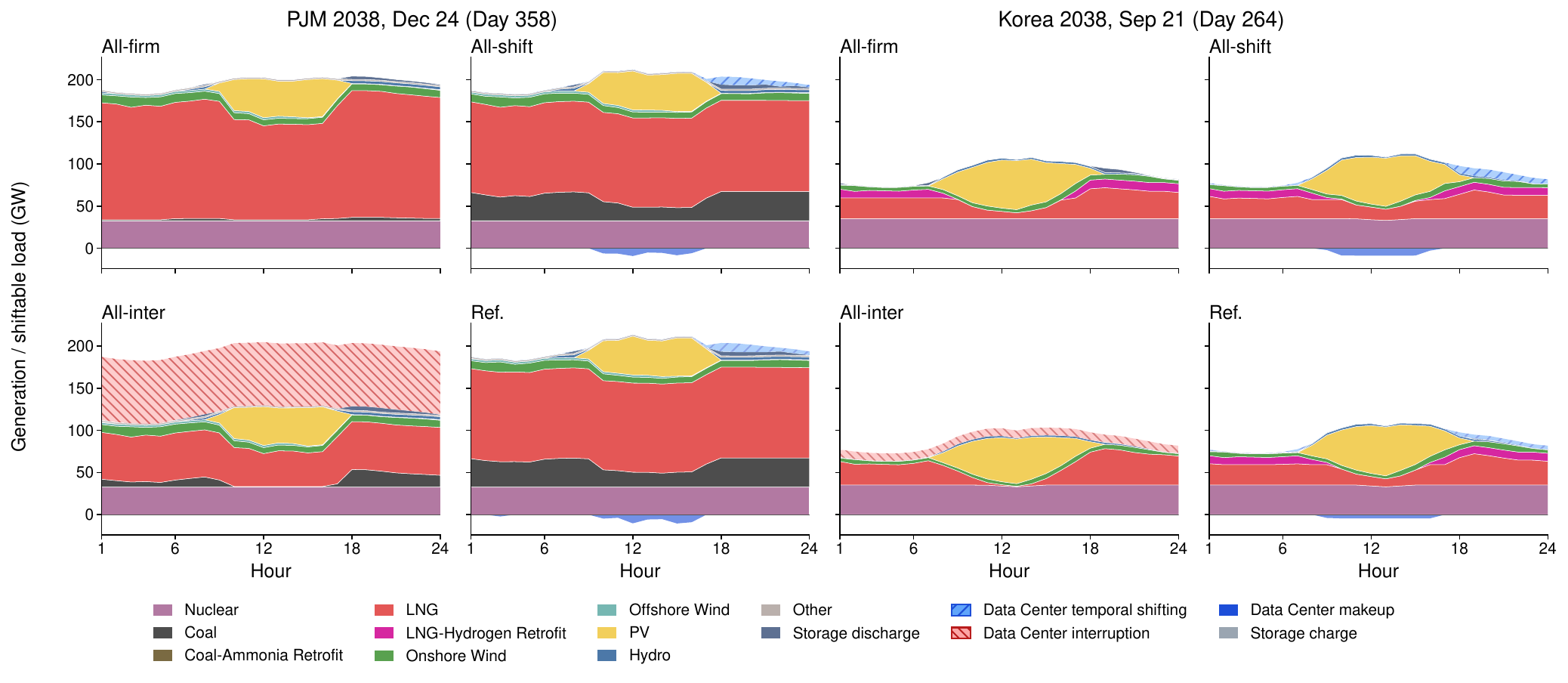}
\caption{Generation dispatch (2038)}
\end{subfigure}
\caption{\textbf{Generation dispatch on a median-load day.} Dispatch stacks for
2028 (a) and 2038 (b) \rev{in both systems}, \rev{on the representative day whose energy is closest to the median,} for comparison with the summer peak-day
stacks in Fig.~\ref{fig:flex_mechanisms_S2} and Fig.~\ref{fig:flex_mechanisms_S1}.
Flexibility is used less on a median day than at the peak, when the binding
constraints are tightest.}
\label{fig:general_dispatch}
\end{figure*}

\begin{figure*}[!htbp]
\centering
\centering
\includegraphics[width=0.95\linewidth]{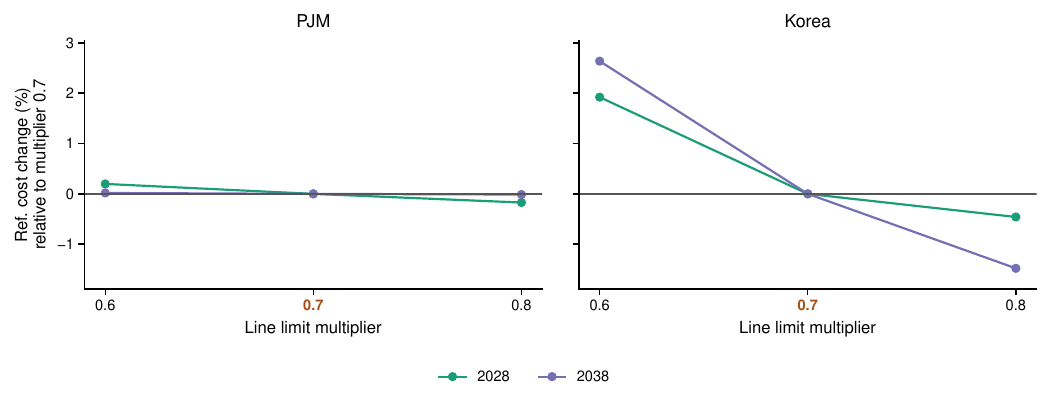}
\caption{\textbf{Transmission derating sensitivities.} Total cost for the
\rev{reference tier} mix as the contingency factor $K$ varies from 0.6 to 0.8\rev{, relative to $K=0.7$}.  Korea's
2038 cost rises by \rev{\res{2.6\%}} as $K$ falls from 0.7 to 0.6, whereas PJM's \rev{cost changes by at most \res{0.2\%}. PJM is represented by 21 zones, so congestion inside a zone is not resolved.}}
\label{fig:tx_sweep}
\end{figure*}

\begin{figure*}[!htbp]
\centering
\begin{subfigure}[t]{0.96\textwidth}
\centering
\includegraphics[width=\linewidth]{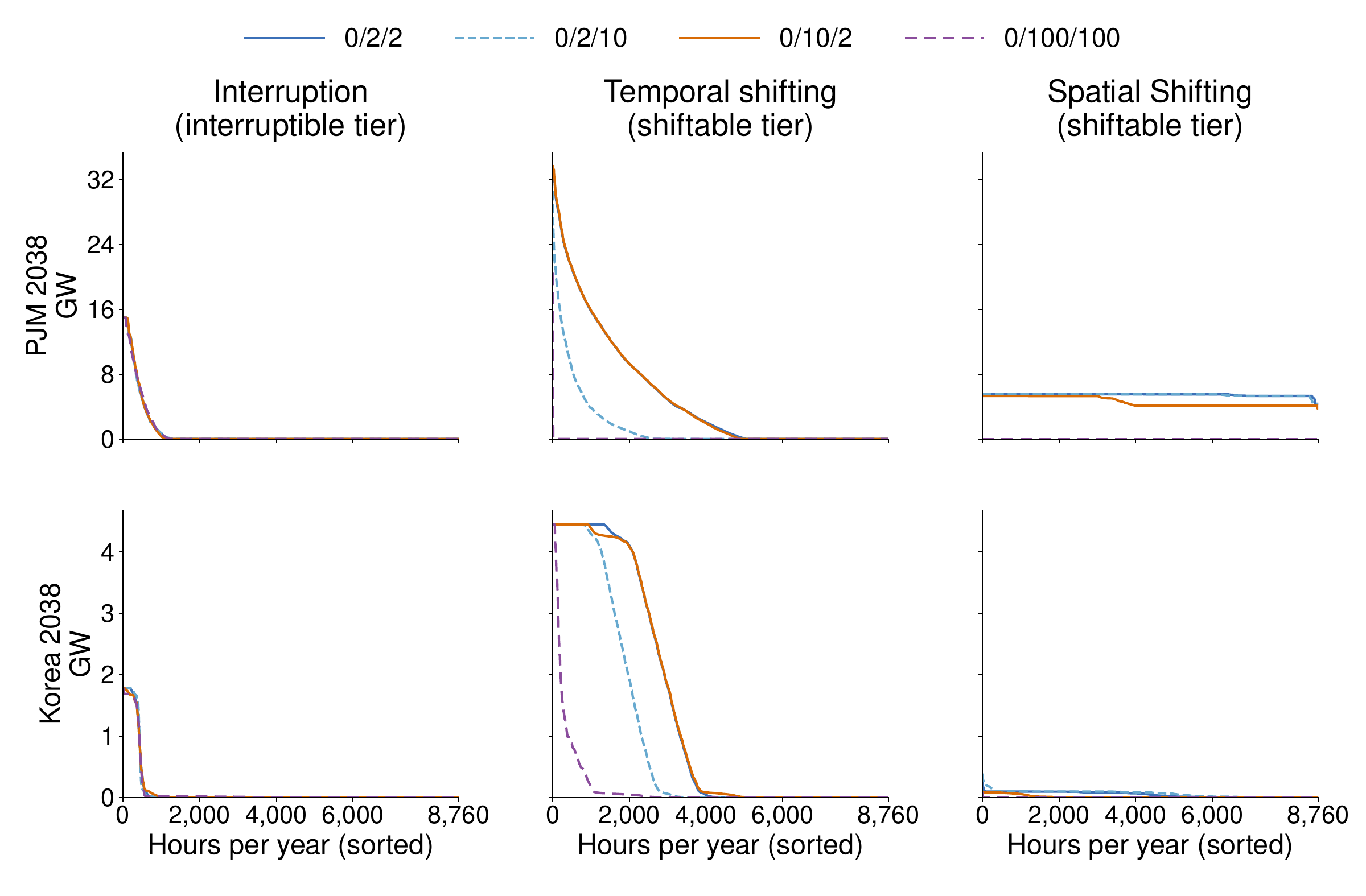}
\end{subfigure}
\caption{\rev{\textbf{Duration curves of data-center flexibility in full-year runs, 2038.} Hourly interruption (interruptible tier) and temporal and spatial shifting (shiftable tier) of the reference tier mix in PJM (top) and Korea (bottom), each sorted from highest to lowest over 8,760 hours. Lines show four sets of flexibility costs, written as interruption/spatial/temporal in \$/MWh: the base case (0/2/10), a lower temporal-shifting cost (0/2/2), a higher spatial-shifting cost (0/10/2) and a high cost for both shifting mechanisms (0/100/100). Each mechanism is sorted independently, temporal shifting counts deferred energy and excludes make-up, and relocated energy is counted once per transfer.}}
\label{fig:duration_curves}
\end{figure*}

\begin{figure*}[!htbp]
\centering
\centering
\includegraphics[width=\linewidth]{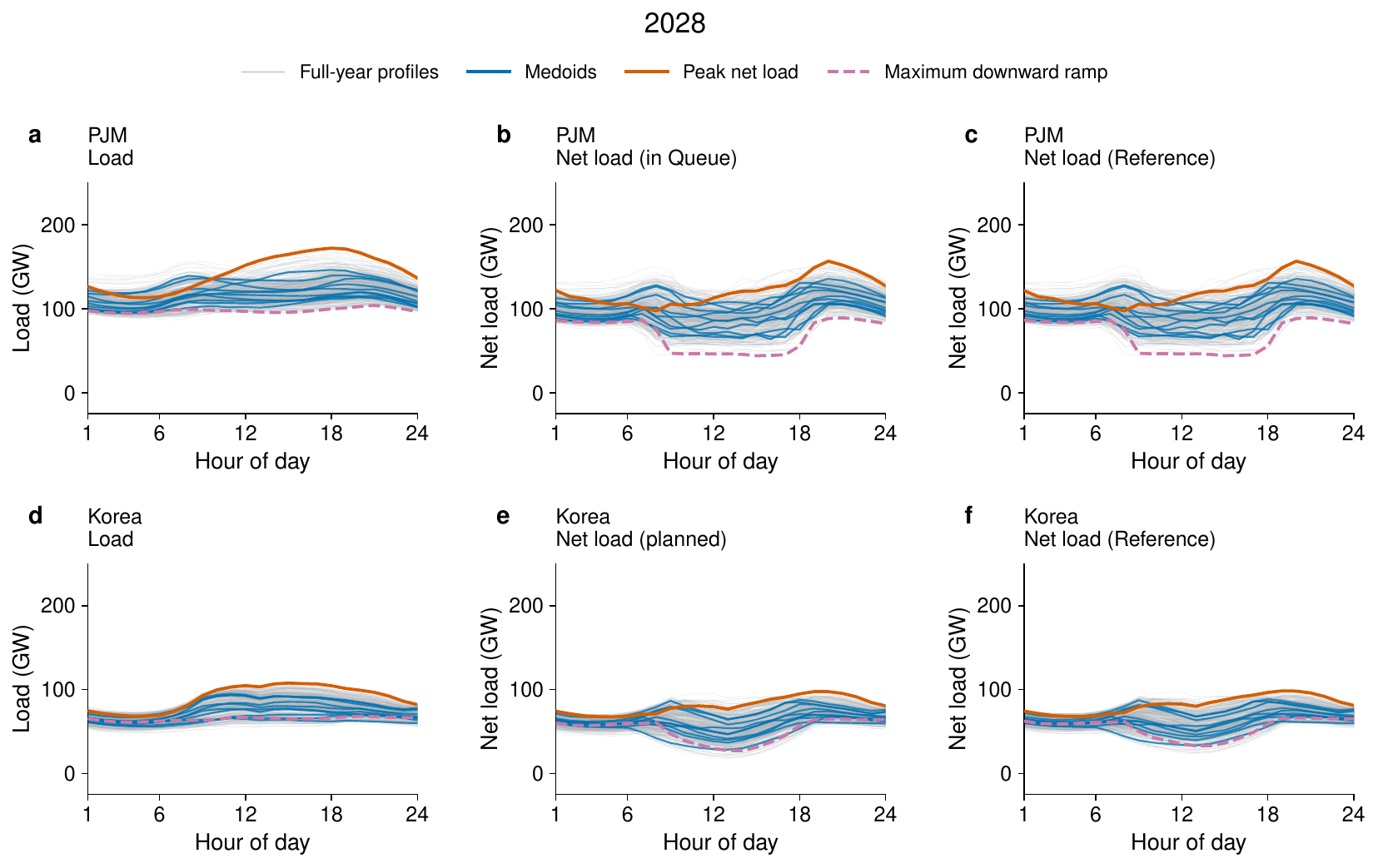}
\caption{\textbf{Representative days in 2028.} \rev{(a,d) Gross load. (b,e) Net load using existing plus queued wind and solar capacity in PJM, or planned capacity in Korea. (c,f) Net load using existing capacity plus optimized Reference case buildouts. Gray lines show all 365 days and blue lines show  medoids, orange the peak-net-load day, and dashed magenta the maximum downward-ramp day. Both extreme days are selected using the capacity in queue (PJM) or planned (Korea).}}
\label{fig:repdays_2028}
\end{figure*}

\end{document}